\documentclass{jfm_modified}

\usepackage{amsmath,amssymb}
\usepackage{mathrsfs}
\usepackage{xcolor}
\usepackage{subfigure}
\usepackage{bbm}
\usepackage{bm}
\usepackage{pict2e} 
\usepackage{booktabs}
\usepackage{tabularx}
\newcolumntype{C}{>{$}c<{$}} 

\usepackage{setspace}
\usepackage{mathtools}
\usepackage{tikz}

\usepackage{graphicx,epstopdf}
\usepackage{soul}
\setstcolor{red}

\usepackage[colorlinks=true, linkcolor=blue, urlcolor=blue, citecolor=blue]{hyperref}
\usepackage{natbib}

\usepackage{multirow,bigdelim}
\usepackage{makecell}
\colorlet{dg}{green!80!black}

\title{$H$-theorem and boundary conditions for the linear R26
equations: 
application to flow past an evaporating droplet}

\shorttitle{$H$-theorem and boundary conditions for the linear R26 equations} 
\shortauthor{A. S. Rana, V. K. Gupta, J. E. Sprittles, M. Torrilhon} 

\author
 {
 {Anirudh S. Rana\aff{1}
\corresp{
 \email{anirudh.rana@pilani.bits-pilani.ac.in}
},
}
 Vinay Kumar Gupta\aff{2},
 James E. Sprittles\aff{3},
 \and
 Manuel Torrilhon\aff{4}
 }

\affiliation
{
\aff{1}
Department of Mathematics, BITS Pilani, Pilani Campus, 333031 Rajasthan, India
\aff{2}
Department of Mathematics, Indian Institute of Technology Indore, Indore 453552, India
\aff{3}
Mathematics Institute, University of Warwick, Coventry CV4 7AL, UK
\aff{4}
Department of Mathematics, RWTH Aachen University, D-52062 Aachen, Germany
}

\begin{document}

\maketitle

\begin{abstract} 
Determining physically admissible boundary conditions for higher moments in an extended continuum model is recognised as a major obstacle.  
Boundary conditions for the regularised 26-moment (R26) equations obtained using Maxwell's accommodation model do exist in the literature; however, we show in this article that these boundary conditions violate the second law of thermodynamics and the Onsager reciprocity relations for certain boundary value problems, and hence are not physically admissible. 
We further prove that the linearised R26 (LR26) equations possess a proper $H$-theorem (second-law inequality) by determining a quadratic form without cross-product terms for the entropy density. 
The establishment of the $H$-theorem for the LR26 equations in turn leads to a complete set of boundary conditions that are physically admissible for all processes and comply with the Onsager reciprocity relations. 
As an application, the problem of a slow rarefied gas flow past a spherical droplet with and without evaporation is considered and solved analytically. 
The results are compared with the numerical solution of the linearised Boltzmann  equation, experimental results from the literature and/or other macroscopic theories to show that 
the LR26 theory with the physically admissible boundary conditions provides an excellent prediction up to Knudsen number $\lesssim 1$ and, consequently, provides transpicuous insights into intriguing effects, such as thermal polarisation. In particular, the analytic results for the drag force obtained in the present work are in an excellent agreement with experimental results even for very large values of the Knudsen number.
\end{abstract}
\section{Introduction}

%
%
The behaviour of aerosols (small droplets
suspended in vapour/gases) covers a wide range of phenomena of interest to
both scientists and engineers \citep{Sergei2014}. 
In the context of airborne disease
transmission \citep{Tang2006, Xie2007}, large droplets ($\gtrsim \mu$m in radius) fall rapidly to the ground due to gravity, and are therefore transmitted over short distances;
smaller droplets ($\lesssim \mu$m in radius), on the other hand, remain suspended in the air for a significant period of time, and hence are
transmitted over large distances. 
The immediate intuitive questions are ``How far does a droplet travel?'' and ``How long does a droplet survive?'' 
The former is answered by evaluating the momentum of the droplet while the latter by the heat and mass transfer properties of the droplet, with the two being inevitably coupled.
At small scales, it is difficult to measure the relevant quantities (e.g. drag forces) on single and multiple droplets; therefore, an accurate modelling is key to understanding transport phenomena (the trajectory and thermal state)
during evaporation. 

When a spherical particle of radius $\hat{R}_0$ moves slowly with a steady translational velocity in a gas, with $\hat{R}_0$ being much larger than the mean free path ($\hat{\lambda}$) of the gas, the flow fields around the particle can be obtained by solving the Stokes equation, and therefore one can obtain the drag force acting on the sphere \citep{Lamb1932}. 
However, the Stokes equation as well as the other classical continuum theories of gas dynamics based on the Euler or Navier-Stokes-Fourier (NSF) equations, which are in general accurate for describing a flow in the so-called hydrodynamic regime (Knudsen number $\mathrm{Kn}\lesssim 0.001$), cannot explain the gas flow past a micro-/nano-sized drop (for which $\mathrm{Kn}=\hat{\lambda}/\hat{R}_0 \sim 1$)
accurately \citep{Cercignani2000, Sone2002, Struchtrup2005}.
Such flows can only be predicted by employing tools of kinetic theory.
%
%
%
%
%
%
%
%
%

Within the framework of
kinetic theory, the vapour/gas is described by the Boltzmann equation \citep{Cercignani1975, Kremer2010}, which is the evolution equation for the (velocity) distribution function of the gas and accurately
describes flows for all Knudsen numbers. 
Notwithstanding, the exact analytical solution of the Boltzmann equation is intractable for a general flow problem while numerical solutions of the Boltzmann equation demand a very high computational cost in general (particularly, for transition-regime flows). 
Although, very recently, some general synthetic iterative schemes have been proposed for solving the Boltzmann equation \citep{SZW2021, ZhuetalJCP2021} that reduce the computational cost of numerical solutions of the Boltzmann equation significantly, it is still appreciable to describe a process with a set of partial differential equations in physical variables (or moments of the distribution function) as they provide a deeper insight to the underlying physics of the process and permit the deployment of classical mathematical techniques.
A variety of approximation techniques, aiming to derive sets of partial differential equations that can describe processes at moderate Knudsen numbers, have been developed; see e.g.~\cite{Struchtrup2005, TorrilhonARFM} and references therein.
Among others, the two well-known approximation techniques are the Chapman--Enskog (CE) expansion \citep{CC1970} and method of moments \citep{Grad1949b}.


The CE expansion is suitable for near-equilibrium gas flows and relies on an asymptotic expansion of the distribution function around the equilibrium distribution function (which is a Maxwellian) in powers of a small parameter---typically the Knudsen number. 
The distribution function in the Boltzmann equation is replaced with this expanded distribution function, and coefficients of each power of the Knudsen number on both sides of the resulting equation are compared. 
The comparison of coefficients of $\mathrm{Kn}^0$ yields the Euler equations, of $\mathrm{Kn}^1$ results into the NSF equations, of $\mathrm{Kn}^2$ leads to the Burnett equations, of $\mathrm{Kn}^3$ gives 
the so-called super-Burnett equations, and so on \citep{Struchtrup2005, Bobylev2018}. 
The Burnett (also the super-Burnett and beyond) equations  are unstable \citep{Bobylev1982}, and due to lack of any coherent entropy inequality these equations lead to physically inadmissible solutions \citep{SN2020}.
%

Grad---in a seminal work \citep{Grad1949b}---proposed an asymptotic solution procedure for the Boltzmann equation through the moment equations.
The method is referred to as Grad's moment method and derives the governing equations for the macroscopic quantities, namely mass density, velocity, temperature, stress, heat flux and other higher-order moments, through the Boltzmann equation since these macroscopic quantities are related to the distribution function via its moments. 
The well-known Grad 13-moment equations, at least in the linearised case, are accompanied with a proper entropy inequality \citep{RanaStruchtrup2016} and are stable.
%
%

Over the last few decades, a significant progress 
has been made towards improving Grad's approach while keeping their advantageous features \citep[see e.g.][]{KGDN1998, Ohr2001, StruchtrupTorrilhon2003, Struchtrup2004, Caietal2014, Karlin2018, CW2020}. 
In particular, Struchtrup \& Torrilhon (\citeyear{StruchtrupTorrilhon2003}) proposed a regularisation procedure that derives a closed set of moment equations---referred to as the regularised 13-moment (R13) equations.
The R13 equations, since their derivation, have paved the way to several developments, including numerical frameworks \citep[see e.g.][]{KST2014, CSRSL2017, CW2020}. 
It can be shown via the order of magnitude approach \citep{Struchtrup2004} that the R13 equations possess the accuracy of order $\mathrm{Kn}^3$ whereas the Euler, NSF and Grad 13-moment equations have the accuracy of order $\mathrm{Kn}^0$, $\mathrm{Kn}^1$ and $\mathrm{Kn}^2$, respectively \citep{Struchtrup2004, Struchtrup2005, TorrilhonARFM}.
%
%
It is generally accepted that to maintain accuracy, the number of moments ought to be increased with the Knudsen number. 
Aiming to increase the applicability of the regularisation procedure to describe processes in the moderate transition regime, Gu \& Emerson (\citeyear{GuEmerson2009}), by following the approach of \cite{StruchtrupTorrilhon2003}, derived the regularised 26-moment (R26) equations. 
The R26 equations are anticipated to possess the accuracy of order $\mathrm{Kn}^5$ in the bulk region (i.e.~outside of the Knudsen layer) \citep{Struchtrup2004, Struchtrup2005}.


Boundary conditions complementing the Grad or regularised moment equations are typically obtained by employing the Maxwell accommodation model \citep{Maxwell1879} along with Grad's closure at the boundary yielding the so-called macroscopic boundary conditions (MBC).
The paper by \cite{Gu&Emerson2007} may be regarded as the pioneering work on wall boundary conditions for the R13 equations using the Maxwell accommodation model. Nevertheless, some inconsistencies and oscillations were reported in this work. 
To overcome these shortcomings, \citet{TorrilhonStruchtrupJCP2008} proposed a theory of boundary conditions for the R13 equations using the Maxwell accommodation model but taking physical and mathematical requirements for the system into account.
Moreover, following the approach of \citet{TorrilhonStruchtrupJCP2008}, the boundary conditions for the R26 equations---based on the Maxwell accommodation model---were obtained in \cite{GuEmerson2009}. 
The boundary conditions derived through the Maxwell accommodation model are generically referred to as the Maxwell boundary conditions.
Solutions of the R13 and R26 equations along with the Maxwell boundary conditions have been obtained for some benchmark problems \citep{Taherietal2009, StruchtrupTaheri2011, GuEmersonTang2010, Gu&Emerson2014, CSRSL2017, gu_barber_john_emerson_2019} in order to investigate rarefaction effects in gas at moderate Knudsen numbers.
Recently, a hybrid approach combining the discrete velocity method with the MBC in the near-wall region, where the  moment method is expected to be less accurate due to strong nonequilibrium effects, has been proposed by \citet{YangetalJCP2020}. 
The approach not only yields more accurate solutions for wall-bounded flows but also reduces the computational cost in numerical simulations.

Thus far, analytical solutions for evaporation and condensation problems have been
obtained from the linearised R13 (LR13) \citep{BRTS2018} and linearised R26 (LR26) \citep{RanaLockerbySprittles2018} equations and compared with
numerical solutions from kinetic theory. 
The R26 equations can capture Knudsen layers---a region of strong non-equilibrium 
within a few mean free paths of the boundary---more accurately than the R13 equations giving a good agreement with kinetic theory for the Knudsen number up to 1. 
The moment equations are now being implemented into state-of-the-art open-source softwares \citep{TorrilhonSarna2017, WT2019},
providing a numerical platform to solve these equations for engineering applications.

Despite the success of the moment equations, there have been concerns over the accuracy of MBC obtained using Grad's closure at the boundary. 
Rana \& Struchtrup (\citeyear{RanaStruchtrup2016}) showed that the MBC for the LR13 equations violate the Onsager reciprocity relations  \citep{Onsager37,Onsager38}, and derived a new set of phenomenological boundary conditions (PBC) for them. 
For PBC, one writes the entropy generation at the boundary
and finds the boundary conditions as phenomenological laws that guarantee positivity of the entropy inequality at the boundary, which in turn leads to stable numerical schemes and convergence of moments methods \citep{Ringhofer2001, ST2018}.  
The PBC for the LR13 equations are thermodynamically admissible for all processes and comply with the Onsager reciprocity relations \citep{RanaStruchtrup2016}. 

In this paper, we first determine a quadratic form without cross-product terms for the entropy density to prove the $H$-theorem (entropy inequality) for the LR26 equations, which in turn leads to the PBC for them.  
We then solve the LR26 equations along with the associated PBC to investigate a slow flow of a gas/vapour past an evaporating/non-evaporating spherical droplet analytically, and compare the results with those obtained from other continuum theories, existing experimental data and/or numerical results obtained from the linearised Boltzmann equation.
Moreover, we estimate the drag force acting on the evaporating/non-evaporating spherical droplet with the LR26 equations. 
The results not only give an excellent match but---thanks to the analytic solutions---also  provide a deeper understanding of the process involved.
Through the examples considered in this paper, the present approach evidently provides the best ever results obtained via macroscopic models.
%

The remainder of the paper is organised as follows.
The LR26 equations in the dimensionless form are presented in \S\,\ref{section:LR26}. 
The $H$-theorem for the LR26 equations is proved in \S\,\ref{section:H-thm}. 
Physically admissible boundary conditions for the LR26 equations are derived 
in \S\,\ref{section:PBC}, where different contributions to the entropy production rate at the interface are also expounded by exploiting the Curie principle and Onsager symmetry relations. 
Canonical boundary value problems are solved with the LR26 equations with the PBC and MBC in \S\,\ref{section:OnsagerRels} to assess the Onsager reciprocity relations and entropy generation at the interface. 
%
The problem of a steady gas flow past a spherical droplet with and without evaporation is investigated in \S\,\ref{section:flowpastdroplet}.
Concluding remarks are made in \S\,\ref{section:conclusion}.

\section{The linearised R26 equations}
\label{section:LR26}
The treatment developed in this paper is based entirely on the linearised 
equations, which result from omitting the nonlinear terms (inertial terms, viscous heating terms, for instance) from the governing equations.
This simplification is justified since many micro- and nano-fluid devices involve sufficiently slow flows.
For succinctness, we shall present only the LR26 equations (and subsystems), which can be obtained from the original R26 equations derived in \cite{GuEmerson2009} in a straightforward way. 
We shall consider small deviations from a constant equilibrium state---given by a constant reference density $\hat{\rho}_{0}$, a constant reference temperature $\hat{T}_{0}$ and all other fields as zero. 
Thereby, all the equations can be linearised with respect to the reference state. 
Moreover, the equations will also be made dimensionless by introducing a length scale $\hat{L}_0$, a velocity scale $\sqrt{\hat{R} \hat{T}_{0}}$, a time scale $\hat{L}_0 / \sqrt{\hat{R} \hat{T}_{0}}$ and the dimensionless deviations in field variables from their respective reference states%
\begin{align}
\label{perturbedvars}
\left.
\begin{gathered}
\rho := \frac{\hat{\rho}-\hat{\rho}_{0}}{\hat{\rho}_{0}},
\quad
v_{i} := \frac{\hat{v}_{i}}{\sqrt{\hat{R} \hat{T}_{0}}},
\quad 
T := \frac{\hat{T}-\hat{T}_{0}}{\hat{T}_{0}},
\\
\sigma_{ij} := \frac{\hat{\sigma}_{ij}}{\hat{\rho}_{0}\hat{R} \hat{T}_{0}},
\quad 
q_{i} := \frac{\hat{q}_{i}}{\hat{\rho}_{0} \big(\hat{R} \hat{T}_{0}\big)^{3/2}},
\\
m_{ijk} := \frac{\hat{m}_{ijk}}{\hat{\rho}_{0} \big(\hat{R} \hat{T}_{0}\big)^{3/2}},
\quad
R_{ij} := \frac{\hat{R}_{ij}}{\hat{\rho}_{0} \big(\hat{R} \hat{T}_{0}\big)^2},
\quad
\Delta := \frac{\hat{\Delta}}{\hat{\rho}_{0} \big(\hat{R} \hat{T}_{0}\big)^2},
\\
\Phi_{ijkl} := \frac{\hat{\Phi}_{ijkl}}{\hat{\rho}_{0} \big(\hat{R} \hat{T}_{0}\big)^2},
\quad
\Psi_{ijk} := \frac{\hat{\Psi}_{ijk}}{\hat{\rho}_{0} \big(\hat{R} \hat{T}_{0}\big)^{5/2}},
\quad
\Omega_{i} := \frac{\hat{\Omega}_{i}}{\hat{\rho}_{0} \big(\hat{R} \hat{T}_{0}\big)^{5/2}},
\end{gathered}
\right\}
\end{align}
%
where $\hat{R}$ is the gas constant; $\rho$, $v_i$, $T$, $\sigma_{ij}$ and $q_i$ are the dimensionless deviations in the density, macroscopic velocity, temperature, stress tensor and heat flux from their respective reference state; $m_{ijk}$, $R_{ij}$, $\Delta$, $\Phi_{ijkl}$, $\Psi_{ijk}$ and $\Omega_i$ are dimensionless deviations in the higher-order moments. 
Note that the notation for the field variables are the same as those in \cite{GuEmerson2009} except for the quantities with hats here are with dimensions. 
As a consequence of the linearisation, the dimensionless deviation in the pressure $p$ is given by linearised equation of state, i.e.~$p = \rho + T$. 
The non-dimensionalisation also introduces a dimensionless parameter---the Knudsen number 
\begin{align}
\mathrm{Kn}= \frac{\hat{\mu}_{0}}{\hat{\rho}_{0} 
\sqrt{\hat{R} \hat{T}_{0}} L_{0}}
\end{align}
where $\hat{\mu}_{0}$ is the viscosity of the gas at the reference state.
Accordingly, the dimensionless and linearised conservation laws for the mass, momentum and energy read
\begin{align}
\label{massBal}
\frac{\partial \rho}{\partial t} +  \frac{\partial v_i}{\partial x_i}&=0,
\\
\label{momBal}
\frac{\partial v_i} {\partial t} + \frac{\partial \sigma_{ij}}{\partial x_j} + \frac{\partial p}{\partial x_i} &= F_i,
\\
\label{energyBal}
\frac{3}{2} \frac{\partial T} {\partial t} + \frac{\partial q_i}{\partial x_i} + \frac{\partial v_i}{\partial x_i} &= 0,
\end{align}
where $F_i$ is the dimensionless external force, $\partial (\cdot)/ \partial t$ denotes the dimensionless time derivative and $\partial (\cdot)/ \partial x_i$ denotes the dimensionless spatial derivative.
In \eqref{massBal}--\eqref{energyBal} and in what follows, indices denote vector and tensor components in Cartesian coordinates, and the Einstein summation convention is assumed over the repeated indices unless stated otherwise. 

The higher moment system also includes the governing equations for the stress and heat flux, which in the linear-dimensionless form read
\begin{align}
\frac{\partial \sigma_{ij}} {\partial t} 
+ \frac{\partial m_{ijk}}{\partial x_k}  
+ \frac{4}{5} \frac{\partial q_{\langle i}}{\partial x_{j\rangle}} 
&=-\frac{1}{\mathrm{Kn}}
\left[\sigma_{ij} + 2 \mathrm{Kn} \frac{\partial v_{\langle i}}{\partial x_{j\rangle}}\right], 
\label{stressBal} 
\\
\frac{\partial q_i} {\partial t} 
+\frac{1}{2} \frac{\partial R_{ij}}{\partial x_j}  
+\frac{1}{6} \frac{\partial \Delta}{\partial x_i}
+\frac{\partial \sigma_{ij}}{\partial x_j}
&=-\frac{\mathrm{Pr}}{\mathrm{Kn}}
\left[q_{i} + \frac{5}{2} \frac{\mathrm{Kn}}{\mathrm{Pr}} \frac{\partial T}{\partial x_i}\right],  
\label{HFBal}
\end{align}%
where $\mathrm{Pr}$ is the Prandtl number and the angular brackets around indices denote the symmetric-tracefree part of a tensor \citep{Struchtrup2005}. 
The right-hand sides of these balance equations contain NSF laws, which can also be obtained at the first order of a CE-like expansion performed on the stress and heat flux balance equations (\eqref{stressBal} and \eqref{HFBal}) in powers of the Knudsen number, see e.g.~\cite{TorrilhonStruchtrup2004, Struchtrup2005}.
The first order of the CE-like expansion on these equations essentially amounts to setting their left-hand sides to zero, yielding the NSF laws. 

The balance equations \eqref{massBal}--\eqref{HFBal} form the so-called 13-moment system. However, this system is not closed 
since they contain additional higher moments $m_{ijk}$, $R_{ij}$ and $\Delta$, which are defined as 
\begin{align}
m_{ijk} &=\int_{\mathbb{R}^{3}} C_{\langle i}C_{j}C_{k\rangle} f \, \mathrm{d}\bm{c},
\\
R_{ij} &=\int_{\mathbb{R}^{3}} C^2 C_{\langle i}C_{j\rangle }f\, \mathrm{d}\bm{c}
- 7 \sigma_{ij},
\\
\Delta  &=\int_{\mathbb{R}^{3}}C^{4}f\, \mathrm{d}\bm{c} - 15,
\end{align}%
where  $C_{i}=c_{i}-v_{i}$ is the dimensionless peculiar velocity with $c_{i}$ being the dimensionless microscopic velocity. 
These higher moments are constructed in such a way that
they vanish on being computed with the Grad 13-moment (G13) distribution function, i.e.~$m_{ijk|\textrm{G13}}=R_{ij|\textrm{G13}}=\Delta_{|\textrm{G13}} =0$.

The (linearised) 26-moment theory is comprised of the conservation laws \eqref{massBal}--\eqref{energyBal}, the stress balance equation \eqref{stressBal}, the heat flux balance equation \eqref{HFBal} and the balance equations for $m_{ijk}$, $R_{ij}$ and $\Delta $. 
The balance equations for $\hat{m}_{ijk}$, $\hat{R}_{ij}$ and $\hat{\Delta}$
have already been obtained from the Boltzmann equation by \citet{GuEmerson2009}; in the  linear-dimensionless form, they read 
\begin{align}
\frac{\partial m_{ijk}} {\partial t} 
+ \frac{\partial \Phi_{ijkl}}{\partial x_l}  
+\frac{3}{7} \frac{\partial R_{\langle ij}}{\partial x_{k\rangle}} &=-\frac{\mathrm{Pr}_{m}}{\mathrm{Kn}}\left[ m_{ijk}+3\frac{\mathrm{Kn}}{\mathrm{Pr}_{m}}\frac{\partial \sigma_{\langle ij}}{\partial x_{k\rangle}} \right] 
\label{mBal} 
\\
\frac{\partial R_{ij}} {\partial t} 
+ \frac{\partial \Psi_{ijk}}{\partial x_k}  
+2 \frac{\partial m_{ijk}}{\partial x_k}  
+\frac{2}{5} \frac{\partial \Omega_{\langle i}}{\partial x_{j\rangle}} &=-\frac{\mathrm{Pr}_{R}}{\mathrm{Kn}}%
\left[ R_{ij} + \frac{28}{5} \frac{\mathrm{Kn}}{\mathrm{Pr}_{R}}\frac{\partial q_{\langle i}}{\partial x_{j\rangle}} \right]   
\label{RBal} 
\\
\frac{\partial \Delta} {\partial t} 
+ \frac{\partial \Omega_i}{\partial x_i}  
&=-\frac{\mathrm{Pr}_{\Delta }}{\mathrm{Kn}}\left[ \Delta + 8 \frac{\mathrm{Kn}}{\mathrm{Pr}_{\Delta}}\frac{\partial q_i}{\partial x_i}\right].   
\label{DeltaBal}
\end{align}%
The coefficients $\mathrm{Pr}$, $\mathrm{Pr}_{m}$, $\mathrm{Pr}_{R}$ and $\mathrm{Pr}_{\Delta}$ in the above equations depend on the choice of intermolecular potential function appearing in the
Boltzmann collision operator. 
For Maxwell molecules (MM), these  parameters read: $\mathrm{Pr}=2/3$, $\mathrm{Pr}_{m}=3/2$, $\mathrm{Pr}_{R}=7/6$ and $\mathrm{Pr}_{\Delta}=2/3$
\citep{GuEmerson2009}.

For a third-order accuracy in $\mathrm{Kn}$, it is sufficient to consider only the terms on the right-hand sides of \eqref{mBal}--\eqref{DeltaBal} and dropping all the terms on the left-hand sides of these equations; 
this gives the linear R13 constitutive relations \citep{StruchtrupTorrilhon2003, Struchtrup2005}. 
Formally, the R13 constitutive relations stem from \eqref{mBal}--\eqref{DeltaBal} on performing a CE-like expansion on these equations in powers of $\mathrm{Kn}$, i.e.~by setting the left-hand sides of \eqref{mBal}--\eqref{DeltaBal} to zero \citep{Struchtrup2004}. 
Thus, the (dimensionless) LR13 equations consist of \eqref{massBal}--\eqref{HFBal} closed with
\begin{align}
\label{R13constitutiveRel}
m_{ijk} = -3\frac{\mathrm{Kn}}{\mathrm{Pr}_{m}}\frac{\partial \sigma_{\langle ij}}{\partial x_{k\rangle}},
\quad
R_{ij} = -\frac{28}{5} \frac{\mathrm{Kn}}{\mathrm{Pr}_{R}}\frac{\partial q_{\langle i}}{\partial x_{j\rangle}}
\quad\textrm{and}\quad
\Delta = -8 \frac{\mathrm{Kn}}{\mathrm{Pr}_{\Delta}}\frac{\partial q_i}{\partial x_i}.
\end{align}

\subsection{The R26 constitutive relations}
The system of 26 moment equations \eqref{massBal}--\eqref{HFBal} and \eqref{mBal}--\eqref{DeltaBal} contains the additional
moments $\Phi_{ijkl}$, $\Psi_{ijk}$ and $\Omega_{i}$:
\begin{align}
\Phi_{ijkl} &= \int_{\mathbb{R}^{3}}
C_{\langle i}C_{j}C_{k}C_{l\rangle }f\, \mathrm{d}\bm{c} ,
\\
\Psi_{ijk} &= \int_{\mathbb{R}^{3}}C^{2}C_{\langle i}C_{j}C_{k\rangle }f\, \mathrm{d}\bm{c} -9 m_{ijk},
\\
\Omega_{i} &= \int_{\mathbb{R}^{3}} C^{4}C_{i}f\, \mathrm{d}\bm{c} - 28 q_{i} ,
\end{align}%
which need to be specified in order to close the system of the 26 moment equations.
The R26 constitutive relations, which provides a closure for the 26 moment equations, follow from a CE-like expansion on the evolution equations for the moments $\Phi_{ijkl}$, $\Psi_{ijk}$ and $\Omega_{i}$ in a 45-moment theory \cite[see][]{GuEmerson2009}. 
The R26 constitutive relations read \citep{GuEmerson2009}
\begin{align}
\label{R26constRel}
\left.
\begin{aligned}
\Phi_{ijkl}&=-4\frac{\mathrm{Kn}}{\mathrm{Pr}_{\Phi}}\frac{\partial m_{\langle ijk}}{\partial x_{l\rangle}},
\\ 
\Psi_{ijk}&=-\frac{27}{7}\frac{\mathrm{Kn}}{\mathrm{Pr}_{\Psi}}\frac{\partial R_{\langle ij}}{\partial x_{k\rangle}},
\\ 
\Omega_{i}&=-\frac{7}{3}\frac{\mathrm{Kn}}{\mathrm{Pr}_{\Omega}}
\left(\frac{\partial \Delta}{\partial x_i} + \frac{12}{7}\frac{\partial R_{ij}}{\partial x_j}\right),
\end{aligned}
\right\}
\end{align}%
where the values of the parameters are $\mathrm{Pr}_{\Phi}=2.097$, $\mathrm{Pr}_{\Psi}=1.698$ and $\mathrm{Pr}_{\Omega}=1$ for MM 
\citep{GuEmerson2009}. 
The system of 26 moment equations \eqref{massBal}--\eqref{HFBal} and \eqref{mBal}--\eqref{DeltaBal} along with the constitutive relations \eqref{R26constRel} form the system of the LR26 equations. 
%
\section{\texorpdfstring{$H$}{}-theorem for the linearised R26 equations}
\label{section:H-thm}
The $H$-theorem plays a salient role in constructing constitutive relations in the bulk \citep{deGrootMazur1962, ST_PRL2007} and in developing physically admissible boundary conditions \citep{ST_PRL2007, RanaStruchtrup2016, RGS2018, ST2018}. Recently, \citet{SN2020} deduced that the Burnett equations, which do not comply with the $H$-theorem, lead to thermodynamically inadmissible solutions (producing work without a driving force).
In this section, we shall---for the first time---show that the LR26 equations are accompanied with an entropy law given as%
\begin{align}
\label{entropylawform}
\frac{\partial \eta}{\partial t} 
+ \frac{\partial \Gamma_i}{\partial x_i} = \Sigma \geqslant 0,
\end{align}%
where $\eta $ denotes the dimensionless entropy density, $\Gamma_{i}$ the dimensionless entropy flux and $\Sigma$ the dimensionless non-negative entropy generation rate.

The Boltzmann entropy density $\hat{\eta}$ and entropy flux $\hat{\Gamma}_i$ (in dimensional form) are given by \citep{MullerandRuggeri1998}
\begin{align}
\label{entropy_withdimension}
\hat{\eta}&=
-\hat{k}_{B}\int \hat{f} \left( \ln \frac{\hat{f}}{\hat{y}}-1\right) 
\mathrm{d}\hat{\bm{c}},
\\
\label{entropyflux_withdimension}
\hat{\Gamma}_i
&=-\hat{k}_{B}\int \hat{f} \left( \ln \frac{\hat{f}}{\hat{y}}-1\right) 
\hat{c}_i \, \mathrm{d}\hat{\bm{c}},
\end{align}
where $\hat{k}_{B}$ is the Boltzmann constant, $\hat{y}$ is a constant having dimensions of the distribution function $\hat{f}$ and $\hat{\bm{c}}$ is the microscopic velocity (with dimensions).
Starting from the Boltzmann entropy density \eqref{entropy_withdimension}, it can be shown that the entropy density of a system of the linearised moment equations---obtained from the Boltzmann equation---is a quadratic form without cross-product terms \cite[see e.g.,][]{ST_PRL2007, RanaStruchtrup2016, SarnaTorrilhon2018, BRTS2018, RGS2018}. 
It is worthwhile noting that the hyperbolic part of the LR26 equations is the same as the linearised Grad 26-moment (G26) equations (in the same way the hyperbolic part of the linearised NSF equations is the same as the linearised Euler equations); therefore, it is sufficient to use the G26 distribution function for computing the entropy density for the LR26 equations.
Using the G26 distribution function for $\hat{f}$ in \eqref{entropy_withdimension} and \eqref{entropyflux_withdimension}, we show that entropy density for the LR26 equations is a quadratic form without cross-product terms  and that the entropy flux obtained with the G26 distribution function contains all the hyperbolic terms present in the expression of the entropy flux for the LR26 equations, although the calculations are relegated to 
appendix~\ref{app:quadratic_entropy} for better readability.
%
The dimensionless entropy density for the LR26 equations is given by (see appendices~\ref{app:quadratic_entropy} and \ref{app:entopyCoeff})
\begin{align}
\label{entropyansatz}
\eta =\underline{a_{0}-\frac{1}{2}\rho ^{2}-\frac{1}{2} v^{2} 
-\frac{3}{4} T^{2}-\frac{1}{4}\sigma ^{2}-\frac{1}{5}q^{2}}  
- \frac{1}{12}m^{2} - \frac{1}{56}R^{2} - \frac{1}{240}\Delta^{2},
\end{align}%
where $a_0$ is just a constant and  the notation $A^{2}:=A_{i_{1}i_{2}\dots i_{k}}A_{i_{1}i_{2}\dots i_{k}}$ has been used to denote the contraction of a $k$-rank tensor $A_{i_{1}i_{2}\dots i_{k}}$ with itself for brevity.
%
The entropy for the LR26 equations \eqref{entropyansatz} comprises of the contribution from the 13-moment theory \citep{RanaStruchtrup2016}---given by the underlined terms in \eqref{entropyansatz}---and additional
contributions from the higher-order moments. 

The dimensionless entropy density \eqref{entropyansatz} can also be obtained in a more comprehensible way by assuming it to be a quadratic form without cross-product terms. 
The approach not only leads to the dimensionless entropy density \eqref{entropyansatz} but also yields the complete entropy flux and entropy production rate for the LR26 equations alongside. 
The approach assumes the following form of dimensionless entropy density:
\begin{align}
\label{entropybasicansatz}
\eta = a_0 + \frac{a_1}{2}\rho ^{2} + \frac{a_2}{2} v^{2} 
+ \frac{a_3}{2} T^{2} + \frac{a_4}{2} \sigma^{2} + \frac{a_5}{2} q^{2}  
+ \frac{a_6}{2} m^{2} + \frac{a_7}{2} R^{2} + \frac{a_8}{2}\Delta^{2}.
\end{align}
Here $a_0,a_1,a_2,\dots,a_8$ are unknown coefficients, which are computed by taking the time derivative of the entropy density \eqref{entropybasicansatz}, then eliminating the time derivatives of the moments by means of the G26 equations, and finally comparing the resulting equation with \eqref{entropylawform}. 
It turns out that one of the variables from $a_1, a_2, \dots, a_8$ can be chosen arbitrarily. 
On taking $a_1 = -1$, the other coefficients $a_2, a_3, \dots, a_8$ turn out to be such that \eqref{entropybasicansatz} is the same as \eqref{entropyansatz}; see appendix~\ref{app:entopyCoeff} for details. 


For the computation of the entropy flux and entropy production rate for the LR26 equations, we therefore can start with the entropy density \eqref{entropyansatz}.
Taking the time derivative of $\eta$ in \eqref{entropyansatz}, and substituting the time
derivatives of the field variables $\rho$, $v_{i}$, $T$ from \eqref{massBal}--\eqref{energyBal}, the time derivative of $\sigma_{ij}$ from \eqref{stressBal}, the time derivative of $q_{i}$ from \eqref{HFBal}, and the time
derivatives of $m_{ijk}$, $R_{ij}$, $\Delta$ from \eqref{mBal}--\eqref{DeltaBal}, one obtains 
\begin{align}
\label{entropylawR26}
\frac{\partial \eta}{\partial t} 
=& \frac{\partial}{\partial x_i} \left(p v_{i} + T q_{i} + v_{j}\sigma_{ij} 
+\frac{2}{5}\sigma_{ij} q_j 
+\frac{1}{2} m_{ijk} \sigma_{jk}
+\frac{1}{5}R_{ij}q_{j} 
+\frac{1}{15}q_{i} \Delta
\right.\nonumber\\
&\left.  
+\frac{1}{14} m_{ijk}R_{jk}
+\frac{1}{6}\Phi_{ijkl}m_{jkl}
+\frac{1}{28}\Psi_{ijk}R_{jk}
+\frac{1}{70}R_{ij}\Omega_{j}
+\frac{1}{120}\Delta \Omega_{i}
\right) 
\nonumber\\
&-\frac{1}{6}\Phi_{ijkl} \frac{\partial m_{jkl}}{\partial x_i}
-\frac{1}{28}\Psi_{ijk} \frac{\partial R_{jk}}{\partial x_i} 
-\frac{1}{70}\Omega_{j}  \frac{\partial R_{ij}}{\partial x_i}
-\frac{1}{120}\Omega_{i} \frac{\partial \Delta}{\partial x_i}
\nonumber\\
&+\frac{1}{2}\frac{1}{\mathrm{Kn}} \sigma^{2}
+\frac{2}{5} \frac{\mathrm{Pr}}{\mathrm{Kn}} q^{2} 
+\frac{1}{6}\frac{\mathrm{Pr}_{m}}{\mathrm{Kn}} m^{2}
+\frac{1}{28}\frac{\mathrm{Pr}_{R}}{\mathrm{Kn}} R^{2}
+\frac{1}{120}\frac{\mathrm{Pr}_{\Delta}}{\mathrm{Kn}} \Delta^{2}.
\end{align}%
Comparison of \eqref{entropylawR26} with \eqref{entropylawform} gives the expression for the entropy flux as
\begin{align}
\label{R26_entropy_flux}
\Gamma_{i} &=  - \left( p v_{i} + T q_{i} + v_{j}\sigma_{ij} 
+\frac{2}{5}\sigma_{ij} q_j 
+\frac{1}{2} m_{ijk} \sigma_{jk}
+\frac{1}{5}R_{ij}q_{j} 
+\frac{1}{15}q_{i} \Delta
\right.
\nonumber\\
&\left.\quad
+\frac{1}{14} m_{ijk}R_{jk}
+\frac{1}{6}\Phi_{ijkl}m_{jkl}
+\frac{1}{28}\Psi_{ijk}R_{jk}
+\frac{1}{70}R_{ij}\Omega_{j}
+\frac{1}{120}\Delta \Omega_{i}\right),
\end{align}
and the entropy production rate as
\begin{align}
\label{EntropyProdRate}
\Sigma &=-\frac{1}{6}\Phi_{ijkl} \frac{\partial m_{jkl}}{\partial x_i}
-\frac{1}{28}\Psi_{ijk} \frac{\partial R_{jk}}{\partial x_i} 
-\frac{1}{70}\Omega_{j}  \frac{\partial R_{ij}}{\partial x_i}
-\frac{1}{120}\Omega_{i} \frac{\partial \Delta}{\partial x_i}
\nonumber\\
&\quad 
+\frac{1}{2}\frac{1}{\mathrm{Kn}} \sigma^{2}
+\frac{2}{5} \frac{\mathrm{Pr}}{\mathrm{Kn}} q^{2} 
+\frac{1}{6}\frac{\mathrm{Pr}_{m}}{\mathrm{Kn}} m^{2}
+\frac{1}{28}\frac{\mathrm{Pr}_{R}}{\mathrm{Kn}} R^{2}
+\frac{1}{120}\frac{\mathrm{Pr}_{\Delta}}{\mathrm{Kn}} \Delta^{2}.
\end{align}
Using the linearised R26 constitutive relations \eqref{R26constRel} in \eqref{EntropyProdRate}, the entropy production rate $\Sigma$ can be written as
\begin{align}
\label{R26 entropy production}
\Sigma &= \frac{1}{24} \frac{\mathrm{Kn}}{\mathrm{Pr}_{\Phi}} \Phi^2
+\frac{1}{108} \frac{\mathrm{Kn}}{\mathrm{Pr}_{\Psi}} \Psi^2
+ \frac{1}{280} \frac{\mathrm{Kn}}{\mathrm{Pr}_{\Omega}} \Omega^2
\nonumber\\
&\quad 
+\frac{1}{2}\frac{1}{\mathrm{Kn}} \sigma^{2}
+\frac{2}{5} \frac{\mathrm{Pr}}{\mathrm{Kn}} q^{2} 
+\frac{1}{6}\frac{\mathrm{Pr}_{m}}{\mathrm{Kn}} m^{2}
+\frac{1}{28}\frac{\mathrm{Pr}_{R}}{\mathrm{Kn}} R^{2}
+\frac{1}{120}\frac{\mathrm{Pr}_{\Delta}}{\mathrm{Kn}} \Delta^{2},
\end{align}
which is non-negative for $\mathrm{Pr}_{\Psi}$,  $\mathrm{Pr}_{\Phi}$, $\mathrm{Pr}_{\Omega }$, $\mathrm{Pr}_{m}$, $\mathrm{Pr}_{R}$, $\mathrm{Pr}_{\Delta}$, $\Pr$, $\mathrm{Kn}\geq 0$.
This completes the proof of the $H$-theorem for the linearised R26 equations. 

\section{Phenomenological boundary conditions for the LR26 equations}
\label{section:PBC}
In this section, we shall demonstrate how the $H$-theorem can be used to generate a proper and complete set of boundary conditions for the LR26 equations. We shall derive the PBC for the LR26 equations from the entropy production rate at the interface exploiting the force-flux relationships.

Let there be two phases, namely the liquid (or solid) and the gaseous (or vapour), connected by a massless interface $I$ of negligible thickness and let there be no surface tension and surface energy at the interface. 
The entropy production rate at such an interface $\Sigma^{I}$ is given by the
difference between the entropy fluxes into and out of an interface \citep{deGrootMazur1962, RanaStruchtrup2016, BRTS2018}, i.e.%
\begin{align}
\label{entropyproductionsurface}
\Sigma^{I}=\left[\Gamma_{i}^{\mathrm{(gas)}} -\Gamma_{i}^{\mathrm{(liquid)}}\right] n_{i}\geq 0.
\end{align}%
where $n_{i}$ is the unit normal to the interface pointing from the liquid into the gas. 
Assuming that the gaseous (or vapour) phase is comprised of an ideal gas,
the entropy flux of the gas (or vapour) medium is $\Gamma_{i}^{\mathrm{(gas)}} = \Gamma_{i}$, given by \eqref{R26_entropy_flux},
and assuming that
the liquid phase is comprised of  an incompressible liquid and can be described by the NSF equations, the entropy flux of the liquid phase is \citep{BRTS2018}
\begin{align}
\label{entropyflux_out}
\Gamma_{i}^{\mathrm{(liquid)}} 
= - p^{\ell} v_{i}^{\ell} 
- T^{\ell} q_{i}^{\ell} 
- v_{j}^{\ell}\sigma_{ij}^{\ell},
\end{align}
where the superscript `$\ell$' on the variables are used to denote that they are the properties of the liquid.
Substituting $\Gamma_{i}^{\mathrm{(gas)}} = \Gamma_{i}$ from \eqref{R26_entropy_flux} and $\Gamma_{i}^{\mathrm{(liquid)}}$ from \eqref{entropyflux_out} into \eqref{entropyproductionsurface},  the entropy flux at the interface---after some algebra---turns out to be (see appendix~\ref{app:entopyfluxI} for details)
\begin{align}
\label{entropyProdwall}
\Sigma^{I}&= - \left(\mathcal{P} \mathcal{V}_i 
+ \mathcal{T} q_{i}
+ \mathscr{V}_j \sigma_{ij}
+\frac{2}{5}\sigma_{ij} q_j 
+\frac{1}{2} m_{ijk} \sigma_{jk}
+\frac{1}{5}R_{ij}q_{j} 
+\frac{1}{15}q_{i} \Delta
\right.
\nonumber\\
&\left.\quad\,
+\frac{1}{14} m_{ijk}R_{jk}
+\frac{1}{6}\Phi_{ijkl}m_{jkl}
+\frac{1}{28}\Psi_{ijk}R_{jk}
+\frac{1}{70}R_{ij}\Omega_{j}
+\frac{1}{120} \Omega_{i}\Delta\right)n_{i},
\end{align}%
where 
\begin{align}
\label{jumpvars}
\mathcal{P} = p - p_{\mathrm{sat}},
\quad
\mathcal{T} = T - T^{\ell},
\quad
\mathcal{V}_{i}=v_{i}-v_{i}^{I}
\quad\text{and}\quad
\mathscr{V}_j = v_j - v_j^{\ell}
\end{align}
with $v_{i}^{I}$ being the velocity of the interface and $p_{\mathrm{sat}} \equiv p_{\mathrm{sat}} (T^{\ell})$ being the saturation pressure corresponding to the temperature $T^{\ell}$.


%
For further simplification, it is imperative to decompose the vectors and tensors into their components in the normal and tangential directions. Such a decomposition for symmetric-tracefree tensors of rank up to three is given in appendix~\ref{app:Decomp}. 
%
%
%
This decomposition allows one to write the entropy production at the wall  \eqref{entropyProdwall} as a sum of the product of the fluxes with an odd degree in $n$ (i.e.~$\mathcal{V}_n$, $q_{n}$, $\Omega_{n}$, $m_{nnn}$, $\Psi_{nnn}$, $\bar{\sigma}_{ni}$, $\bar{R}_{ni}$, $\bar{\Phi}_{nnni}$, $\tilde{m}_{nij}$, $\tilde{\Psi }_{nij}$, $\check{\Phi}_{nijk}$) 
and the moments with an even degree in $n$ as
\begin{align}
\label{entropyProdInterface}
\Sigma^{I} &=-\mathcal{V}_n (\mathcal{P}+\sigma_{nn}) -q_{n}\left( \mathcal{T}+\frac{2}{5}\sigma_{nn}+\frac{1}{5}R_{nn} +\frac{1}{15}\Delta \right) 
\nonumber\\
&\quad -m_{nnn}\left( \frac{3}{4}\sigma_{nn}+\frac{3}{28}%
R_{nn}+\frac{5}{12}\Phi_{nnnn}\right)
-\Psi_{nnn} \left(\frac{3}{56} R_{nn}\right)
\nonumber\\
&\quad
-\Omega_{n}\left( \frac{1}{120} \Delta + \frac{1}{70} R_{nn} \right)
-\bar{\sigma}_{ni} \left( \bar{\mathscr{V}}_{i}+\frac{2}{5}\bar{q}_{i} +\bar{m}_{nni}\right) 
\nonumber\\
&\quad 
-\bar{R}_{ni}\left( \frac{1}{5}\bar{q}_{i} + \frac{1}{7} \bar{m}_{nni} +\frac{1}{14} \bar{\Psi}_{nni} +\frac{1}{70} \bar{\Omega}_{i}\right) 
- \bar{\Phi}_{nnni} \left(\frac{5}{8}\bar{m}_{nni}\right)
\nonumber\\
&\quad 
-\tilde{m}_{nij} 
\left(\frac{1}{2} \tilde{\sigma}_{ij}+\frac{1}{14}\tilde{R}_{ij} + \frac{1}{2} \tilde{\Phi}_{nnij} \right)
 -\tilde{\Psi}_{nij} \left(\frac{1}{28} \tilde{R}_{ij}\right)
 -\check{\Phi}_{nijk} \left(\frac{1}{6} \check{m}_{ijk}\right).
\end{align}
While writing \eqref{entropyProdInterface}, 
the relation $\sigma_{nn} \mathscr{V}_n \approx \sigma_{nn} \mathcal{V}_n$ in light of \eqref{normalslipvelocity} and the Curie principle---which states that only forces and fluxes of the same tensor type (scalars, vectors, 2-tensors, etc.) can be combined \citep{deGrootMazur1962}---have been exploited. 
In the above equation, $\lbrace \mathcal{V}_n$, $q_{n}$, $m_{nnn}$, $\Psi_{nnn}$, $\Omega_{n}\rbrace$ are the odd (in $n$) scalar fluxes, $\lbrace\bar{\sigma}_{ni}$, $\bar{R}_{ni}$, $\bar{\Phi}_{nnni}\rbrace$ are the odd vector fluxes, $\lbrace\tilde{m}_{nij}$, $\tilde{\Psi }_{nij}\rbrace$ are the odd tensor fluxes of rank two and $\check{\Phi}_{nijk}$ is the odd tensor flux of rank three. 
The positive entropy production at the interface can be attained by taking the odd fluxes proportional to the even (in $n$) driving forces (written in brackets in \eqref{entropyProdInterface})---with proportionality constants being positive. This step to obtain the positive entropy production at the interface yields the required boundary conditions. Mathematically, the positive entropy production, and hence the thermodynamically consistent PBC at the interface are achieved by writing a linear force-flux relationship of the form
\begin{align}
\label{PBCfluxes}
\mathcal{J}_{r}=-
\mathcal{L}_{rs}\mathcal{F}_{s}
\end{align} 
for scalar, vector and tensor fluxes of each rank, successively. 
Here, $\mathcal{J}_{r}$ is the $r^{\mathrm{th}}$ component of a vector $\bm{\mathcal{J}}$ containing the fluxes, $\mathcal{F}_{s}$ is the $s^{\mathrm{th}}$ component of a vector $\bm{\mathcal{F}}$ containing the forces, $\mathcal{L}_{rs}$ is the $(r,s)^{\mathrm{th}}$ element of an unknown matrix $\bm{\mathcal{L}}$ and Einstein summation is assumed over $s$ in \eqref{PBCfluxes}. 
The unknown matrix $\bm{\mathcal{L}}$ consists of phenomenological coefficients as its elements. 
The matrix $\bm{\mathcal{L}}$ must be symmetric in order to comply with the Onsager reciprocity relations \citep{Onsager37,Onsager38} and must be positive semidefinite to ensure a non-negative entropy production.  
\subsection{Boundary conditions for the scalar fluxes}
%
The scalar fluxes in \eqref{entropyProdInterface} and their corresponding forces are collected in $\bm{\mathcal{J}}$ and $\bm{\mathcal{F}}$ as
\begin{align}
\label{JFscalar}
\renewcommand{\arraystretch}{1.5}
\bm{\mathcal{J}} =
\begin{bmatrix}
\mathcal{V}_n
\\ 
q_{n}
\\ 
m_{nnn}
\\ 
\Psi_{nnn}
\\
\Omega_{n}
\end{bmatrix}
\qquad\textrm{and}\qquad
\bm{\mathcal{F}} =
\begin{bmatrix}
\mathcal{P}+\sigma_{nn}
\\
\mathcal{T}+\frac{2}{5}\sigma_{nn}+\frac{1}{5}R_{nn} +\frac{1}{15}\Delta
\\
\frac{3}{4}\sigma_{nn}+\frac{3}{28}R_{nn}
+\frac{5}{12}\Phi_{nnnn}
\\
\frac{3}{56} R_{nn}
\\
\frac{1}{120} \Delta + \frac{1}{70} R_{nn}
\end{bmatrix}.
\end{align}
Substituting $\bm{\mathcal{J}}$ and $\bm{\mathcal{F}}$ from \eqref{JFscalar} in \eqref{PBCfluxes}, one obtains the PBC for the scalar fluxes. 
Nevertheless, the matrix $\bm{\mathcal{L}}$ is still unknown. 
The elements of the matrix $\bm{\mathcal{L}}$ can be obtained by comparing the coefficients of $\mathcal{P}$, $\mathcal{T}$, $\sigma_{nn}$, $R_{nn}$ and $\Delta$ in the corresponding terms of the PBC and MBC for the scalar fluxes. 
The MBC for the scalar fluxes (obtained through the Maxwell accommodation model) read \citep{RanaLockerbySprittles2018}
\begin{align}
\label{BCscalar}
\left.
\begin{aligned}
\mathcal{V}_n &= - \varsigma \left(\mathcal{P} - \frac{1}{2} \mathcal{T} + \frac{1}{2} \sigma_{nn} - \frac{1}{28} R_{nn} - \frac{1}{120} \Delta 
+b_1 \Phi_{nnnn}\right),
\\
q_n &= -\frac{\mathcal{V}_n}{2} - \varkappa \left(2 \mathcal{T} + \frac{1}{2} \sigma_{nn} + \frac{5}{28} R_{nn} + \frac{1}{15} \Delta 
+b_2 \Phi_{nnnn} \right),
\\
m_{nnn} &= -\frac{2}{5} \mathcal{V}_n + \varkappa \left(\frac{2}{5} \mathcal{T} - \frac{7}{5} \sigma_{nn} - \frac{1}{14} R_{nn} + \frac{1}{75} \Delta 
+b_3 \Phi_{nnnn}\right),
\\
\Psi_{nnn} &= \frac{6}{5} \mathcal{V}_n + \varkappa \left(\frac{6}{5} \mathcal{T} + \frac{9}{5} \sigma_{nn} - \frac{93}{70} R_{nn} + \frac{1}{5} \Delta 
+b_4 \Phi_{nnnn}\right),
\\
\Omega_n &= 3 \mathcal{V}_n + \varkappa \left(8 \mathcal{T} + 2 \sigma_{nn} - R_{nn} - \frac{4}{3} \Delta
+b_5 \Phi_{nnnn}\right),
\end{aligned}
\right\}
\end{align}
where
\begin{align}
\label{biMBCscalar}
b_1 = -\frac{1}{24},
\quad
b_2 = - \frac{1}{12},
\quad
b_3 = - \frac{13}{30},
\quad
b_4 = - \frac{1}{2},
\quad
b_5 = 0,
\end{align}
\begin{align}
\varsigma = \frac{\vartheta}{2-\vartheta} \sqrt{\frac{2}{\pi}} 
\qquad\textrm{and} \qquad
\varkappa = \frac{\vartheta + \chi (1-\vartheta)}{2-\vartheta - \chi (1-\vartheta)} \sqrt{\frac{2}{\pi}}
\end{align}
with $\vartheta$ being the evaporation/condensation coefficient and $\chi$ being the accommodation coefficient. Clearly, for $\vartheta=0$, $\mathcal{V}_n=0$ from (\ref{BCscalar}), hence one obtains the boundary conditions for a non-evaporating boundary \citep{GuEmerson2009}.
It is worthwhile noting that the MBC \eqref{BCscalar} with \eqref{biMBCscalar} cannot be expressed in the force-flux formalism \eqref{PBCfluxes}, and we shall show later that this leads to violation of the Onsager reciprocity relations and/or the second law at the interface.

Comparison of the coefficients of $\mathcal{P}$, $\mathcal{T}$, $\sigma_{nn}$, $R_{nn}$ and $\Delta$ in the PBC for the scalar fluxes (obtained on substituting $\bm{\mathcal{J}}$ and $\bm{\mathcal{F}}$ from \eqref{JFscalar} in \eqref{PBCfluxes}) and MBC \eqref{BCscalar} yields
\begin{align}
\label{Lscalar}
\renewcommand{\arraystretch}{1.5}
\bm{\mathcal{L}} = \varsigma
\begin{bmatrix*}[r]
1 & -\frac{1}{2} & -\frac{2}{5} & \frac{6}{5} & 3 
\\
-\frac{1}{2} & \frac{1}{4} & \frac{1}{5} & -\frac{3}{5} & -\frac{3}{2} \\
-\frac{2}{5} & \frac{1}{5} & \frac{4}{25} & -\frac{12}{25} & -\frac{6}{5} 
\\
\frac{6}{5} & -\frac{3}{5} & -\frac{12}{25} & \frac{36}{25} & \frac{18}{5} 
\\
3 & -\frac{3}{2} & -\frac{6}{5} & \frac{18}{5} & 9
\end{bmatrix*}
+\varkappa
\begin{bmatrix*}[r]
 0 & 0 & 0 & 0 & 0 
 \\
 0 & 2 & -\frac{2}{5} & -\frac{6}{5} & -8 
 \\
 0 & -\frac{2}{5} & \frac{52}{25} & -\frac{44}{25} & \frac{8}{5} 
 \\
 0 & -\frac{6}{5} & -\frac{44}{25} & \frac{916}{25} & -\frac{72}{5} 
 \\
 0 & -8 & \frac{8}{5} & -\frac{72}{5} & 224
\end{bmatrix*}.
\end{align}
Both the matrices on the right-hand side of \eqref{Lscalar} are symmetric and positive semidefinite. 
Therefore, the matrix $\bm{\mathcal{L}}$ in \eqref{Lscalar} is also symmetric and positive semidefinite as $\varsigma$ and $\varkappa$ are non-negative.
Thus, the PBC for the scalar fluxes are obtained from \eqref{PBCfluxes} with $\bm{\mathcal{J}}$, $\bm{\mathcal{F}}$ and $\bm{\mathcal{L}}$ being given by \eqref{JFscalar} and \eqref{Lscalar}. 
On simplification, the PBC for the scalar fluxes are given by \eqref{BCscalar} with
\begin{align}
\label{biPBCscalar}
b_1 = -\frac{1}{6},
\quad
b_2 = - \frac{1}{6},
\quad
b_3 = - \frac{13}{15},
\quad
b_4 =  \frac{11}{15},
\quad
b_5 = - \frac{2}{3}.
\end{align}
\subsection{Boundary conditions for the vector fluxes}
The vector fluxes in \eqref{entropyProdInterface} and their corresponding forces are collected in $\bm{\mathcal{J}}$ and $\bm{\mathcal{F}}$ as
\begin{align}
\label{JFvector}
\renewcommand{\arraystretch}{1.5}
\bm{\mathcal{J}} =
\begin{bmatrix}
\bar{\sigma}_{ni} 
\\ 
\bar{R}_{ni}
\\ 
\bar{\Phi}_{nnni}
\end{bmatrix}
\qquad\textrm{and}\qquad
\bm{\mathcal{F}} =
\begin{bmatrix}
\bar{\mathscr{V}}_{i}+\frac{2}{5}\bar{q}_{i} +\bar{m}_{nni}
\\
\frac{1}{5}\bar{q}_{i} + \frac{1}{7} \bar{m}_{nni} +\frac{1}{14} \bar{\Psi}_{nni} +\frac{1}{70} \bar{\Omega}_{i}
\\
\frac{5}{8}\bar{m}_{nni}
\end{bmatrix}.
\end{align}
Substituting $\bm{\mathcal{J}}$ and $\bm{\mathcal{F}}$ from \eqref{JFvector} in \eqref{PBCfluxes}, one obtains the PBC for the vector fluxes. 
Nevertheless, the matrix $\bm{\mathcal{L}}$ is still unknown. 
The elements of the matrix $\bm{\mathcal{L}}$ can be obtained by comparing the coefficients of $\bar{\mathscr{V}}_{i}$, $\bar{q}_{i}$ and $\bar{m}_{nni}$ in the corresponding terms of the PBC and MBC. 
The MBC for the vector fluxes (obtained through the Maxwell accommodation model) read \citep{GuEmerson2009,SBRF2017}
\begin{align}
\label{BCvector}
\left.
\begin{aligned}
\bar{\sigma}_{ni} &=- \varkappa
\left( \bar{\mathscr{V}}_{i}
+\frac{1}{5}\bar{q}_{i} +\frac{1}{2}\bar{m}_{nni} 
+ b_6 \bar{\Psi}_{nni} 
+ b_7 \bar{\Omega}_{i}\right),
\\
\bar{R}_{ni} &=- \varkappa \left( -\bar{\mathscr{V}}_{i} 
+ \frac{11}{5} \bar{q}_{i} + \frac{1}{2} \bar{m}_{nni} + 
b_8 \bar{\Psi}_{nni} + 
b_9 \bar{\Omega}_{i}\right), 
\\
\bar{\Phi}_{nnni} &=- \varkappa \left( -\frac{4}{7}\bar{\mathscr{V}}_{i}
-\frac{12}{35} \bar{q}_{i}+\frac{9}{7} \bar{m}_{nni} + 
b_{10} \bar{\Psi}_{nni}
+
b_{11} \bar{\Omega}_{i}\right),
\end{aligned}
\right\}
\end{align}
where
\begin{align}
\label{biMBCvector}
b_6 = -\frac{1}{36},
\quad 
b_7 = -\frac{1}{280},
\quad 
b_8 = \frac{5}{12},
\quad 
b_9 = \frac{5}{56},
\quad 
b_{10} = \frac{1}{14},
\quad 
b_{11} = - \frac{3}{490}.
\end{align}
Comparison of the coefficients of $\bar{\mathscr{V}}_{i}, \bar{q}_{i}, \bar{m}_{nni}$ in the PBC for the vector fluxes (obtained on substituting $\bm{\mathcal{J}}$ and $\bm{\mathcal{F}}$ from \eqref{JFvector} in \eqref{PBCfluxes}) and MBC \eqref{BCvector} yields
\begin{align}
\renewcommand{\arraystretch}{1.5}
\label{Lvector}
\bm{\mathcal{L}} 
= \varkappa
\begin{bmatrix*}[r]
1 & -1 & -\frac{4}{7} 
\\ 
-1 & 13 & -\frac{4}{7} 
\\ 
-\frac{4}{7} & -\frac{4}{7} & \frac{152}{49}%
\end{bmatrix*}
\end{align}
The matrix $\bm{\mathcal{L}}$ in \eqref{Lvector} is symmetric and positive semidefinite as $\varkappa$ is non-negative.
Thus, the PBC for the vector fluxes are obtained from \eqref{PBCfluxes} with $\bm{\mathcal{J}}$, $\bm{\mathcal{F}}$ and $\bm{\mathcal{L}}$ being given by \eqref{JFvector} and \eqref{Lvector}. 
On simplification, the PBC for the vector fluxes are given by \eqref{BCvector} with
\begin{align}
\label{biPBCvector}
b_6 = -\frac{1}{14},
\quad 
b_7 = -\frac{1}{70},
\quad 
b_8 = \frac{13}{14},
\quad 
b_9 = \frac{13}{70},
\quad 
b_{10} = - \frac{2}{49},
\quad 
b_{11} = - \frac{2}{245}.
\end{align}
\subsection{Boundary conditions for the tensor fluxes}
\subsubsection{Boundary conditions for the rank-2 tensor fluxes}
The rank-2 tensor fluxes in \eqref{entropyProdInterface} and their corresponding forces are collected in $\bm{\mathcal{J}}$ and $\bm{\mathcal{F}}$ as
\begin{align}
\label{JFtensor2}
\renewcommand{\arraystretch}{1.5}
\bm{\mathcal{J}} =
\begin{bmatrix}
\tilde{m}_{n ij}
\\ 
\tilde{\Psi}_{n ij}
\end{bmatrix}
\qquad\textrm{and}\qquad
\bm{\mathcal{F}} =
\begin{bmatrix}
\frac{1}{2} \tilde{\sigma}_{ij} + \frac{1}{14}\tilde{R}_{ij} + \frac{1}{2} \tilde{\Phi}_{nn ij} 
\\
\frac{1}{28} \tilde{R}_{ij}
\end{bmatrix}.
\end{align}
Substituting $\bm{\mathcal{J}}$ and $\bm{\mathcal{F}}$ from \eqref{JFtensor2} in \eqref{PBCfluxes}, one obtains the PBC for the rank-2 tensor fluxes. 
Nevertheless, the matrix $\bm{\mathcal{L}}$ is still unknown. 
The elements of the matrix $\bm{\mathcal{L}}$ can be obtained by comparing the coefficients of $\tilde{\sigma}_{ij}$ and $\tilde{R}_{ij}$ in the corresponding terms of the PBC and MBC for the rank-2 tensor fluxes. 

The MBC for the rank-2 tensor fluxes (obtained through the Maxwell accommodation model) read 
\begin{align}
\label{BCtensor2}
\left.
\begin{aligned}
\tilde{m}_{nij} &= - \varkappa \left( \tilde{\sigma}_{ij} + \frac{1}{14}\tilde{R}_{ij} + b_{12} \tilde{\Phi}_{nn ij} \right),
\\
\tilde{\Psi}_{nij} &= \varkappa \left( \tilde{\sigma}_{ij} - \frac{15}{14}\tilde{R}_{ij} + b_{13} \tilde{\Phi}_{nn ij} \right),
\end{aligned}
\right\}
\end{align}
%
where
\begin{align}
\label{biPBCtensor2}
b_{12} = \frac{1}{2} 
\quad\textrm{and}\quad
b_{13} = - \frac{1}{2}. 
\end{align}
Comparison of the coefficients of $\tilde{\sigma}_{ij}$ and $\tilde{R}_{ij}$ in the PBC for the rank-2 tensor fluxes (obtained on substituting $\bm{\mathcal{J}}$ and $\bm{\mathcal{F}}$ from \eqref{JFtensor2} in \eqref{PBCfluxes}) and MBC \eqref{BCtensor2} yields
\begin{align}
\renewcommand{\arraystretch}{1.5}
\label{Ltensor2}
\bm{\mathcal{L}} 
= \varkappa
\begin{bmatrix*}[r]
2 & -2 
\\ 
-2 & 34 
\end{bmatrix*}
\end{align}
The matrix $\bm{\mathcal{L}}$ in \eqref{Lvector} is symmetric and positive semidefinite as $\varkappa$ is non-negative.
Thus, the PBC for the rank-2 tensor fluxes are obtained from \eqref{PBCfluxes} with $\bm{\mathcal{J}}$, $\bm{\mathcal{F}}$ and $\bm{\mathcal{L}}$ being given by \eqref{JFtensor2} and \eqref{Ltensor2}. 
On simplification, the PBC for the rank-2 tensor fluxes are given by \eqref{BCtensor2} with
\begin{align}
\label{biPBCtensor2a}
b_{12} = 1 
\quad\textrm{and}\quad
b_{13} = 1. 
\end{align}
\subsubsection{Boundary conditions for the rank-3 tensor fluxes}
The rank-3 tensor flux in \eqref{entropyProdInterface} and its corresponding forces are
\begin{align}
\label{JFtensor3}
\renewcommand{\arraystretch}{1.5}
J_r = \check{\Phi}_{n ijk}
\qquad\textrm{and}\qquad
\mathcal{F}_s = \frac{1}{6} \check{m}_{ijk}.
\end{align}
Substituting $\mathcal{J}_r$ and $\mathcal{F}_s$ from \eqref{JFtensor3} in \eqref{PBCfluxes}, one obtains the PBC for the rank-3 tensor flux. Nevertheless, the coefficient $\mathcal{L}_{rs}$ in \eqref{PBCfluxes} is still unknown but can be obtained by comparing the coefficients of $\check{m}_{ijk}$ in the PBC and MBC for the rank-3 tensor flux. 

The MBC for rank-3 tensor flux (obtained through the Maxwell accommodation model) read 
\begin{align}
\label{BCtensor3}
\check{\Phi}_{nijk} = -\varkappa \left(\check{m}_{ijk} + b_{14} \check{\Psi}_{ijk}\right)
\end{align}
where
\begin{align}
\label{biPBCtensor3}
b_{14} = \frac{1}{18}. 
\end{align}
Comparison of the coefficients of $\check{m}_{ijk}$ in the PBC for rank-3 tensor fluxes  (obtained on substituting $\mathcal{J}_r$ and $\mathcal{F}_s$ from \eqref{JFtensor3} in \eqref{PBCfluxes}) and MBC \eqref{BCtensor3} yields
\begin{align}
\label{Ltensor3}
\mathcal{L}_{rs} = 6 \varkappa
\end{align}
%
Thus, the PBC for the rank-3 tensor flux is obtained from \eqref{PBCfluxes} with $\mathcal{J}_r$, $\mathcal{F}_s$ and $\mathcal{L}_{rs}$ being given by \eqref{JFtensor3} and \eqref{Ltensor3}. 
On simplification, the PBC for the rank-3 tensor flux is given by \eqref{BCtensor3} with
\begin{align}
\label{biPBCtensor2b}
b_{14} = 0. 
\end{align}

%
\section{Onsager reciprocity relations in some canonical problems}
\label{section:OnsagerRels}
In this section, we consider two canonical boundary value problems, namely (i) the Poiseuille and thermal transpiration gas flows and (ii) evaporation of a spherical liquid droplet, for which the Onsager reciprocity relations have been shown to hold for the linearised Boltzmann equation (LBE) \citep{WU2017431, CM1989}.
We shall employ the LR26 equations with the PBC and MBC to these problems to demonstrate that the solutions of the LR26 equations with the MBC for these problems either do not satisfy the Onsager reciprocity relations or do not respect the second law of thermodynamics or do not comply with both. 
On the other hand, the solutions of the LR26 equations with the PBC satisfy the Onsager reciprocity relations and always respect the second law of thermodynamics.
%
For comparison purpose, we shall also include the results obtained through the LR13 equations with the MBC and PBC.%
\subsection{Poiseuille flow and thermal transpiration flow}
\label{sec:channel}

\begin{figure}
\centering
\includegraphics[scale=1]{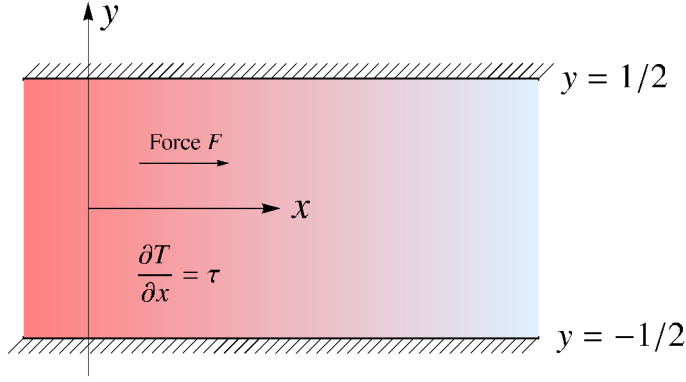}
\caption{\label{fig:channel}Schematic of the Poiseuille flow ($F=1$, $\tau=0$) and thermal transpiration flow ($F=0$, $\tau=1$) in a channel.}
\end{figure}

Let us first consider the steady state [$\partial_t(\cdot)=0$] of a gas confined between two stationary ($v^{w}_{i}=0$), isothermal ($T^{w}=1$) and fully diffusive ($\chi=1$) walls of a channel located at $y=\mp 1/2$; see figure~\ref{fig:channel}. 
The flow is assumed to be fully
developed and driven either by a uniform  external force $F$ (Poiseuille flow) in the
positive $x$-direction or by a constant temperature gradient parallel to the walls  $\tau$ (thermal transpiration flow). 
The normals for the bottom and top walls are pointing in $\pm y$-directions, respectively.  

The solution for the velocity and heat flux---for the problem under consideration---from the LR26 equations read 
\begin{subequations}
\label{velocity and heat LR26}
\begin{align}
\label{vel_LR26}
v_{x} &= C_{1}+\underline{\frac{F}{2\mathrm{Kn}}\left( \frac{1}{4}-y^{2}\right)} -\frac{2}{5}%
\sum_{i=1}^{4}\left( C_{i}^{K}-\frac{5}{2}\beta_{i}\right)\exp \left( \frac{y\gamma _{i}}{\mathrm{Kn}}\right),
\\ 
\label{HF_LR26}
q_{x} &= \underline{-\frac{5\mathrm{Kn}}{2\Pr }\tau} -\frac{\mathrm{Kn}}{\Pr }%
F+\sum_{i=1}^{4}C_{i}^{K}\exp \left( \frac{y\gamma _{i}}{\mathrm{Kn}}\right),
\end{align}
\end{subequations}
where $C_{1}$ and $C^{K}_{1}$, $C^{K}_{2}$, $C^{K}_{3}$ and $C^{K}_{4}$ are the constants of integration. 
Moreover, due to symmetry about the $x$-axis,
$C^{K}_{2}=C^{K}_{1}$, $C^{K}_{4}=C^{K}_{3}$, $\beta_2 = \beta_1$ and $\beta_4 = \beta_3$. The unspecified three constants of integration, i.e.~$C_1$, $C^{K}_{1}$ and  $C^{K}_{3}$, are to be evaluated from the three (vector) boundary conditions at $y=-1/2$.

\begin{figure}
\centering
\includegraphics[height=45mm]{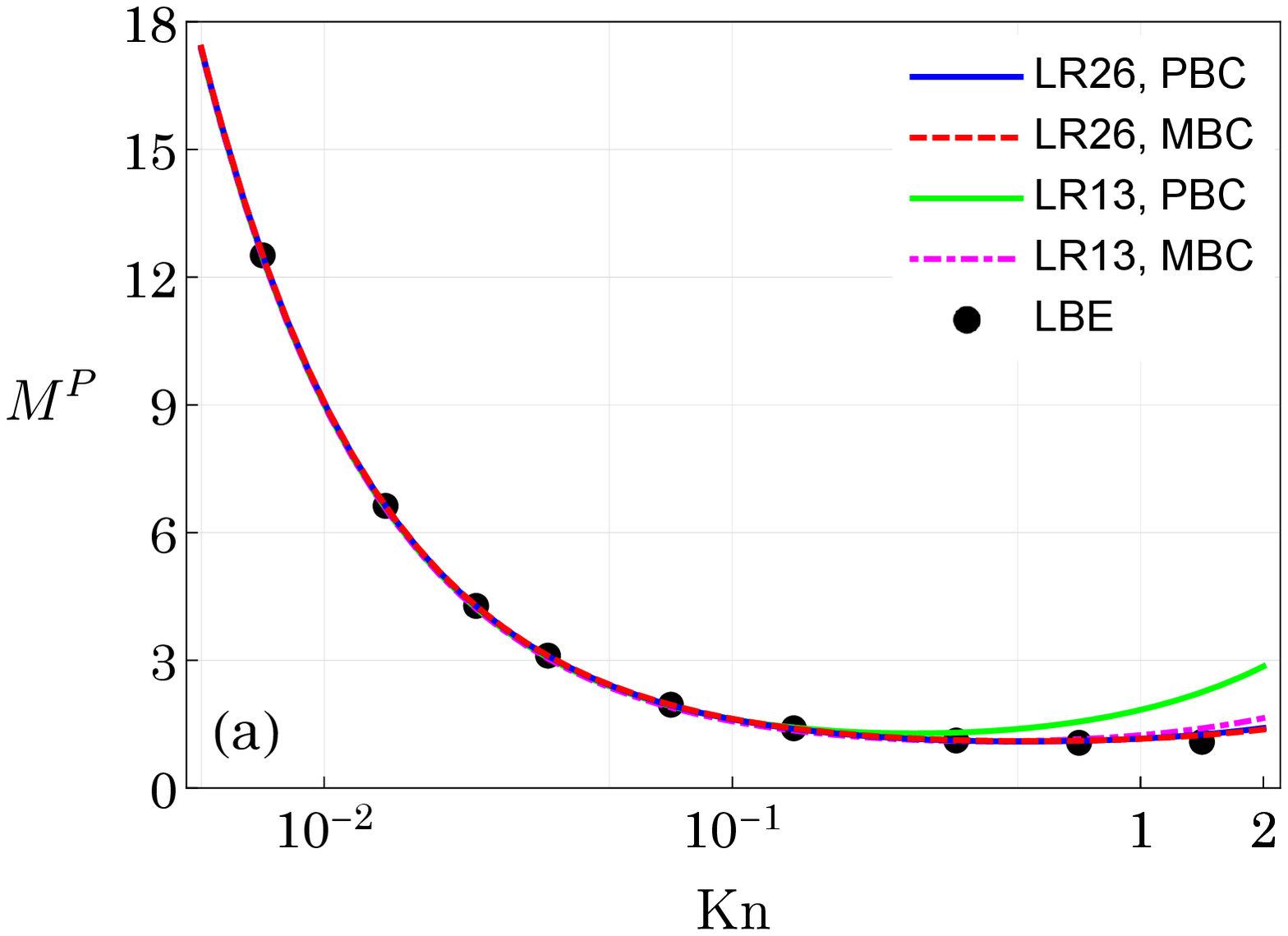}\hfill
\includegraphics[height=45mm]{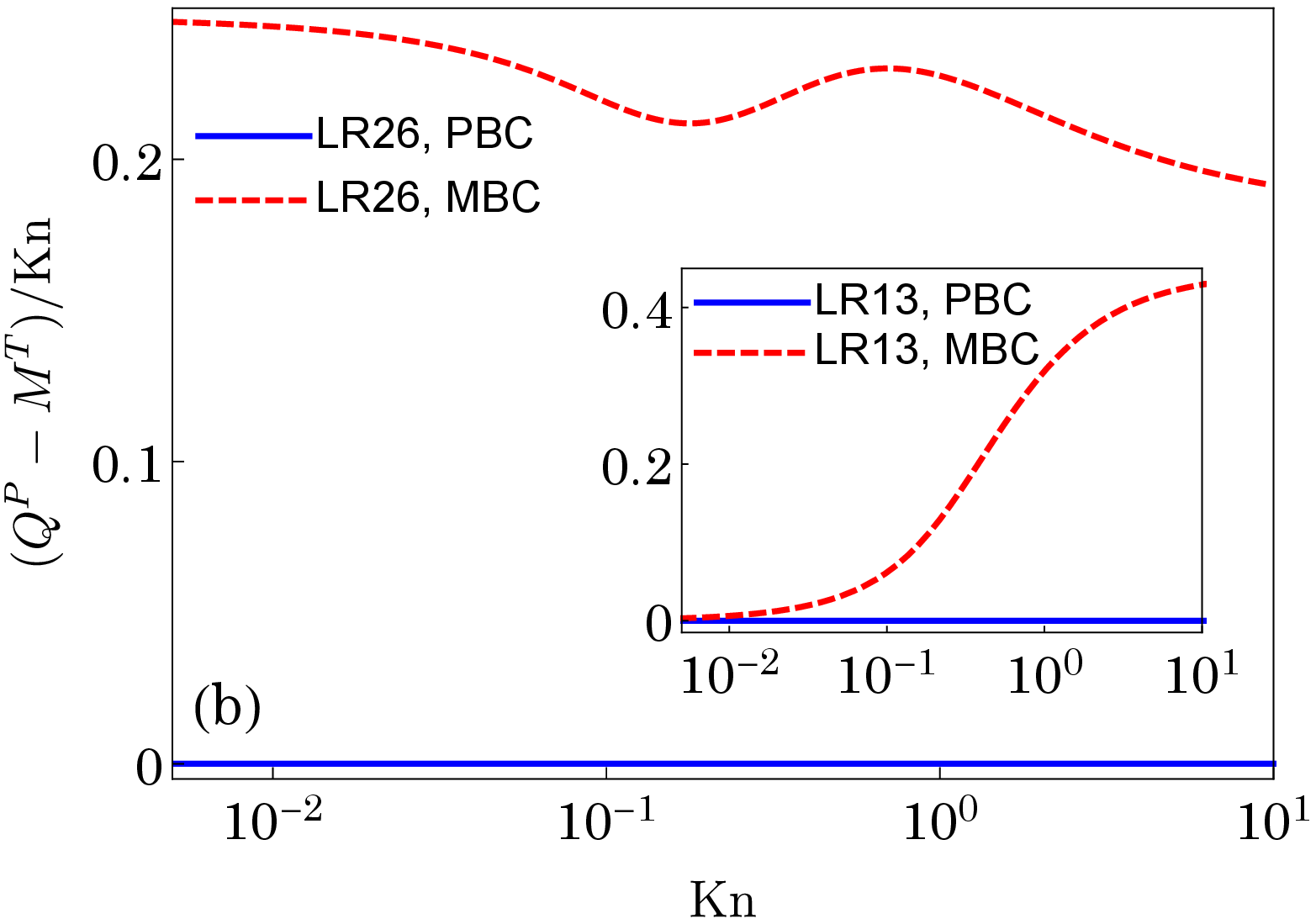}
\caption{\label{fig:PandTplots}
Poiseuille flow and thermal transpiration flow: (a) the total mass flux $M^{P}$ in the Poiseuille flow and (b) violation of the  Onsager reciprocity relation, $M^{T}=Q^{P}$, for the MBC in the case of the LR26 equations (main panel) and in the case of the LR13 equations (inset).
The triangles in the left panel denote the results obtained from the LBE in \cite{WU2017431}.}
\end{figure}

The underlined terms in \eqref{velocity and heat LR26} represent the classical solution of the NSF equations without velocity slip for the Poiseuille flow.
%
The constant term $C_{1}$ in \eqref{velocity and heat LR26} accounts for the velocity slip and terms with exponential functions describe the Knudsen
layer contributions to the velocity and heat flux. 
In classical hydrodynamics,
the velocity profile for the Poiseuille flow is parabolic; 
the LR26 equations add velocity slip $C_{1}$ and Knudsen layer contributions. Similarly, the Fourier's contribution to the heat flux is given by the underlined term in \eqref{HF_LR26}.
The coefficients $\gamma_{i}$'s ($i\in\{1,2,3,4\}$) in \eqref{velocity and heat LR26} determine the thickness of the Knudsen layer; hence the LR26 equations describe the Knudsen layer as a superposition of four exponential functions. 
The coefficients $\beta_1$, $\beta_3$ and $\gamma_{i}$'s ($i\in\{1,2,3,4\}$) for MM read $\beta_1 = -0.24697$, $\beta_3 = 2.25494$, $\gamma_1 = -0.510199$, $\gamma_2 = 0.510199$, $\gamma_3 = -1.26497 $ and $\gamma_4 = 1.26497$ \citep{GuEmersonTang2010, Gu&Emerson2014}.


The total mass flow and heat flow rates of the gas are, respectively, defined as 
\begin{equation}
\label{mfrCCR}
M = \int_{-1/2}^{1/2}v_{x}\,\mathrm{d}y \quad\textrm{and}\quad Q = \int_{-1/2}^{1/2}q_{x}\,\mathrm{d}y.
\end{equation}
We shall use the superscripts `$P$' and `$T$' for the Poiseuille ($F=1$, $\tau=0$) and transpiration ($F=0$, $\tau=1$) flows, respectively.


Figure~\ref{fig:PandTplots}(a) exhibits the mass flow rate, $M^{P}$, in the Poiseuille flow plotted over the Knudsen number. 
The results computed from the MBC  applied to the LR26 equations with the MM collision model are compared with the LBE data from \cite{WU2017431}.
The Knudsen minimum is observed in the Poiseuille flow, in which the mass flow rate $M^{P}$ first
decreases with the Knudsen number, attains a minimum value around a critical
Knudsen number and then increases; see figure~\ref{fig:PandTplots}(a). The LR26 equations with PBC closely follow the mass flow rate profile from the LBE up to $\mathrm{Kn}\lesssim 5$. 

Due to the microscopic reversibility, the Onsager reciprocity relations should hold, which in this case state that $M^{T}=Q^{P}$ \citep{WU2017431}. 
It is evident from figure~\ref{fig:PandTplots}(b) that the MBC do not comply with the Onsager reciprocity relations for the LR26 (and also for the LR13; shown in the inset) equations, while the PBC do.

\begin{figure}
\centering
\includegraphics[height=45mm]{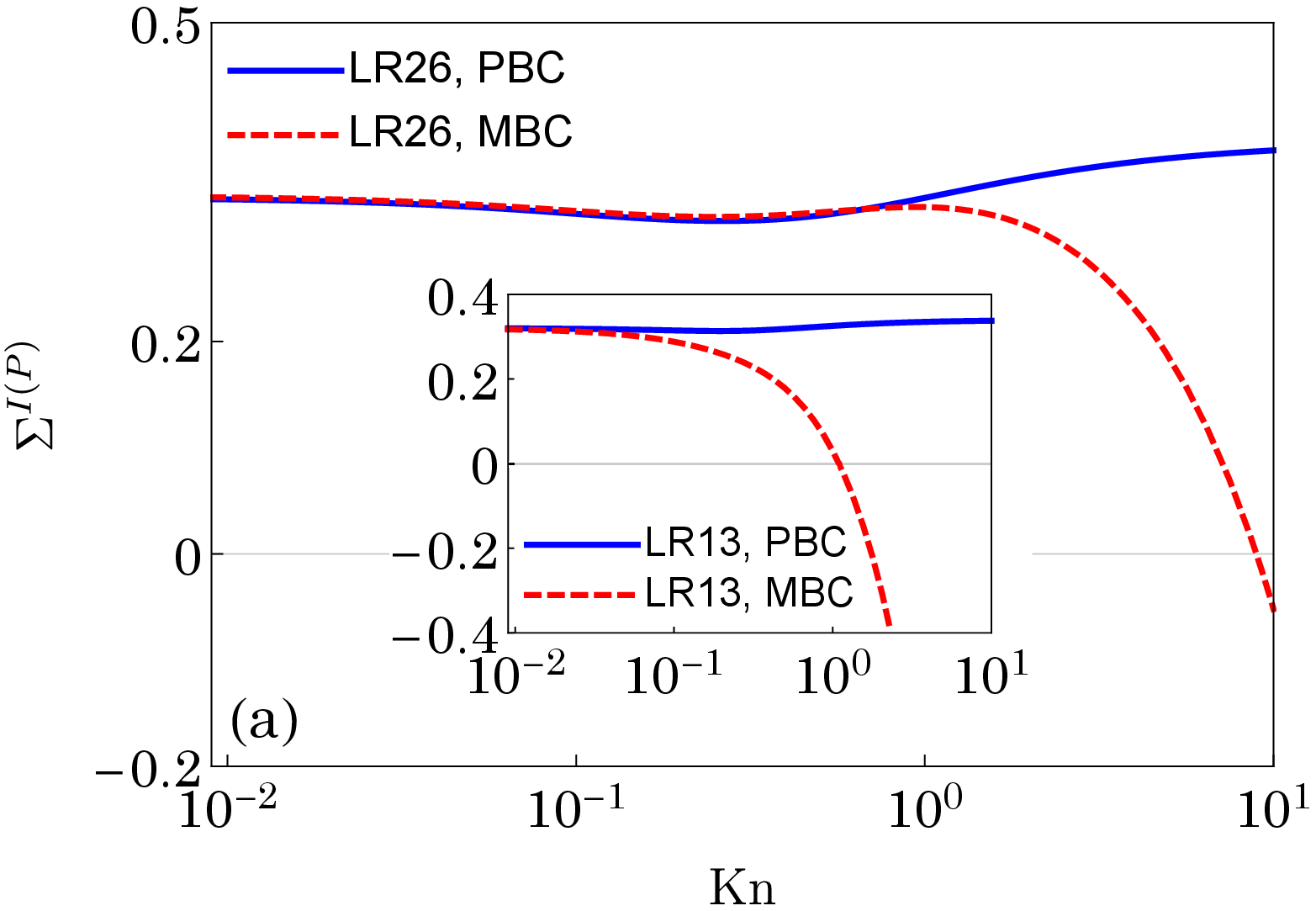}\hfill
\includegraphics[height=45mm]{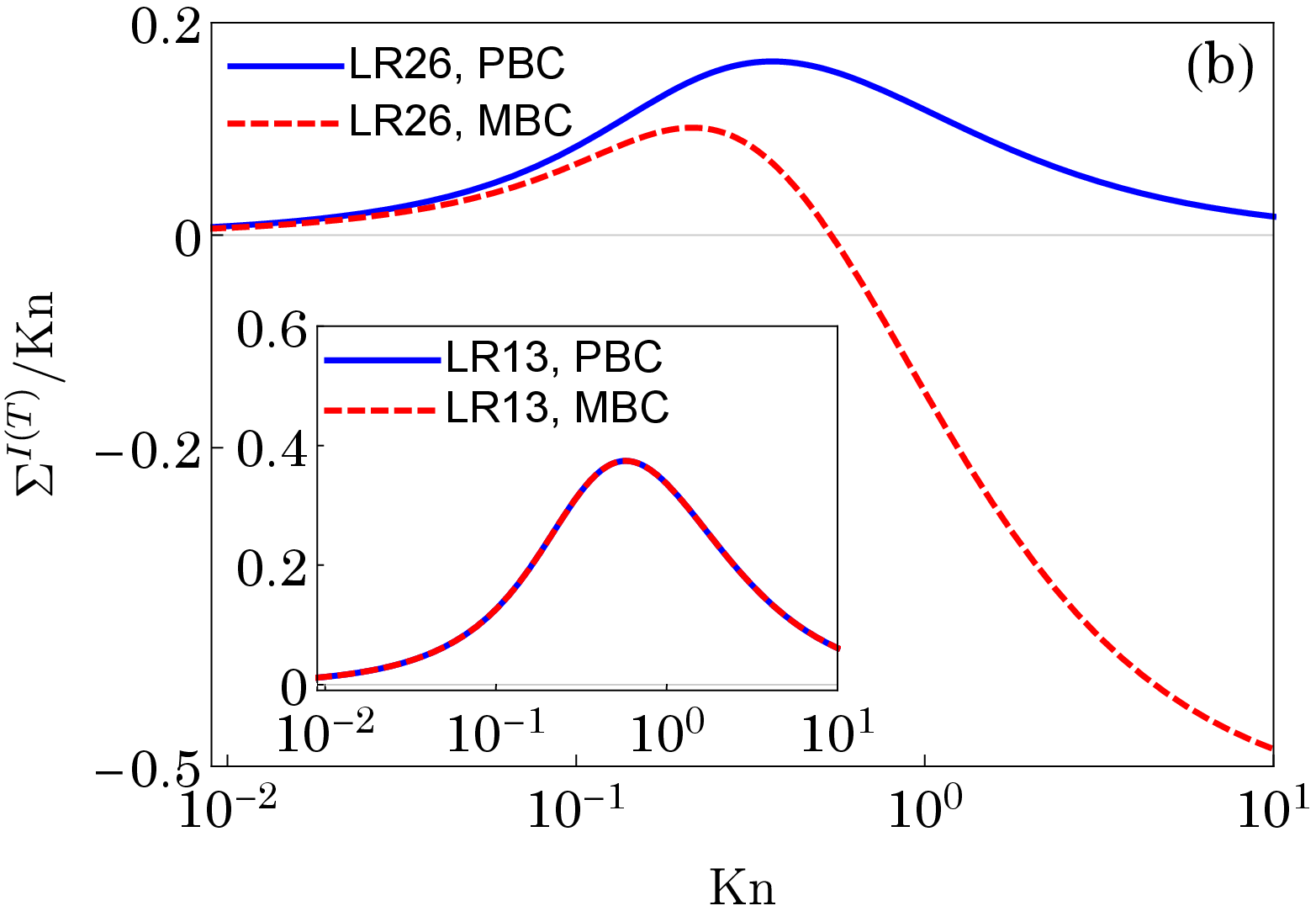}
\caption{\label{fig:PandT2ndlaw}
The entropy generation rates at the boundary for (a) the Poiseuille flow and (b) the thermal transpiration flow. The main panels and insets exhibit the results computed with the LR26 and LR13 equations, respectively.} 
\end{figure}

Figure~\ref{fig:PandT2ndlaw} illustrates the entropy generation rates at the interface computed with the LR26 equations (main panels) and the LR13 equations (insets) in the case of (a) the Poiseuille flow and (b) the transpiration flow.
It is clear from figure~\ref{fig:PandT2ndlaw}(a) that the MBC for the LR26 equations (as well as for the LR13 equations) always violate the second law of thermodynamics (i.e.~the entropy production $\Sigma$ becomes negative) for large Knudsen numbers. 
For the transpiration flow case (figure~\ref{fig:PandT2ndlaw}(b)),  the violation of the second law of thermodynamics by the MBC for the LR26 equations occurs already at a relatively small Knudsen number ($\mathrm{Kn}<1$), which is within the transition regime. 
Surprisingly, the LR13 equations with MBC in the transpiration case  do not contravene with the second law of thermodynamics whilst they do in the Poiseuille flow case around $\mathrm{Kn}=1$; see the insets in figure~\ref{fig:PandT2ndlaw}. 
Nevertheless, the PBC always predict a positive entropy production in both the cases for both models in contrast to the conventional MBC.
\subsection{Evaporation of a spherical liquid droplet}
\begin{figure}
\centering
\includegraphics[scale=0.6]{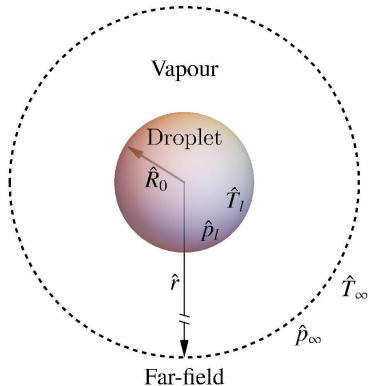}
\caption{\label{fig:evp_drop} Schematic of an evaporating liquid droplet surrounded by its own vapour.}
\end{figure}

We consider a liquid droplet of (fixed) initial radius $\hat{R}_0$ with a given interface temperature $\hat{T}_l$ and the corresponding saturation pressure $\hat{p}_{\mathrm{sat}}$, immersed in
its own vapour; see figure~\ref{fig:evp_drop}. 
The far-field (dimensional) temperature $\hat{T}_{\infty}$, pressure $\hat{p}_{\infty}$ and the initial radius of the droplet $\hat{R}_0$ are used for the purpose of non-dimensionalisation.

We consider two cases: (i) the flow is driven by a unit pressure difference while the temperature of the liquid is equal to the far-field temperature [i.e.~$\Delta p := (\hat{p}_{\mathrm{sat}}-\hat{p}_{\infty})/\hat{p}_{\infty}=1$ and $\Delta T := (\hat{T}_{l}-\hat{T}_{\infty})/\hat{T}_{\infty}=0$] and (ii)  the flow is driven by a unit temperature difference while the saturation pressure is the same as the far-field pressure (i.e.~$\Delta p =0 $ and $\Delta T =1$). 
The problem has been investigated in detail by \citet{RanaLockerbySprittles2018}. 
We skip the details here for the sake of succinctness and focus mainly on looking at the Onsager reciprocity relations; for computational (and other) details, the reader is referred to \cite{RanaLockerbySprittles2018}.
The Onsager reciprocity relations for this problem lead to $Q^{\Delta p}=M^{\Delta T}$ \citep{CM1989},
where $Q^{\Delta p}$ is the  (dimensionless) heat flux in the pressure-driven process, $M^{\Delta T}$ is the (dimensionless) mass flux in the temperature-driven process, and 
the superscripts `$\Delta p$' and `$\Delta T$' are used to denote the pressure- and temperature-driven processes, respectively. 

\begin{figure}
\centering
\includegraphics[height=45mm]{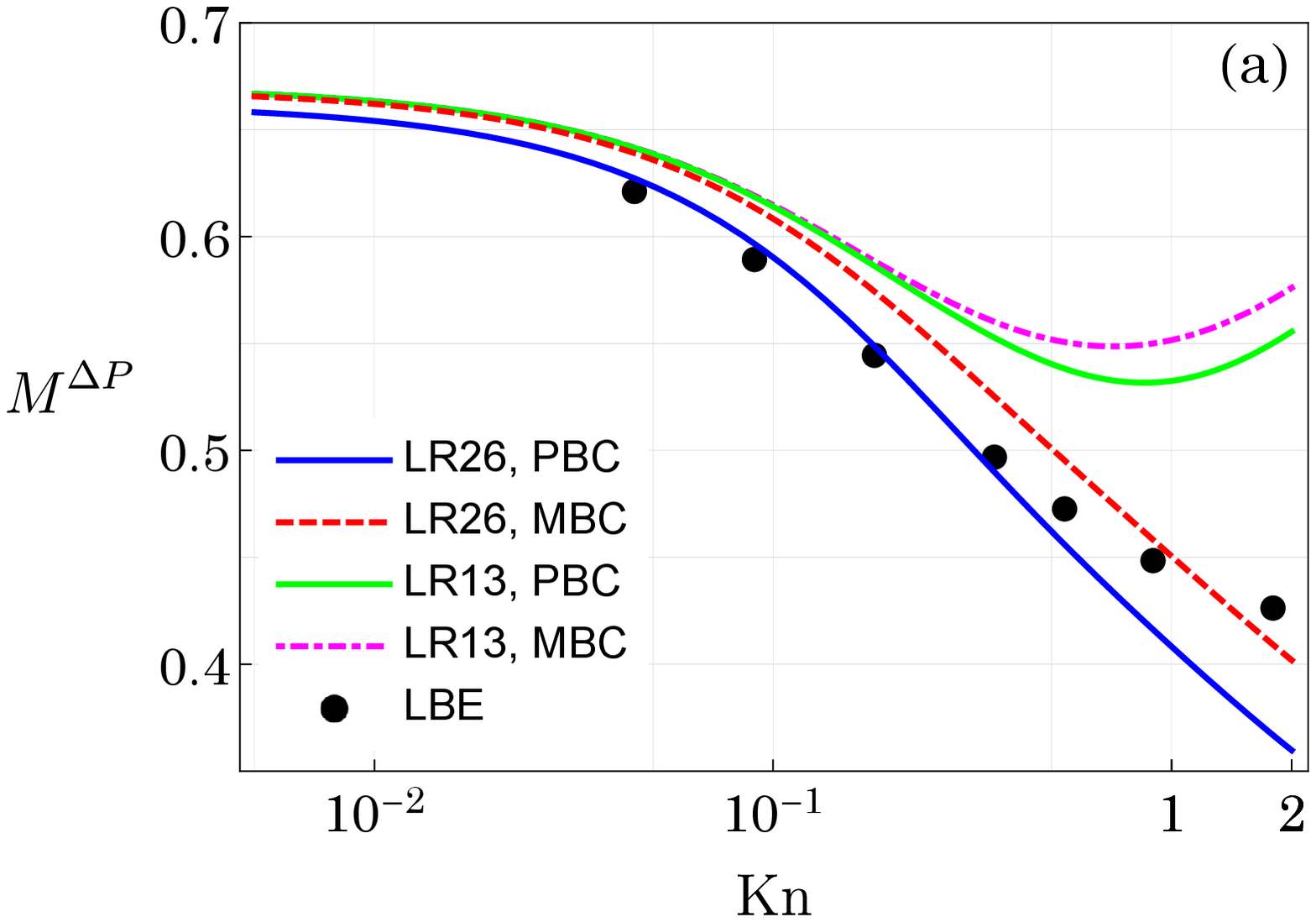}
\hfill
\includegraphics[height=45mm]{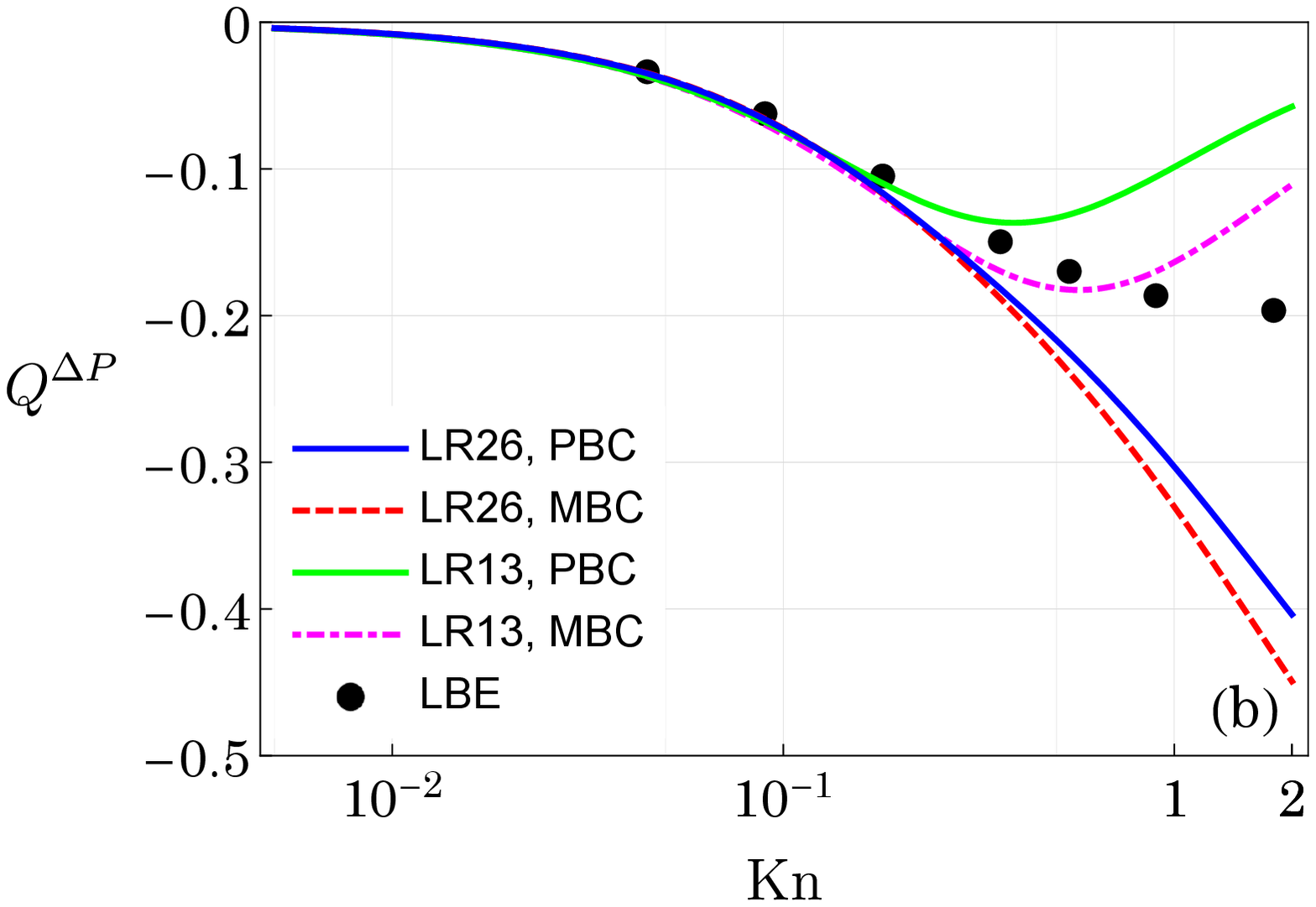}
\\
\includegraphics[height=45mm]{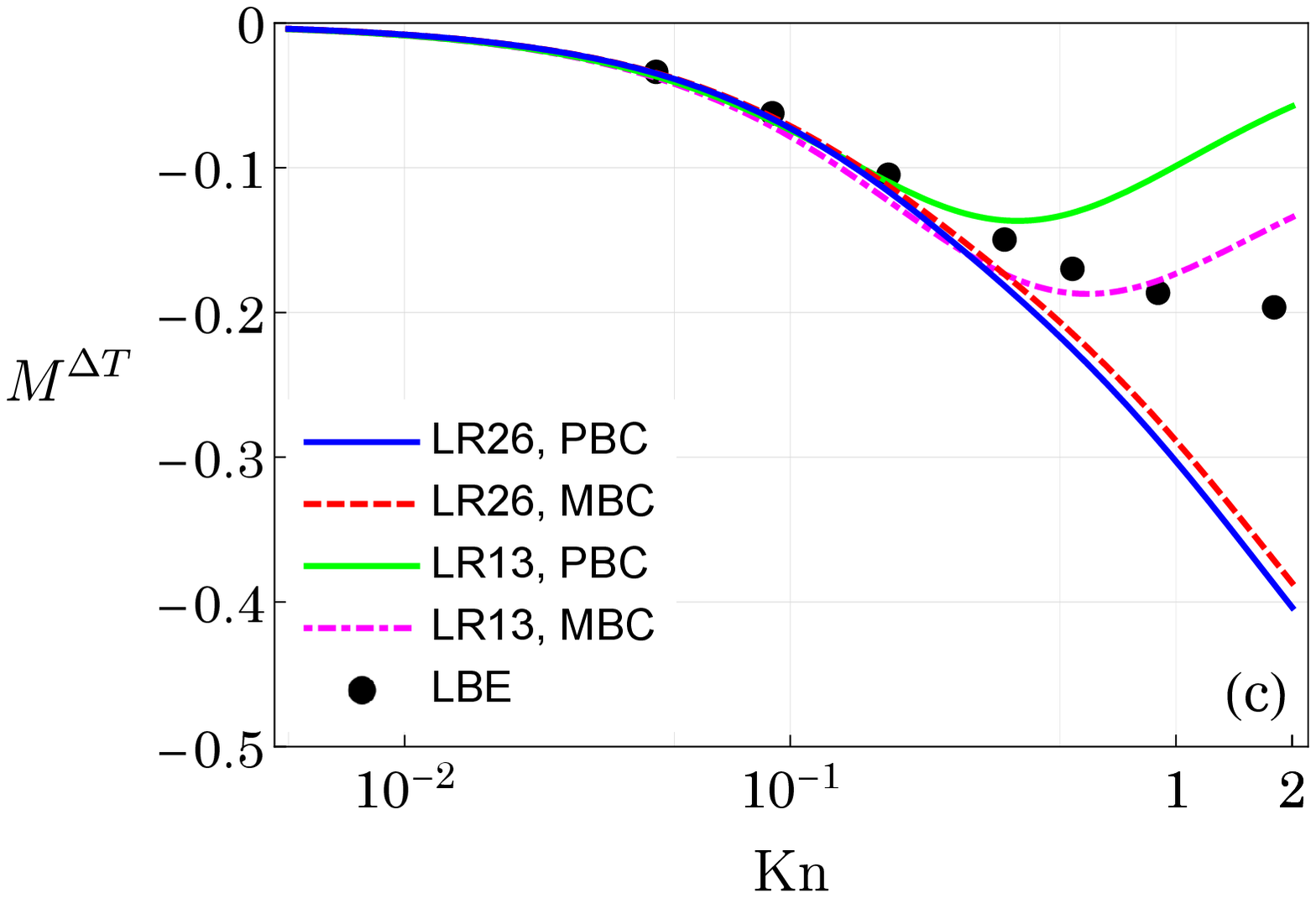}
\hfill
\includegraphics[height=45mm]{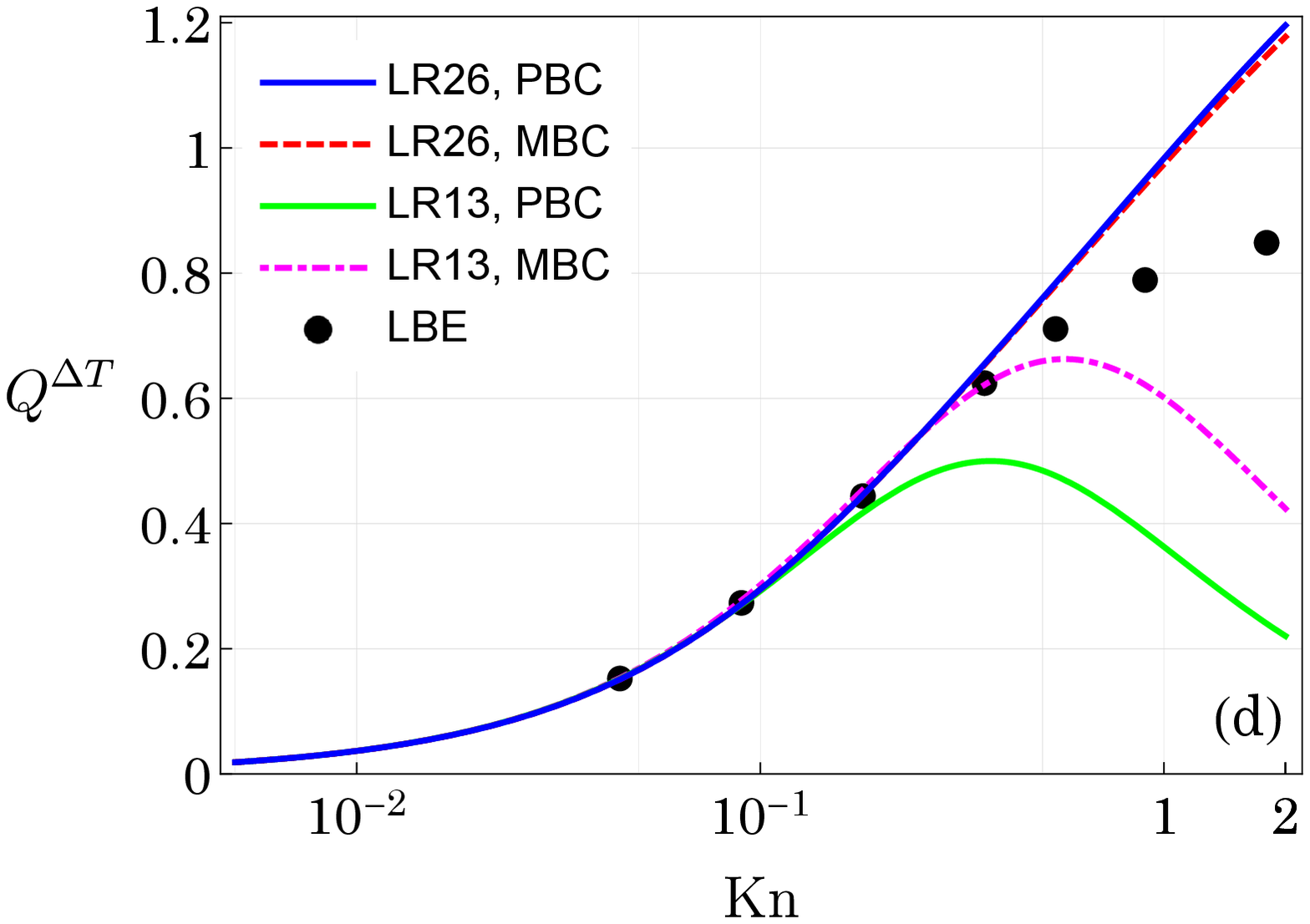}
\caption{\label{fig:evp_droplet_massheatfluxes}Profiles of the mass and heat fluxes plotted over the Knudsen number for the problem of evaporation of a spherical liquid droplet: (a) mass flux in the pressure-driven case, (b) heat flux in the pressure-driven case, (c) mass flux in the temperature-driven case and (d) heat flux in the temperature-driven case. 
The LBE data (symbols) are taken from \cite{Sone1994}.}
\end{figure}
\begin{figure}
\centering
\includegraphics[scale=0.5]{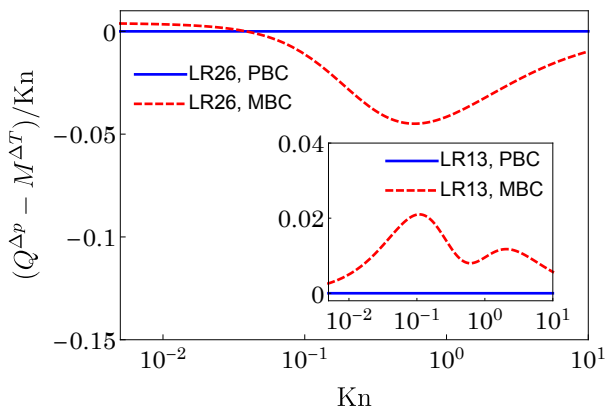}
\caption{\label{fig:dropletOnsager} Violation of the Onsager reciprocity relations for an evaporating droplet.}
\end{figure}

The mass and heat fluxes in the pressure- and temperature-driven processes computed from the R13 and R26 theories using the PBC and MBC are illustrated in figure~\ref{fig:evp_droplet_massheatfluxes} for different values of the Knudsen number. 
The top and bottom rows of figure~\ref{fig:evp_droplet_massheatfluxes} exhibit the results for the pressure- and temperature-driven processes, respectively, while the left and right columns display the mass and heat fluxes, respectively.
For comparison purpose, the results obtained by solving the LBE \citep{Sone1994} are also included and they are depicted by symbols in the figure.
In general, the results from both the R13 and R26 theories agree with those from the LBE for $\mathrm{Kn} \lesssim 0.5$. 
However, for the mass flux in the pressure-driven case [figure~\ref{fig:evp_droplet_massheatfluxes}(a)], the R26 equations on using the PBC (blue solid lines) give an excellent match with the LBE results (symbols) for $\mathrm{Kn} \lesssim 0.5$ whereas the results from the other theories are comparatively higher than the LBE results even for small Knudsen numbers. 
Let us focus on panels (b) and (c) of figure~\ref{fig:evp_droplet_massheatfluxes}, which illustrate the cross-effects, i.e.~the heat flux in the pressure-driven case and mass flux in the temperature-driven case, respectively, and account for the verification of the Onsager reciprocity relations. 
The LBE predicts the same values the heat flux in the pressure-driven case and mass flux in the temperature-driven case, thanks to the Onsager reciprocity relations \citep{CM1989, Sone1994}. 
Evidently from panels (b) and (c) of figure~\ref{fig:evp_droplet_massheatfluxes}, the heat flux in the pressure-driven case computed with the LR26/LR13 theory using the PBC (blue solid line for the LR26 equations and green solid line for the LR13 equations) seems to be the same as the mass flux in the temperature-driven case computed with the LR26/LR13 theory using the PBC (blue solid line for the LR26 equations and green solid line for the LR13 equations) whereas the heat flux in the pressure-driven case computed with the LR26/LR13 theory using the MBC (red dashed line for the LR26 equations and magenta dashed line for the LR13 equations) seems to be different from the mass flux in the temperature-driven case computed with the LR26/LR13 theory using the MBC (red dashed line for the LR26 equations and magenta dashed line for the LR13 equations). 
In order to have a clear picture of the Onsager reciprocity relations, the difference of $Q^{\Delta p}$ and $M^{\Delta T}$ is plotted in figure~\ref{fig:dropletOnsager}, which clearly shows that both the LR13 and LR26 equations on using the MBC for this problem violate the Onsager reciprocity relations while both the LR13 and LR26 equations on using the PBC respect the Onsager reciprocity relations for all Knudsen numbers.

We have also computed the entropy generation rate at the interface for this problem and found (but not shown here for the sake of conciseness) that it remains positive for the R13 and R26 equations with the PBC in both pressure- and temperature-driven cases, also remains positive for the R13 equations with the MBC in the pressure-driven case but does not remain positive for the R13 equations with the MBC in the temperature-driven case for all Knudsen numbers. The R26 equations with the MBC yield a positive entropy generation rate for this problem.

%

%
\section{Gas flow past an evaporating droplet: a new analytic solution}
\label{section:flowpastdroplet}

As an application of the above findings, we compute the analytic solution of the LR26 equations for a steady vapour passing over an evaporating spherical droplet. 
Furthermore, we also consider a special case of this problem---when there is no evaporation;
this is essentially the problem of a streaming flow of a gas past a rigid sphere, which is a classical problem in fluid dynamics.
The results presented in this section are determined using the PBC only, since we have established that they are the physically admissible boundary conditions. 
Nevertheless, we show in the supplementary material that the difference between the results obtained using the PBC and MBC for this problem is negligibly small.

%

\begin{figure}
\centering
\includegraphics[scale=0.5]{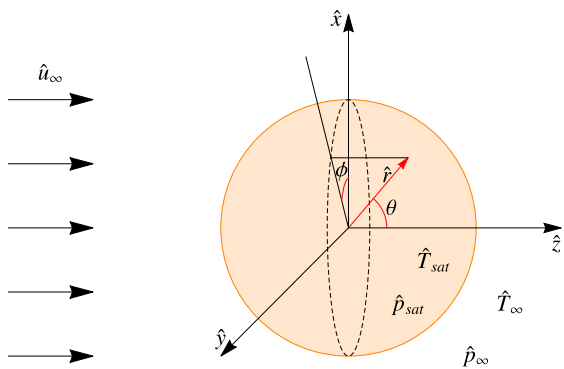}
\caption{\label{fig:flowoversphere}
Schematic of a flow past a sphere.}
\end{figure}

We consider a rarefied gas flow past a spherical droplet as illustrated in figure~\ref{fig:flowoversphere}. It is convenient to study the problem in the spherical coordinate system $(r,\theta,\phi)$ that relates to the Cartesian coordinate system $(x,y,z)$ via $(x,y,z) \equiv (r \sin{\theta} \cos{\phi}, r \sin{\theta} \sin{\phi}, r \cos{\theta})$, where $r\in[0,\infty)$, $\theta\in[0,\pi]$ and $\phi\in[0,2\pi)$. 
The flow is assumed to be in the $(r,\theta)$-plane coming from the left as shown in the figure. 
Owing to spherical symmetry, the flow is independent of the direction $\phi$. 
Consequently, all the field variables can depend only on $r$ and $\theta$. 
All the vectors and tensors involved in the problem need to be converted to spherical coordinates. 
To this end, it is worthwhile noting that a symmetric-tracefree $k$-rank tensor has $2k+1$ independent components in three dimensions \citep{Gupta2020}.  

For the considered problem, any vector $a_i \in \{v_i,q_i,\Omega_i\}$ in the spherical coordinate system is given by
\begin{align}
\label{vectorsph}
a_i=(a_r, a_\theta, 0)^\mathsf{T}.
\end{align}
A symmetric-tracefree second-rank tensor $a_{ij} \in \{\sigma_{ij}, R_{ij}\}$ in the spherical coordinate system is given by
\begin{align}
\label{tensor2sph}
a_{ij}=\begin{bmatrix}
       a_{rr} & a_{r\theta} & 0
       \\
       a_{r\theta} & - \frac{1}{2} a_{rr} & 0
       \\
       0 & 0 & - \frac{1}{2} a_{rr} 
       \end{bmatrix}.
\end{align}
Note that a symmetric-tracefree second-rank tensor has five independent components; however, for the problem under consideration, it is clear from \eqref{tensor2sph} that all components of $a_{ij}$ can be expressed in terms of only two components, namely $a_{rr}$ and $a_{r\theta}$.
A symmetric-tracefree third-rank tensor has seven independent components; however, for the problem under consideration, all components of $a_{ijk}$ can again be expressed in terms of $a_{rrr}$ and $a_{rr\theta}$.
The five independent components (other than $a_{rrr}$ and $a_{rr\theta}$) of $a_{ijk}$, for the problem under consideration, read
\begin{align}
a_{r\theta\theta} = - \frac{1}{2} a_{rrr},
\quad
a_{\theta\theta\theta} = - \frac{3}{4} a_{rr\theta},
\quad 
a_{rr\phi} = a_{\theta\theta\phi} = a_{r\theta\phi} = 0.
\end{align}
A symmetric-tracefree fourth-rank tensor $a_{ijkl}$ has nine independent components; however, for the problem under consideration, all components of $a_{ijkl}$ can again be expressed in terms of $a_{rrrr}$ and $a_{rrr\theta}$.
The seven independent components (other than $a_{rrrr}$ and $a_{rrr\theta}$) of $a_{ijkl}$, for the problem under consideration, read
\begin{align}
\left.
\begin{gathered}
a_{rr\theta\theta} = - \frac{1}{2} a_{rrrr},
\quad
a_{\theta\theta\theta\theta} = \frac{3}{8} a_{rrrr},
\quad
a_{r\theta\theta\theta} = - \frac{3}{4} a_{rrr\theta},
\\ 
a_{rrr\phi} = a_{rr\theta\phi} = a_{r\theta\theta\phi}  = a_{\theta\theta\theta\phi} =0.
\end{gathered}
\right\}
\end{align}
Accordingly, the conservation laws \eqref{massBal}--\eqref{energyBal} for this problem reads
\begin{subequations}
\label{eqn:conservationlaws}
\begin{align}
\frac{\partial v_r}{\partial r}
+\frac{2 v_r}{r}
+\frac{\mathscr{D} v_\theta}{r} 
&= 0,
\\
\frac{\partial p}{\partial r}
+\frac{\partial \sigma_{rr}}{\partial r}
+\frac{3 \sigma_{rr}}{r}
+\frac{\mathscr{D} \sigma_{r\theta}}{r}
&= 0,
\\
\frac{\partial \sigma_{r\theta}}{\partial r}
+\frac{3 \sigma_{r\theta}}{r}
-\frac{1}{2 r} \frac{\partial \sigma_{rr}}{\partial \theta}
+\frac{1}{r} \frac{\partial p}{\partial \theta}
&= 0,
\\
\frac{\partial q_r}{\partial r}
+\frac{2 q_r}{r}
+\frac{\mathscr{D} q_\theta}{r}
&= 0,
\end{align}
\end{subequations}
where $\mathscr{D} \equiv \cot {\theta} + \frac{\partial}{\partial \theta}$. 
The stress balance equation \eqref{stressBal} reduces to
\begin{subequations}
\label{eqn:stressbalance}
\begin{align}
\frac{\partial m_{rrr}}{\partial r}
&+\frac{4 m_{rrr}}{r}
+\frac{4}{5} \frac{\partial q_r}{\partial r}
+2 \frac{\partial v_r}{\partial r}
+\frac{\mathscr{D} m_{rr\theta}}{r}
=-\frac{1}{\mathrm{Kn}} \sigma_{rr},
\\
\frac{\partial m_{rr\theta}}{\partial r}
&+\frac{4 m_{rr\theta}}{r}
+\frac{2}{5} \left(\frac{\partial q_\theta}{\partial r}
-\frac{q_\theta}{r}
\right)
+\frac{\partial v_\theta}{\partial r}
-\frac{v_\theta}{r}
\nonumber\\
&-\frac{1}{2 r} \frac{\partial m_{rrr}}{\partial \theta}
+\frac{1}{r} \frac{\partial v_r}{\partial \theta}
+\frac{2}{5r} \frac{\partial q_r}{\partial \theta}
= - \frac{1}{\mathrm{Kn}} \sigma_{r\theta}.
\end{align}
\end{subequations}
The heat flux balance equation \eqref{HFBal} gives
\begin{subequations}
\label{eqn:heatbalance}
\begin{align}
\frac{1}{2} \left(\frac{\partial R_{rr}}{\partial r}
+\frac{\mathscr{D} R_{r\theta}}{r}
+\frac{3 R_{rr}}{r}\right)
+\frac{1}{6} \frac{\partial \Delta}{\partial r}
-\frac{\partial p}{\partial r}
+\frac{5}{2} \frac{\partial T}{\partial r}
&= - \frac{\mathrm{Pr}}{\mathrm{Kn}} q_r,
\end{align}
\begin{align}
\frac{1}{2} \left(\frac{\partial R_{r\theta}}{\partial r}
+\frac{3 R_{r\theta}}{r}\right)
+\frac{1}{6r} \frac{\partial \Delta}{\partial \theta}
-\frac{1}{4 r} \frac{\partial R_{rr}}{\partial \theta}
-\frac{1}{r}\frac{\partial p}{\partial \theta}
+\frac{5}{2 r} \frac{\partial T}{\partial \theta}
&= - \frac{\mathrm{Pr}}{\mathrm{Kn}} q_\theta.
\end{align}
\end{subequations}
The $m_{ijk}$ balance equation \eqref{mBal} simplifies to
\begin{subequations}
\label{eqn:mbalance}
\begin{align}
\frac{1}{r}\mathscr{D} %
\bigg[-\frac{6}{5}\sigma_{r\theta}&+\Phi_{rrr\theta} - \frac{6}{35}R_{r\theta}\bigg] +\frac{9}{5}\left( \frac{\partial }{\partial r}%
-\frac{2}{r}\right) \sigma_{rr}
\nonumber \\
&+\left( \frac{\partial }{\partial r}+\frac{5}{r}\right) \Phi_{rrrr}
+\frac{9}{35}\left( \frac{\partial }{\partial r}-\frac{2}{r}\right) R_{rr} = -\frac{\mathrm{Pr}_{m}}{\mathrm{Kn}}m_{rrr}, 
\\
\frac{6}{5r}\frac{\partial \sigma_{rr}}{\partial \theta} 
&+\frac{6}{35r} \frac{\partial R_{rr}}{\partial \theta }-\frac{1}{2r}\frac{\partial \Phi_{rrrr}}{\partial \theta} +\frac{8}{5}\left( \frac{\partial }{\partial r}-\frac{2}{r}\right) \sigma_{r\theta}
\nonumber\\
&+\frac{8}{35} \left( \frac{\partial }{\partial r}-\frac{2}{r}\right) R_{r\theta}+\left( \frac{\partial }{\partial r}+\frac{5}{r}\right) \Phi_{rrr\theta }=-\frac{\mathrm{Pr}_{m}}{\mathrm{Kn}}m_{rr\theta}.
\end{align}
\end{subequations}

The $R_{ij}$ balance equation \eqref{RBal} reduces to
\begin{subequations}
\label{eqn:T=Rbalance}
\begin{align}
\frac{1}{r}\mathscr{D}%
\bigg[ 2m_{rr\theta}
&-\frac{2}{15}\Omega_{\theta }+\psi_{rr\theta}-\frac{28}{15}q_\theta \bigg] +\frac{56}{15}\left( 
\frac{\partial }{\partial r}-\frac{1}{r}\right) q_r  
+2\left( \frac{\partial }{\partial r}+\frac{4}{r}\right) m_{rrr}
\nonumber\\
&+\left( \frac{\partial }{\partial r}+\frac{4}{r}\right) \psi_{rrr}+%
\frac{4}{15}\left( \frac{\partial }{\partial r}-\frac{1}{r}\right) \Omega_{r}=-\frac{\mathrm{Pr}_{R}}{\mathrm{Kn}}R_{rr},
\\
2\left( \frac{\partial }{\partial r}+\frac{4}{r}\right) m_{rr\theta}
&+\left( \frac{\partial }{\partial r}+\frac{4}{r}\right) \psi_{rr\theta}+\frac{1}{5}\left( \frac{\partial }{\partial r}-\frac{1}{r}%
\right) \Omega _{\theta} + \frac{14}{5}\left( \frac{\partial }{\partial r}-%
\frac{1}{r}\right) q_\theta  
\nonumber\\
&-\frac{1}{r}\frac{\partial m_{rrr}}{\partial \theta }+\frac{14}{5r}%
\frac{\partial q_r }{\partial \theta }-\frac{1}{2r}\frac{\partial 
\psi_{rrr}}{\partial \theta }+\frac{2}{5r}\frac{\partial 
\Omega_{r}}{\partial \theta }=-\frac{\mathrm{Pr}_{R}}{\mathrm{Kn}}R_{r\theta }.
\end{align}
\end{subequations}
The $\Delta$ balance equation \eqref{DeltaBal} gives
\begin{equation}
\frac{1}{r}\mathscr{D}
\left( 8 q_{\theta}+\Omega _{\theta }\right) +\left( \frac{\partial 
}{\partial r}+\frac{2}{r}\right) \left( 8 q_{r}+\Omega_{r}\right) =-\frac{\mathrm{Pr}_{\Delta }}{\mathrm{Kn}}\Delta.
\label{eqn:deltabalance}
\end{equation}
Finally, the LR26 closure (\ref{R26constRel}) simplifies to
\begin{subequations}
\label{eqn:closuretensors}
\begin{align}
\frac{4}{7}\left( \frac{\partial }{\partial r}-\frac{3}{r}\right) m_{rrr} 
-\frac{3}{7}\frac{\mathscr{D} m_{rr\theta}}{r} 
&= - \frac{1}{4} \frac{\mathrm{Pr}_{\Phi}}{\mathrm{Kn}}\Phi_{rrrr} ,
\\
\frac{15}{28}\left( \frac{\partial }{\partial r}-\frac{3}{r}\right) m_{rr\theta}
+\frac{5}{14 r} \frac{\partial m_{rrr}}{\partial \theta} 
&= -\frac{1}{4} \frac{\mathrm{Pr}_{\Phi}}{\mathrm{Kn}} \Phi_{rrr\theta } ,
\\
\frac{3}{5}\left( \frac{\partial }{\partial r}-\frac{2}{r}\right) R_{rr} 
-\frac{2}{5} \frac{\mathscr{D} R_{r\theta}}{r}
&=-\frac{7}{27} \frac{\mathrm{Pr}_{\psi}}{\mathrm{Kn}}\psi_{rrr}  ,
\\
\frac{8}{15}\left( \frac{\partial }{\partial r}-\frac{2}{r}\right) \text{$%
R_{r\theta }$}+\frac{2}{5r}\frac{\partial R_{rr}}{\partial \theta }
&=-\frac{7}{27} \frac{\mathrm{Pr}_{\psi}}{\mathrm{Kn}}\psi_{rr\theta} ,
\end{align}
\end{subequations}
and
\begin{subequations}
\begin{align}
\frac{\partial \Delta }{\partial r}+\frac{12}{7r}\mathscr{D} R_{r\theta}+\frac{12}{7}%
\left( \frac{\partial }{\partial r}+\frac{3}{r}\right) R_{rr} &=-\frac{3}{7}
\frac{\mathrm{Pr}_{\Omega}}{\mathrm{Kn}} \Omega_{r}, 
\\
\frac{1}{r}\frac{\partial \Delta}{\partial \theta }+\frac{12}{7}\left( 
\frac{\partial R_{r\theta}}{\partial r}+\frac{3 R_{r\theta}}{r}\right) -\frac{6}{7r}\frac{\partial R_{rr}}{\partial \theta }
&=-\frac{3}{7} 
\frac{\mathrm{Pr}_{\Omega}}{\mathrm{Kn}} \Omega_{\theta}.
\end{align}%
\label{eqn:omegabalance}
\end{subequations}

The mathematical structure of the LR26 equation is the same as that of the LR13 equations, with which \cite{Torrilhon2010} obtained an analytic solution for the problem of gas flow past a rigid sphere. Following \cite{Torrilhon2010}, we shall adopt the same solution method to solve \eqref{eqn:conservationlaws}--\eqref{eqn:omegabalance}. 
In the next subsection, we shall outline the steps involved in solving the LR26 equations for the considered problem. For more  details, the reader is referred to \cite{Torrilhon2010}.
\subsection{Analytic solution of the LR26 equations}
The analytic solutions of the LR26 equations \eqref{eqn:conservationlaws}--\eqref{eqn:omegabalance} for the aforesaid problem can be obtained by converting these partial differential equations to ordinary differential equations via assuming an explicit dependency on the azimuthal angle $\theta$. 
To this end, the vectorial and tensorial components with an odd number of indices in $\theta$ are taken to be proportional to $\sin {\theta}$ while scalars and tensorial components with an even number of indices in $\theta$ are taken to be proportional to $\cos{\theta}$ \citep{Torrilhon2010}. 
The components for vector fields, namely the velocity $\bm{v}(r,\theta)$ and the heat flux $\bm{q}(r,\theta)$, are taken to be
\begin{align}
\label{eqn:vectorfields}
\left.
\begin{aligned}
v_r(r, \theta) &= u_{\infty} [1+a_{1}(r)] \cos{\theta}, &
\quad
v_{\theta}(r, \theta) &= -u_{\infty}[1+a_{2}(r)] \sin{\theta},
\\
q_r(r, \theta) &= \alpha_1(r) \cos{\theta}, &
\quad
q_\theta(r, \theta) &= \alpha_2(r) \sin{\theta}.
\end{aligned}
\right\}
\end{align}
The scalars, namely $T(r,\theta)$, $p(r,\theta)$ and $\Delta(r,\theta)$, are taken to be
\begin{align}
\label{eqn:scalarfields}
T(r,\theta) = c(r) \cos{\theta},
\quad
p(r,\theta) = d(r) \cos{\theta},
\quad
\Delta(r,\theta) = \delta(r)\cos{\theta}.
\end{align}
The components of the second- and third-rank tensors are taken to be 
\begin{align}
\label{eqn:tensorrfields}
\left.
\begin{aligned}
\sigma_{rr}(r, \theta) &= b_1(r)\cos{\theta}, &
\quad
\sigma_{r\theta}(r, \theta) &= b_2(r)\sin{\theta},
\\
R_{rr}(r, \theta) &= \beta_1(r)\cos{\theta}, &
\quad
R_{r\theta}(r, \theta) &= \beta_2(r) \sin{\theta},
\\
m_{rrr}(r, \theta) &= \kappa_1(r)\cos{\theta}, &
\quad
m_{rr\theta}(r, \theta) &= \kappa_2(r) \sin{\theta}.
\end{aligned}
\right\}
\end{align}
In \eqref{eqn:vectorfields}--\eqref{eqn:tensorrfields}, $a_{1}(r)$, $a_{2}(r)$,  $\alpha_{1}(r)$, $\alpha_{2}(r)$,  $c(r)$, $d(r)$, $\delta(r)$, $b_{1}(r)$, $b_{2}(r)$, $\beta_{1}(r)$, $\beta_{2}(r)$, $\kappa_{1}(r)$ and $\kappa_{2}(r)$ are thirteen unknowns in total, which need to be determined from thirteen equations \eqref{eqn:conservationlaws}--\eqref{eqn:deltabalance}. 

Substitution of constitutive relations \eqref{eqn:closuretensors}--\eqref{eqn:omegabalance} along with solution ansatz \eqref{eqn:vectorfields}--\eqref{eqn:tensorrfields} in  \eqref{eqn:conservationlaws}--\eqref{eqn:deltabalance} results into a set of eight first-order ordinary differential equations (emanating from \eqref{eqn:conservationlaws}--\eqref{eqn:heatbalance}) and five second-order ordinary differential equations (from \eqref{eqn:mbalance}--\eqref{eqn:deltabalance}). 
The latter include second derivatives of $\delta(r)$, $\beta_1(r)$, $\beta_2(r)$, $\kappa_1(r)$ and $\kappa_2(r)$ due to regularised constitutive relations \eqref{eqn:closuretensors}--\eqref{eqn:omegabalance}.  
%
%
With this, one obtains an analytic solution for the unknowns $a_{1}(r)$, $a_{2}(r)$,  $\alpha_{1}(r)$, $\alpha_{2}(r)$,  $c(r)$, $d(r)$, $\delta(r)$, $b_{1}(r)$, $b_{2}(r)$, $\beta_{1}(r)$, $\beta_{2}(r)$, $\kappa_{1}(r)$ and $\kappa_{2}(r)$, which has been relegated to appendix \ref{app:sol} for better readability.
Expressions \eqref{eqn:vectors}--\eqref{eqn:kappa2} contain eight integration constants $C_{1,2,3}$ and $K_{1,2,3,4, 5}$ that are determined from boundary conditions \eqref{BCscalar}--\eqref{BCvector}. 

%
%
Let us investigate a few qualitative features of these solutions. 
These solutions are the sums of the bulk contributions (which are polynomials in $1/r$) and the Knudsen layer contributions (terms containing the exponential functions $\Theta_i(r)$); see  appendix \ref{app:sol}. 
The bulk solutions contain three constants of integration $C_{1,2,3}$ while Knudsen layers introduce five additional constants---$K_{1,2}$ for vector variables and $K_{3,4,5}$ for scalar variables. 
The bulk part can further be divided into the classical NSF contribution and second-order contribution (in the Knudsen number) that is neglected in the classical NSF theory. 
In the hydrodynamic limit (i.e.~Stokes flow without velocity slip and temperature jump), $C_1=-3$ and $C_2=1/2$ while $C_3=0$ \citep{Lamb1932}. 
The last relation ($C_3=0$) implies an isothermal flow condition. 
Hence, the non-zero bulk contributions to the temperature and heat flux stem from the second-order terms, i.e.~the terms with $C_3$ in equations (\ref{eqn:scalars}a) and (\ref{eqn:vectors}c,d). 
The NSF equations with velocity-slip boundary condition predict $C_1$ and $C_2$ as a function of the Knudsen number $\mathrm{Kn}$ and the accommodation coefficient $\chi$ \citep{Torrilhon2010}.


For the Knudsen layer contributions in \eqref{eqn:vectors}--\eqref{eqn:kappa2}, the LR13 theory \citep{Torrilhon2010} approximates Knudsen layers with two exponential functions whilst the LR26 theory give superposition of five exponential functions to describe the Knudsen layers. On the contrary, the NSF theory does not predict Knudsen layers at all. 
In the following, we shall show that the LR26 theory gives a remarkably improved flow description with added details to the Knudsen layers.
%
\subsection{
Gas flow past a rigid sphere}
First, let us consider the classical problem of a gas flow past a rigid sphere. 
The same problem was investigated by \citet{Torrilhon2010} using the LR13 and linear NSF equations with velocity slip.

\begin{figure}
\centering
\includegraphics[height=45mm]{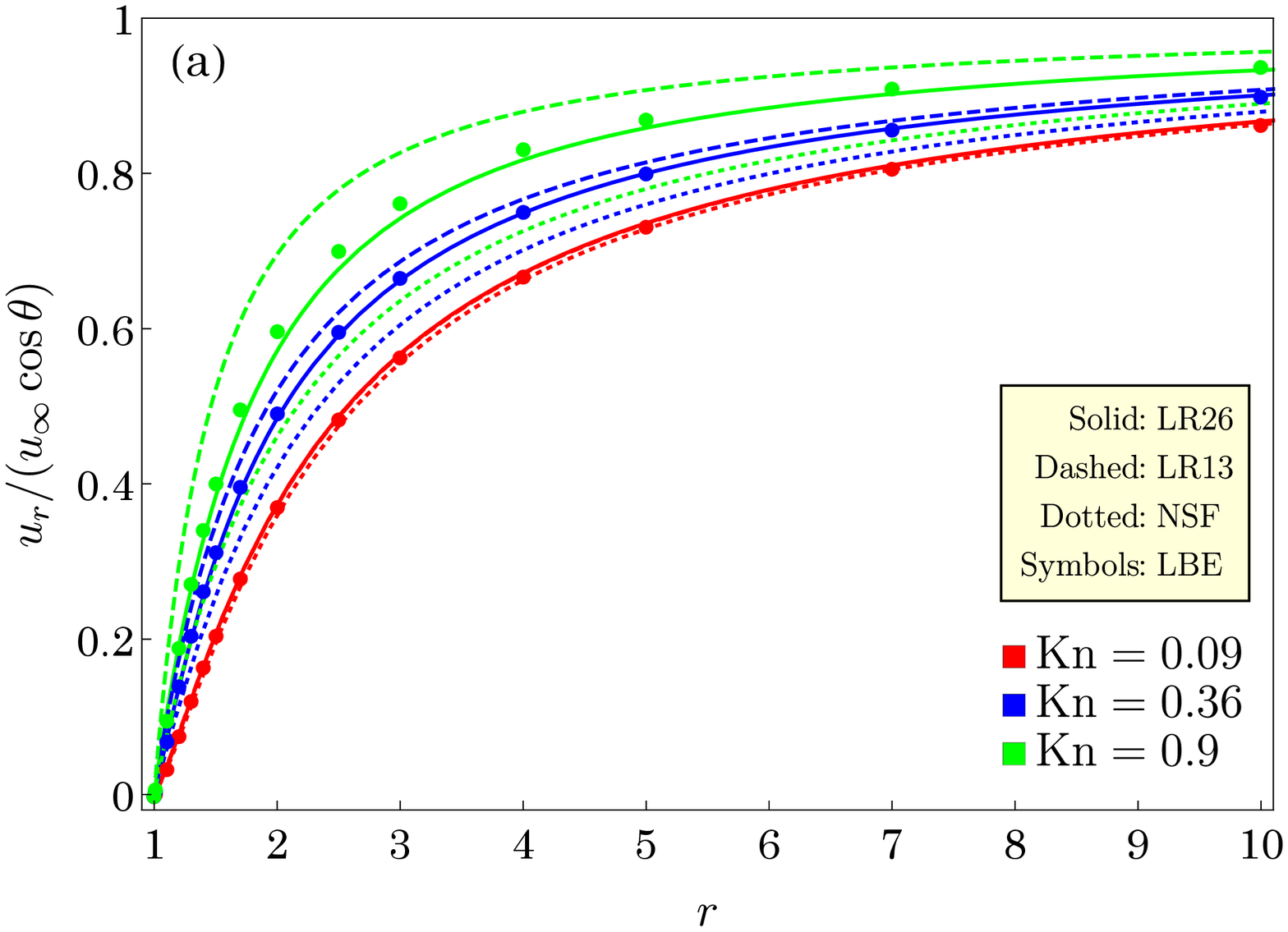}
\hfill
\includegraphics[height=45mm]{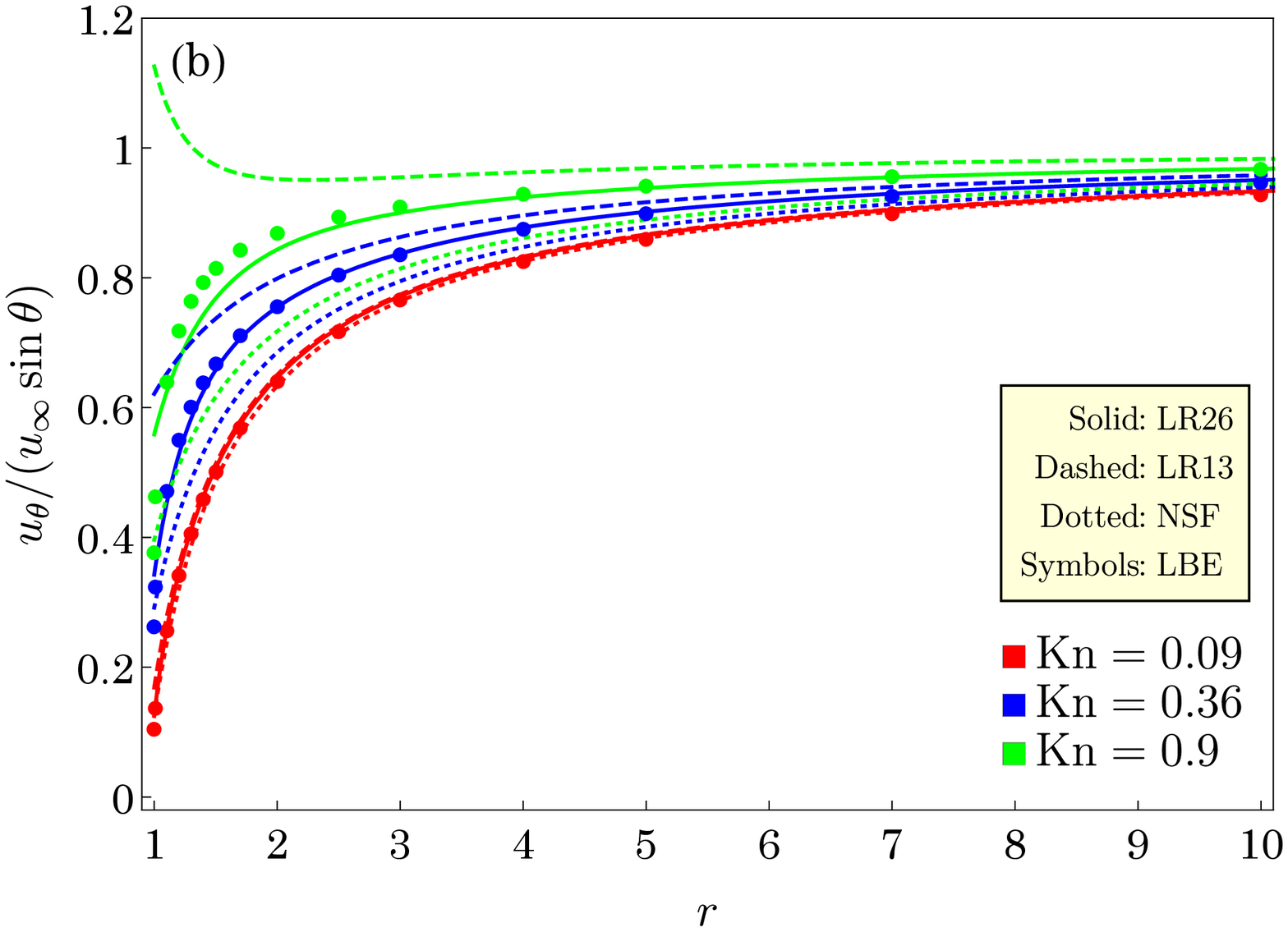}
\\
\includegraphics[height=45mm]{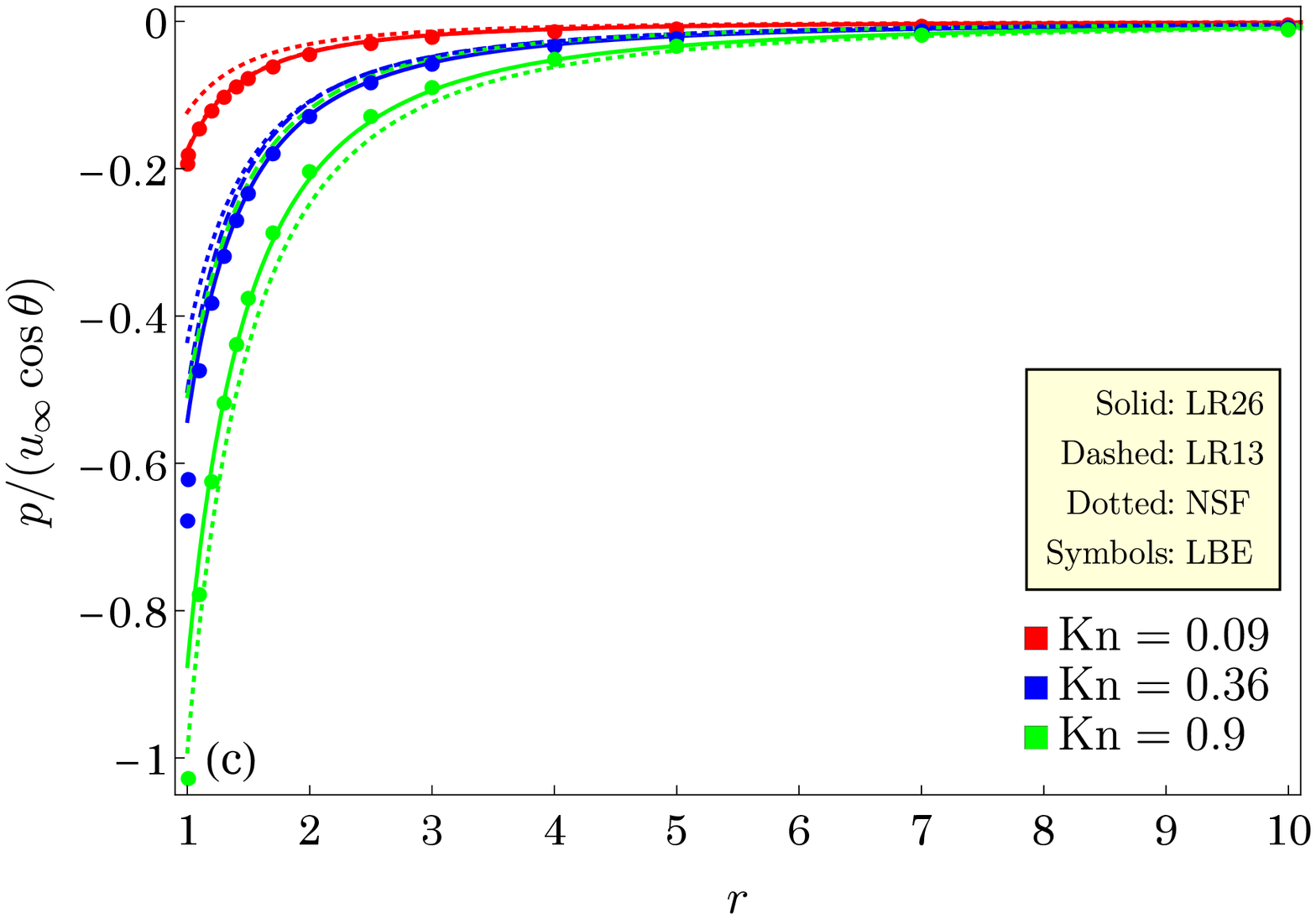}
\hfill
\includegraphics[height=45mm]{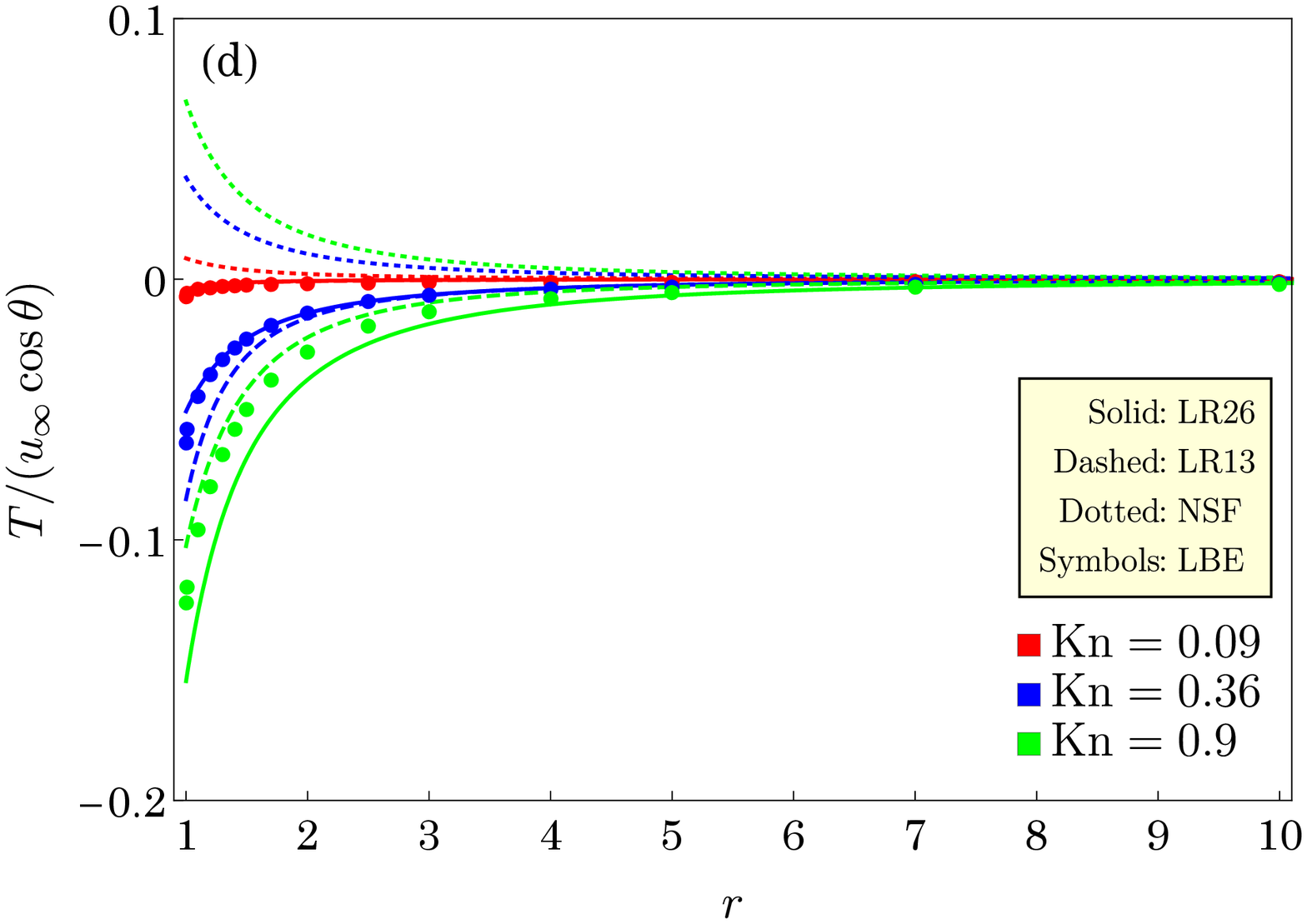}
\caption{\label{fig:profiles_non_evp}Profiles of the (a) radial velocity, (b) polar velocity (c) pressure and (d) temperature of the gas past a rigid sphere as functions of the radial distance from the surface of the sphere for various Knudsen numbers. 
Results obtained from the LR26 theory (solid lines) are compared with those obtained from the LR13 (dashed lines) and NSF (dotted lines) theories \citep{Torrilhon2010} and with the numerical solution of the LBE (symbols) from \cite{Takata1993a}.}
\end{figure}

Figure \ref{fig:profiles_non_evp} illustrates the profiles of the radial and polar components of the velocity, pressure and temperature as functions of the radial distance $r$ starting from the surface of the sphere $r=1$. 
The exact solutions for the problem under consideration obtained with the LR26 equations in the present work are compared with those obtained with the LR13 and linear NSF solutions in \cite{Torrilhon2010} and with the numerical solutions of the LBE for three values of the Knudsen number $\mathsf{Kn} = 0.1, 0.4, 1$ given in \cite{Takata1993a}. 
Note that the Knudsen number $\mathsf{Kn}$ used in \cite{Takata1993a} relates to the Knudsen number $\mathrm{Kn}$ used in the present work via $\mathrm{Kn}= 1.27\,\mathsf{Kn}/\sqrt{2}$.
Therefore, the Knudsen number values $\mathsf{Kn} = 0.1, 0.4, 1$ in \cite{Takata1993a}  correspond to the Knudsen number values $\mathrm{Kn} \approx 0.09, 0.36, 0.9$, respectively, used in the present work.
As expected, for small values of the Knudsen number [red curves and symbols ($\mathrm{Kn} = 0.09$)], all theories, including the NSF theory, give a good match with the LBE data for the velocities.
However, for moderate values of the Knudsen number [blue curves and symbols ($\mathrm{Kn} = 0.36$)],
the results from the NSF theory show deviations from those from kinetic theory (LBE results) while the LR13 theory still gives reasonably good results. 
The mismatch between the solutions from the NSF equations and LBE is attributed to the lack of Knudsen layer description by the former. 
For even larger values of the Knudsen number  [green curves and symbols ($\mathrm{Kn} = 0.9$)], even the results from the LR13 theory start deviating from the LBE results.
Notably, the LR13 theory in this case predicts $u_\theta / (u_\infty \sin{\theta}) >1$ near the surface of the sphere, which means that the slip velocity on the top of the sphere ($\theta=\pi/2$) is larger than the far field velocity---a result which is clearly unphysical and contradicted by more accurate theories (LBE and LR26). Hence, one may conclude that for large values of the Knudsen number, the results from LR13 theory cannot be trusted. On the other hand, the LR26 theory offers an excellent match upto $\mathrm{Kn} \sim 1$. 
Interestingly, the LR26 theory  also predicts $u_\theta / (u_\infty \sin{\theta}) > 1$ for $\mathrm{Kn}\gtrsim 5$, which deduces that the LR26 theory can predict reasonably good results for the Knudsen numbers $0<\mathrm{Kn}\lesssim 5$.

Another remarkable feature in figure \ref{fig:profiles_non_evp}(d) is that the NSF theory predicts a non-negative temperature profile whereas all other theories, including the LBE, give a non-positive temperature profile. 
\begin{figure}
\centering
\includegraphics[width=0.31\textwidth]{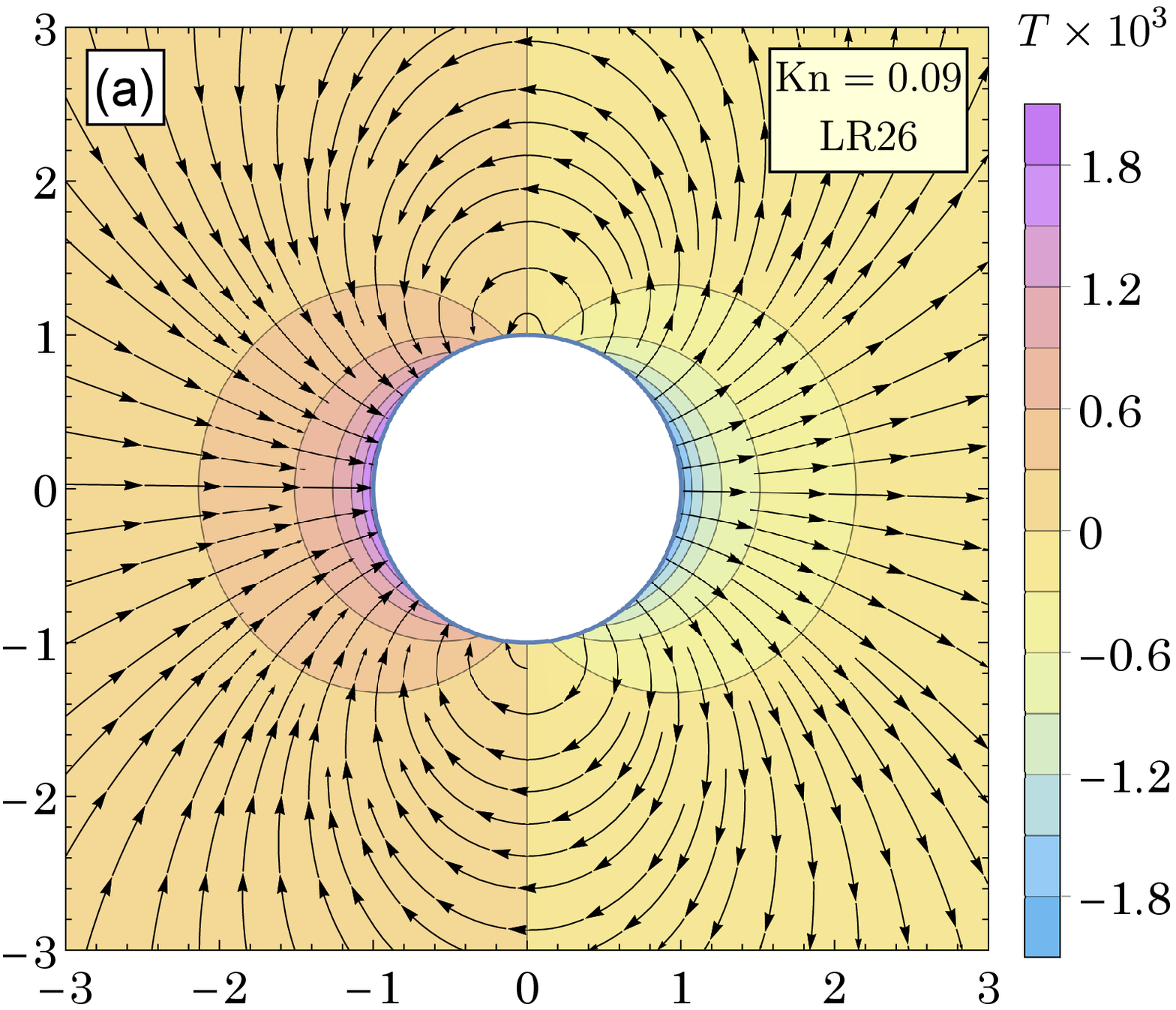}
\includegraphics[width=0.31\textwidth]{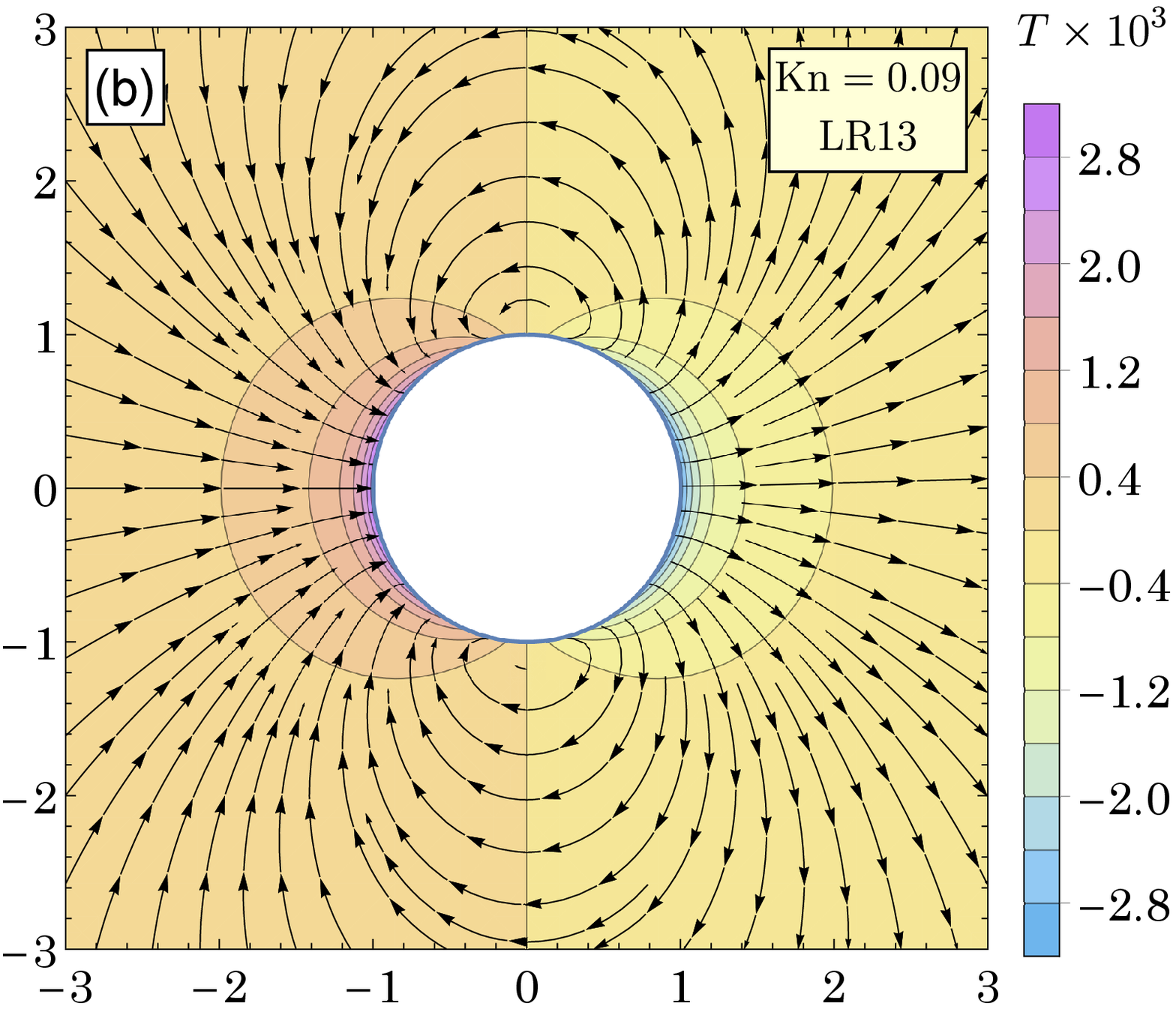}
\includegraphics[width=0.31\textwidth]{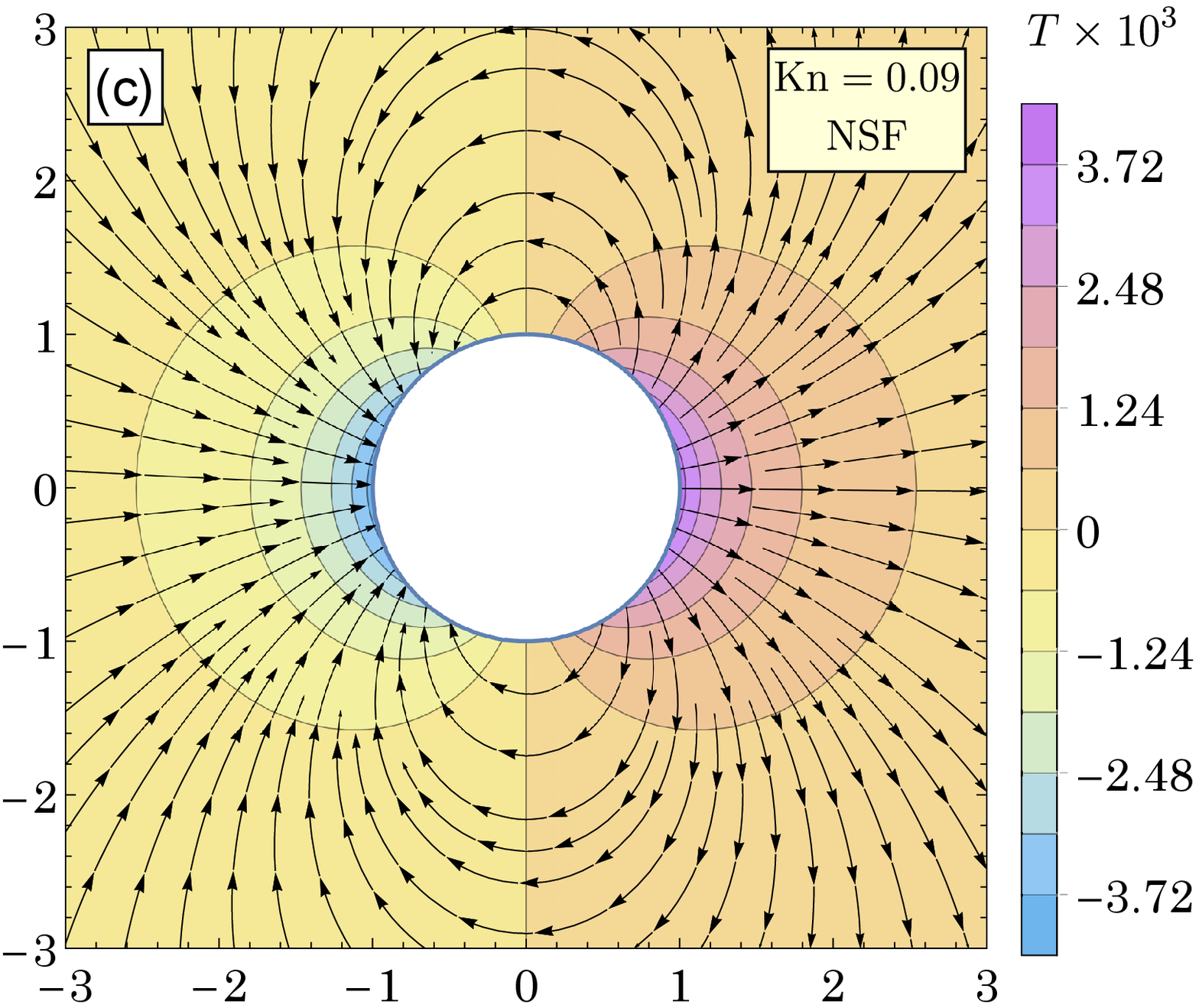}
\\
\includegraphics[width=0.31\textwidth]{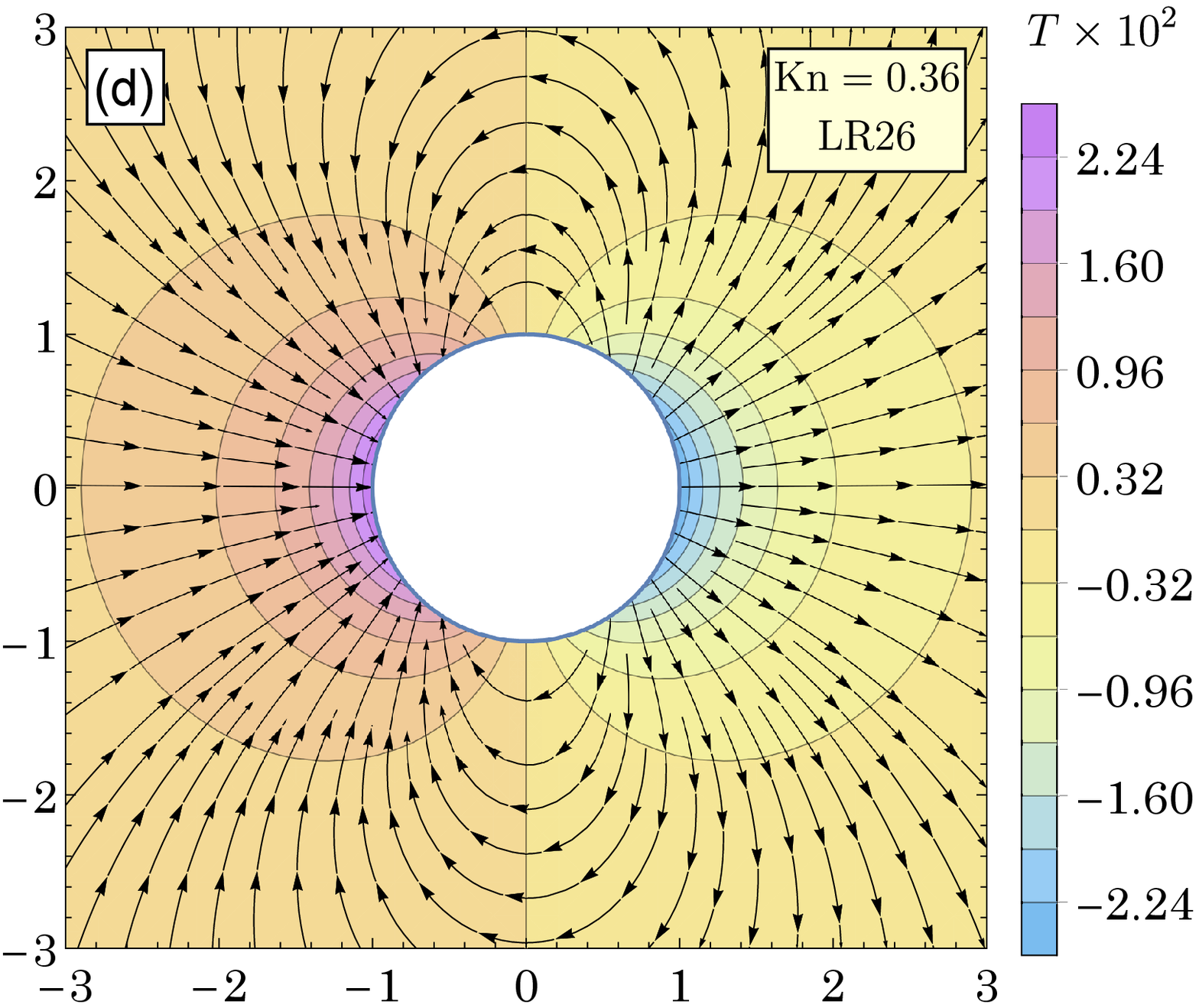}
\includegraphics[width=0.31\textwidth]{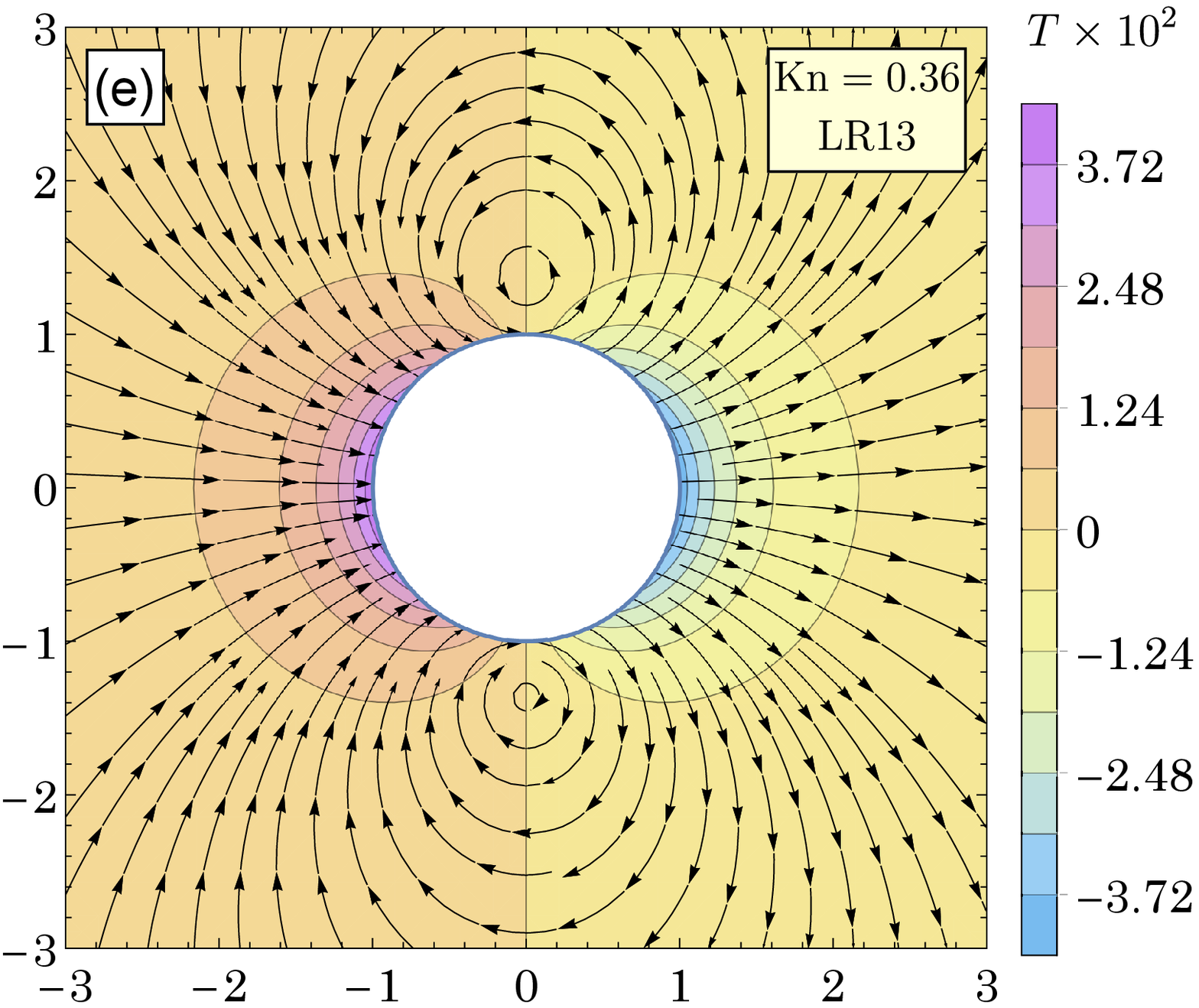}
\includegraphics[width=0.31\textwidth]{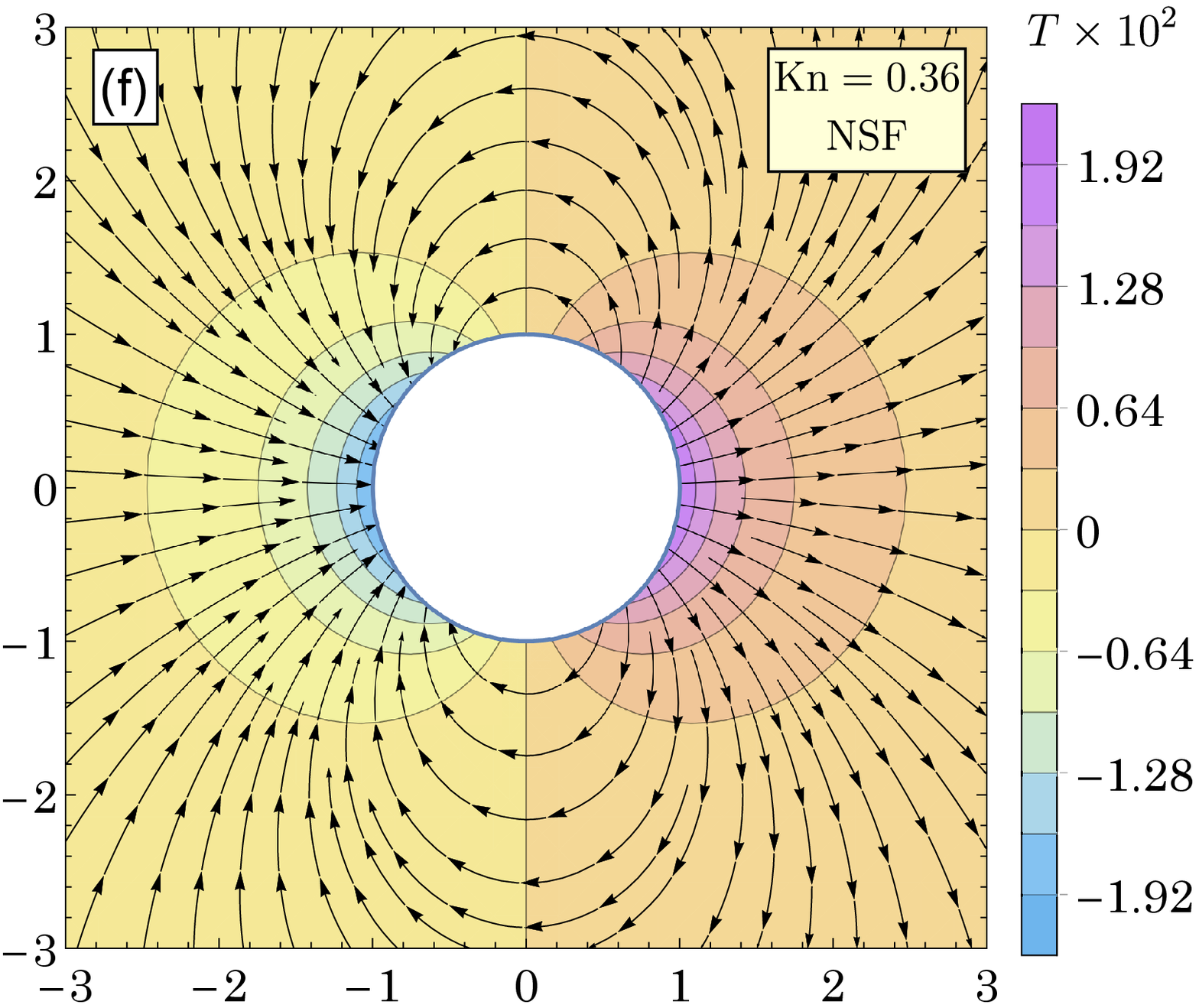}
\caption{\label{fig:2d_plots} Heat flux lines drawn on top of the temperature contours for $\mathrm{Kn} = 0.09$ (top row) and $\mathrm{Kn} = 0.36$ (bottom row) from the LR26, LR13 and NSF theories.
}
\end{figure}
To demonstrate the consequences of this, the heat flux lines and temperature contours obtained from LR26, LR13 and NSF theories are illustrated in figure~\ref{fig:2d_plots}. 
It is evident from the figure that all the theories, including the NSF theory, do predict the thermal polarisation---an effect that,  for the problem under consideration, shows different temperatures in the front and back sides of the sphere in the absence of nonlinear dissipations and external temperature differences \citep{AokiSone1987, Takata1993a, Torrilhon2010}. 
Nevertheless, the LR13 and LR26 theories show a hot region in the front side of the sphere and a cold region in the back side of the sphere while the temperature prediction by the NSF theory is the other way round. 
The temperature predictions by the LR13 and LR26 theories are in agreement with the findings of \cite{Takata1993a} based on the LBE and 
of \cite{Torrilhon2010} based on the LR13 equations, and also make sense because the temperature is likely to be more on the front side of the sphere due to compression of the gas at the front side of the sphere and is likely to be comparatively lesser on the back  side of the sphere due to expansion of the gas at the back side of the sphere. 
Furthermore, the LR13 and LR26 theories predict that the heat flows from the cold region (back side of the sphere) to the hot region (front side of the sphere), which is a non-Fourier effect. 
The NSF theory cannot capture this effect and shows the heat flowing from the hot region to the cold region. 
Such a non-Fourier effect is well established in lid-driven cavity problems, \citep[see e.g.][and references therein]{Ranaetal2013}. 
By comparing the top and bottom rows of figure~\ref{fig:2d_plots}, it is also clear that the non-uniformity in temperature (i.e., the temperature difference between the temperatures at the front and back sides of the sphere) is more for a larger Knudsen number. This is also in agreement with the findings of \cite{Takata1993a}.




\begin{figure}
\centering
\includegraphics[scale=0.7]{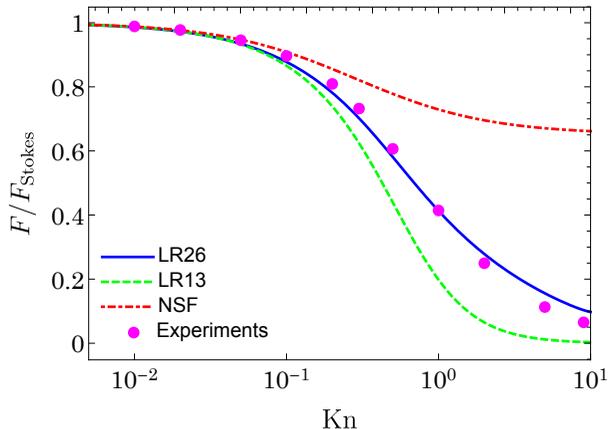}
\caption{\label{fig:drag_non_evp} Normalised drag force from the NSF (dot-dashed red line), R13 (dashed green line) and R26 (solid blue line) theories as a function of the Knudsen number.
The circles denote the experimental data of \cite{Goldberg1954} (fitted by \citet{Bailey2005}). 
The results are normalised with the Stokes drag $F_{\mathrm{Stokes}}=6\pi\mathrm{Kn} u_\infty$.}
\end{figure}

Figure \ref{fig:drag_non_evp} illustrates the drag force  normalised with the Stokes drag, $F_{\mathrm{Stokes}} = 6\pi \mathrm{Kn}u_{\infty}$, plotted over the Knudsen number. 
The experimental data from \cite{Goldberg1954} (fitted by \citet{Bailey2005}) is included for comparison. 
As expected, all theories agree with the experimental data for small values of the Knudsen number. 
As the Knudsen number increases, the drag decreases due to slip, which in not embedded in the Stokes formula and hence the Stokes formula gives vanishing drag force for all values of the Knudsen number.  
Notably, the drag force from the NSF theory with velocity-slip boundary conditions attains a nonzero positive value as $\mathrm{Kn} \to \infty$ while the other theories (LR13 and LR26) predict a vanishing drag in this limit---an observation tantamount to experimental results. 
Furthermore, the LR26 theory provides an excellent prediction for the drag force for values of the Knudsen number beyond unity (upto $\mathrm{Kn} \approx 10$). 
On the contrary, the LR13 theory  underpredicts the drag force significantly for $\mathrm{Kn} \gtrsim 0.2$.

\subsection{
Gas flow past an evaporating spherical droplet}

Finally, we consider the case of an evaporating droplet placed in a saturated vapour ($p_{\mathrm{sat}} = p_\infty = 0$) flowing past the droplet with velocity $u_\infty$, i.e. phase change is driven by the flow field, not by the far field. 
The assumptions made in the previous section remain true here, i.e.~the shape and size of the droplet remain fixed and the internal motion of the droplet is ignored. 
In addition, the evaporation coefficient $\vartheta$ is assumed to be unity.

\begin{figure}
\centering
\includegraphics[height=45mm]{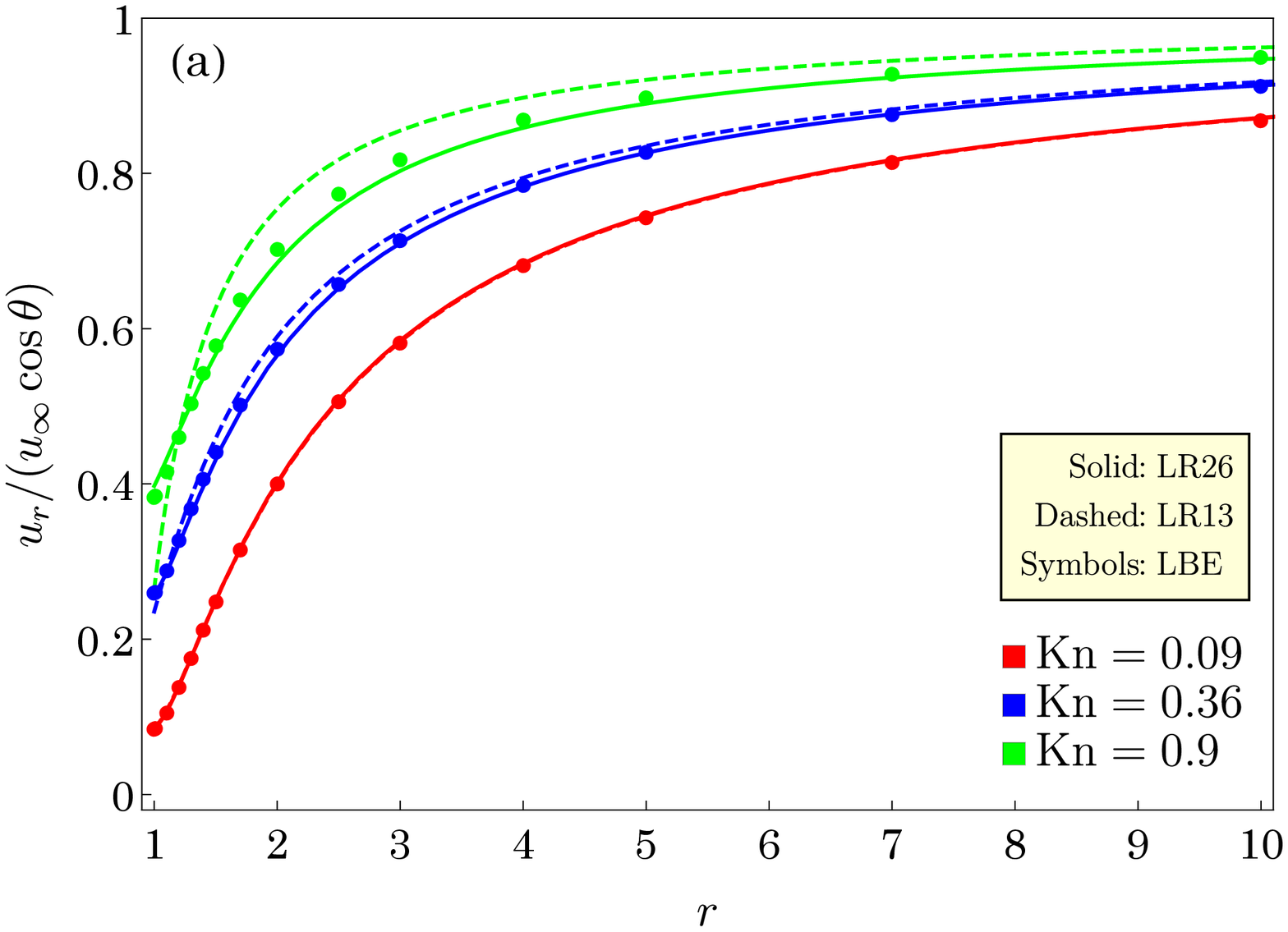}
\hfill
\includegraphics[height=45mm]{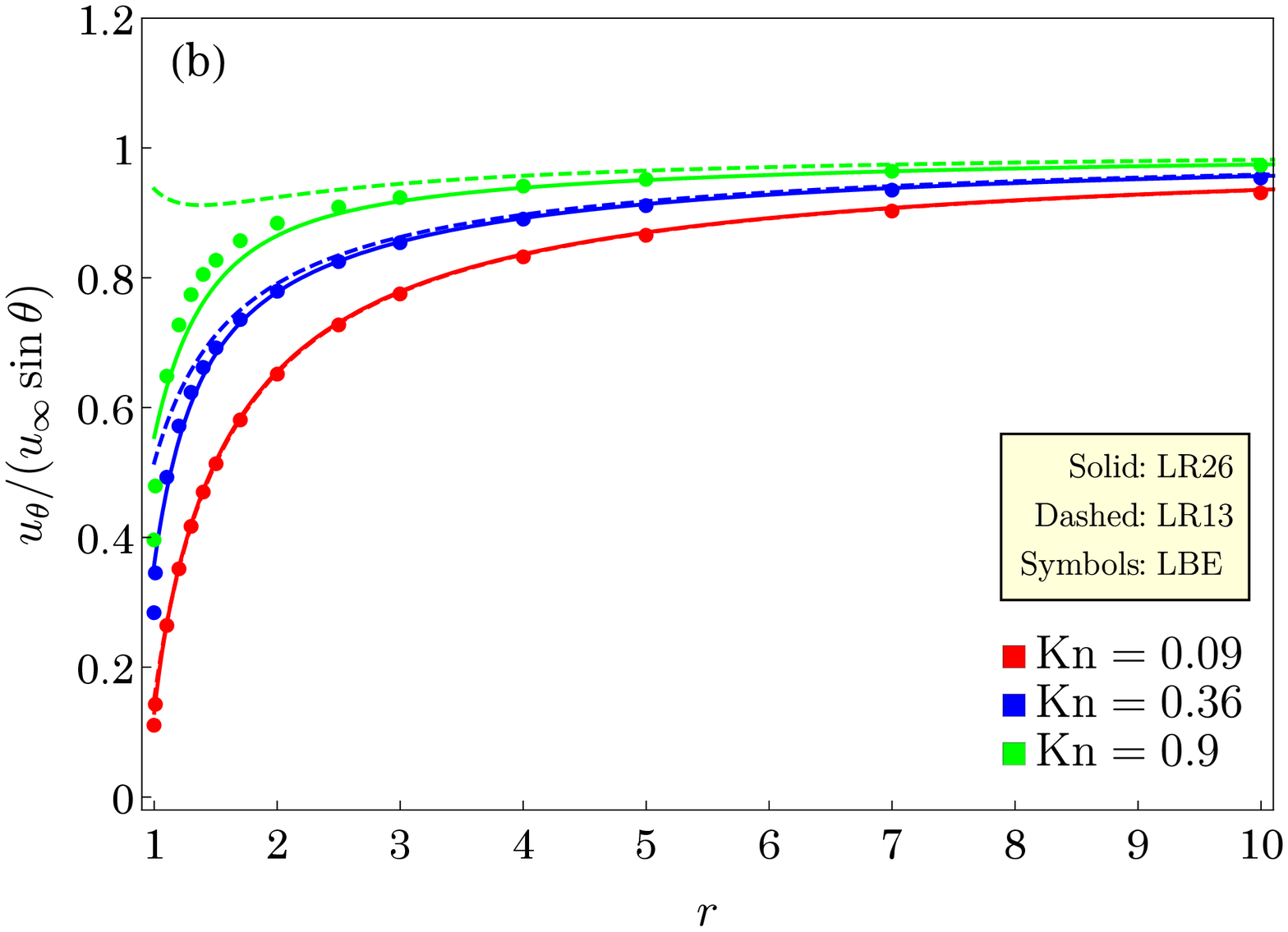}
\\
\includegraphics[height=45mm]{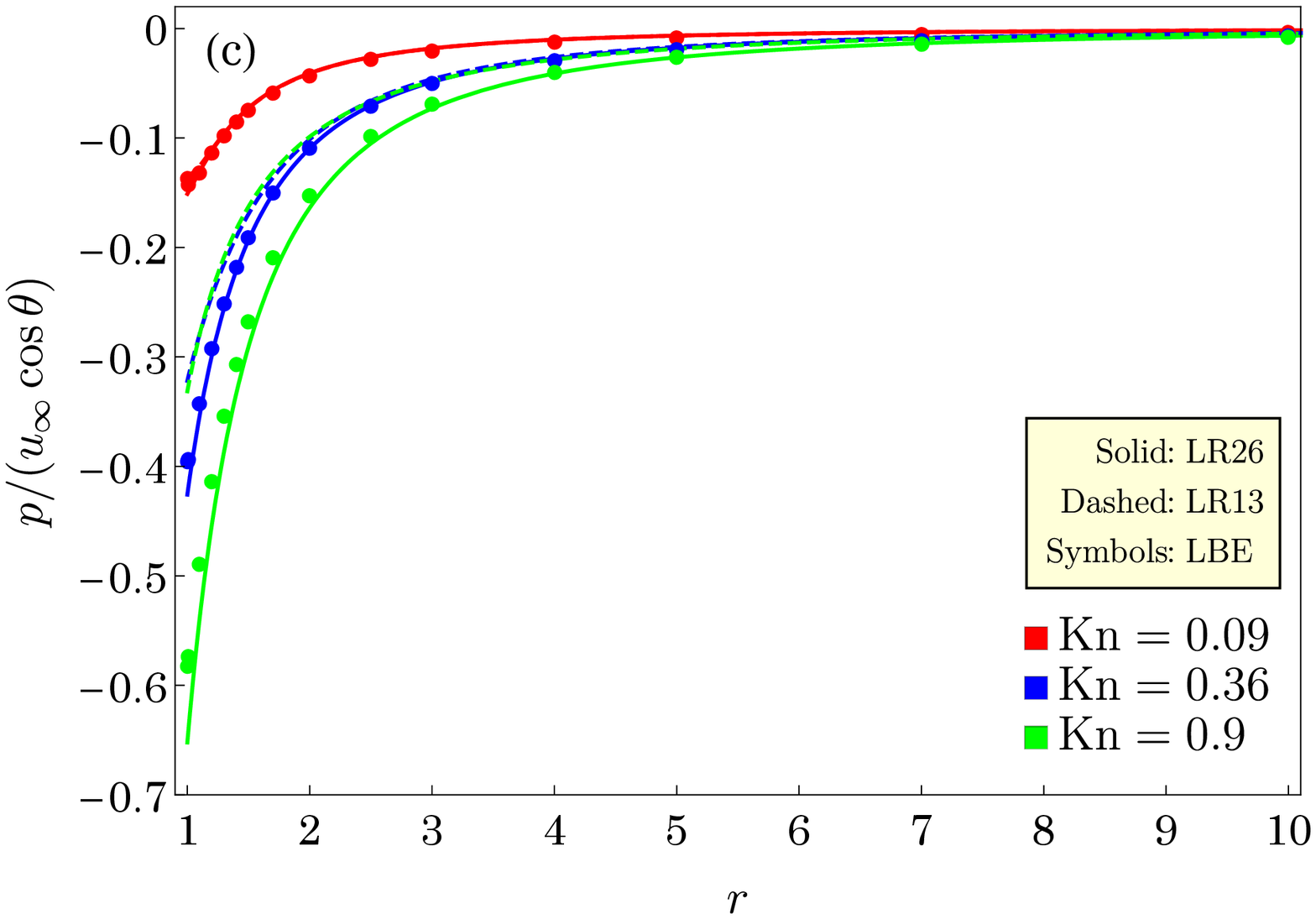}
\hfill
\includegraphics[height=45mm]{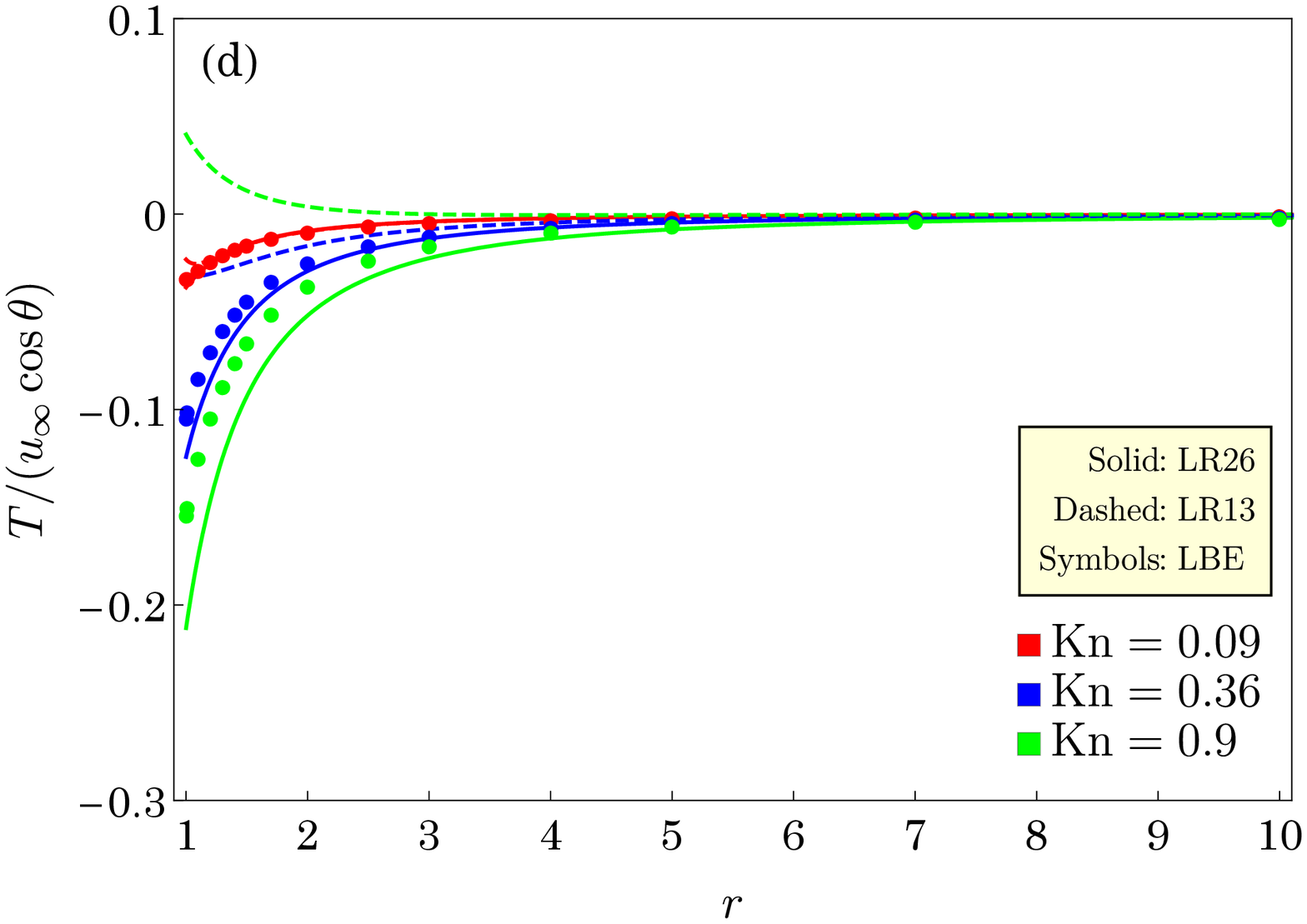}
\caption{\label{fig:profiles_evp}Profiles of the (a) radial velocity, (b) polar velocity (c) pressure and (d) temperature of the gas past an evaporating spherical droplet as functions of the radial distance from the surface of the droplet for various Knudsen numbers. 
Results obtained from the LR26 theory (solid lines) are compared with the LR13 theory (dashed lines) and with those obtained from the numerical solution of the LBE (symbols) from \cite{Sone1994}.}
\end{figure}

Figures \ref{fig:profiles_evp} again illustrates the profiles of the velocity components, pressure and temperature as functions of the radial
direction $r$ starting from the surface of the sphere $r=1$. 
The exact solutions for the problem under consideration obtained with the LR26 equations (solid lines) and LR13 equations (dashed lines) are compared with those obtained with the numerical solutions of the LBE for three values of the Knudsen number $\mathsf{Kn} = 0.1, 0.4, 1$ given in \cite{Sone1994}. Notice that the Knudsen number $\mathsf{Kn}$ is the same as that used in \cite{Takata1993a}. 
Therefore, similarly to the above, these Knudsen number values correspond to $\mathrm{Kn} = 0.09, 0.26, 0.9$, respectively.

As expected, when the Knudsen number is small [red lines and symbols ($\mathrm{Kn} = 0.09$)],
both the LR13 and LR26 theories (dashed and solid lines) agree well with the results from the LBE (symbols). However, as the Knudsen number increases, the differences between the results from the two theories become noticeable---in particular, for the radial velocity and temperature.  
At higher Knudsen numbers  ($\mathrm{Kn}\gtrsim 0.5$), the Knudsen layer description by the LR13 theory is inaccurate whilst the LR26 theory remains valid even for $\mathrm{Kn}= 1$.



\begin{figure}
\centering
\includegraphics[scale=0.7]{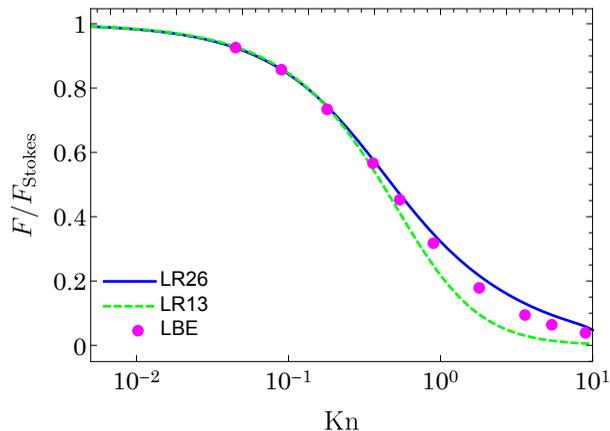}
\caption{\label{fig:drag_evp} Normalised drag force from the LR13 (dashed green line) and LR26 (solid blue line) theories and from the numerical solution of the LBE (circles) from \citet{Sone1994} as a function of the Knudsen number. 
The results are normalised with the Stokes drag $F_{\mathrm{Stokes}}=6\pi\mathrm{Kn} u_\infty$.}
\end{figure}

The normalised drag force in the case of a gas flow past an evaporating droplet is illustrated in figure \ref{fig:drag_evp}. 
The solutions obtained from the NSF, LR13 and LR26 theories are compared against the numerical solutions of the LBE given in  \citet{Sone1994}. 
In contrast to the non-evaporating droplet case, all theories in this case predict vanishing drag as $\mathrm{Kn}\to \infty$. 
For large values of the Knudsen number, the NSF theory overpredicts the drag while the LR13 theory underpredicts it. 
Again, the LR26 theory gives accurate predictions for the drag force even for large values of the Knudsen number ($\mathrm{Kn} \lesssim 10$).
\section{Conclusion}
\label{section:conclusion}
 In this article, we have addressed the question of thermodynamically admissible  boundary conditions for the LR26 equations. It has been shown that these equations are accompanied with a proper entropy inequality with non-negative entropy generation (the $H$-theorem). 
 
The entropy production rate at the interface has been written as a sum of the products of the fluxes with odd degree in $n$ and the requirement of the non-negative entropy generation at the wall has been imposed, which have eventually yielded the desired boundary conditions for the LR26 equations. 
Furthermore, considering the coupling between the cross-terms of the same tensorial structure (the Curie principle) a positive entropy production has been achieved by setting the unknown boundary values for the odd fluxes proportional to their driving forces.
The phenomenological coefficients appearing in these boundary conditions have been obtained by comparing them with those obtained via the Maxwell accommodation model (with Grad's approximation) to the leading order, thus allowing us to formulate the PBC aligned with the Maxwell kinetic-wall model.
We have demonstrated that the solutions of the LR26 equations with the MBC either do not satisfy the Onsager reciprocity relations or do not respect the second law of thermodynamics or do not comply with both. On the other hand, the solutions of the LR26 equations with the PBC satisfy the Onsager reciprocity relations and always respect the second law of thermodynamics.

As an application of the PBC, we have computed the analytic solution of the LR26 equations for a steady vapour passing over an evaporating spherical droplet. To the best of authors' knowledge, this is the first time this problem has been considered within the extended fluid dynamics framework. 
These analytic solution have been found to be useful in understanding some intriguing rarefaction effects, such as the Knudsen layers and thermal polarisation. 
A  special  case  of  this  problem, when there is no evaporation, has also been considered, which is a classic problem in fluid dynamics.
It has been found that the LR26 equations with the PBC are able to provide a more detailed and accurate description of the flow features emanating from rarefaction effects up to $\mathrm{Kn}\lesssim 1$. 
Furthermore, the analytic results for the drag force obtained from the LR26 theory with the PBC have been found to be in excellent agreement with the theory/experiments even for very large values of the Knudsen number---in the case of both an evaporating droplet and a non-evaporating droplet.


In \S\,\ref{section:PBC}, it was noticed that matching the coefficients of the lower-order moments (i.e.~moments involved in the G26 equations) in the PBC and MBC yields the matrices $\mathcal{L}$ to be symmetric and semi-positive definite.
The reason behind it is presently unclear and will be explored in the future work.
Furthermore, it is worthwhile noting that the results in the present work are valid only for slow moving droplets, i.e.~when the Reynolds number $\mathrm{Re}$ ($=\mathrm{Ma}\sqrt{\gamma}/\mathrm{Kn})$ is much less than unity. 
Here $\gamma$ (=5/3 for monatomic gases) is the ratio of specific heat capacities and $\mathrm{Ma}$ is the Mach number. 
For a relatively large value of the Reynolds number, the nonlinear convective terms play an important role in evaporation dynamics, which have been neglected in the present work. 
In classical fluid dynamics texts \citep[see e.g.][]{Lamb1932}, Oseen correction terms are utilised to obtain solutions for the flow past various bodies at moderate Reynolds numbers [$\mathcal{O}(\mathrm{Re}) \sim 1$]. Oseen correction terms are linear in nature as oppose to the non-linear Navier--Stokes operator, making these equations analytically tractable.
The present work can be extended to flows with a small but finite Reynolds number by including the Oseen correction terms (with the partial inclusion of convective terms) in the future to study the combined effects of inertia and rarefaction on evaporation dynamics.
Motivated by the present findings, the present work will also be extended to investigate unsteady flow problems and to incorporate coupling with liquid dynamics \citep[see e.g.][]{RanaPRL2019} elsewhere in the future. 
Thermophoresis on a sphere is a classic problem in rarefied gas flows, which has been studied by many researchers; see e.g.~\cite{YI1988, TAS1994, TY1995, BC1995, Takata2009, PSL2019, KS2020}.
The model developed in the present work can also be used to investigate thermophoresis on an evaporating droplet in the future.


\bigskip
\noindent
\paragraph{\textbf{Acknowledgments.}}
ASR acknowledges the funding from the BITS, Pilani (India) through the Research Initiation Grant. 
VKG gratefully acknowledges financial support through the `MATRICS' project MTR/2017/000693 funded by the SERB, India. This work has also been financially supported in the UK by EPSRC Grants No. EP/N016602/1,
EP/P031684/1, EP/S029966/1. 

\smallskip
\noindent
\paragraph{\textbf{Declaration of Interests.}} The authors report no conflict of interest.

\appendix

\section{The entropy density for the LR26 equations from the Boltzmann entropy density}
\label{app:quadratic_entropy}
The Boltzmann entropy density $\hat{\eta}$ and entropy flux $\hat{\Gamma}_i$ (in dimensional form) are given by \eqref{entropy_withdimension} and \eqref{entropyflux_withdimension}. We replace $\hat{f}$ in \eqref{entropy_withdimension} and \eqref{entropyflux_withdimension} with the G26 distribution function \citep{Struchtrup2005, GT2012, Gupta2015}
\begin{align*}
\hat{f}_{\mathrm{G26}}=\hat{f}_{\mathrm{M}} \left( 1+\Phi \right) ,
\end{align*}
where 
\begin{align}
\label{hatfM}
\hat{f}_{\mathrm{M}}=\frac{\hat{\rho}}{(2\pi \hat{\theta})^{3/2}}\exp
\left( -\frac{\hat{C}^{2}}{2\hat{\theta}}\right), 
\end{align}
is the Maxwellian distribution function and
\begin{align*}
\Phi &=\frac{\hat{\sigma}_{ij}}{2}\frac{\hat{C}_{i}\hat{C}_{j}}{\hat{\rho}%
\hat{\theta}^{2}}-\frac{\hat{q}_{i}\hat{C}_{i}}{\hat{\rho}\hat{\theta}^{2}}%
\left( 1-\frac{\hat{C}^{2}}{5\hat{\theta}}\right) +\frac{\hat{m}_{ijk}}{6%
\hat{\rho}\hat{\theta}^{3}}\hat{C}_{i}\hat{C}_{j}\hat{C}_{k}  
\\
&\quad+\frac{\hat{\Delta}}{8\hat{\rho}\hat{\theta}^{2}}\left( 1-\frac{2\hat{C}^{2}}{3\hat{%
\theta}}+\frac{\hat{C}^{4}}{15\hat{\theta}^{2}}\right)
-\frac{\hat{R}_{ij}}{4\hat{\rho}\hat{\theta}^{3}}\left( 1-\frac{\hat{C}^{2}%
}{7\hat{\theta}}\right) \hat{C}_{i}\hat{C}_{j}
\end{align*}%
is the dimensionless deviation of the distribution function from the equilibrium state.
Here, $\hat{C}_{i}=\hat{c}_{i} -\hat{v}_{i}$ is the (dimensional) peculiar velocity with $\hat{c}_{i}$ and $\hat{v}_{i}$ being the  (dimensional) microscopic and macroscopic velocities, respectively.

For the entropy density and entropy flux of the LR26 equations, 
the integrations in \eqref{entropy_withdimension} and \eqref{entropyflux_withdimension} need to be evaluated by expending $\hat{f}_{\mathrm{G26}}$ up to second order in deviations. 
For this purpose, we introduce a smallness parameter $\varepsilon$, whose power in a term tells the order of the deviation.
The parameter is introduced just for the bookkeeping purpose and will be replaced with unity at the end of computations. 
Each moment is replaced with its value in the equilibrium plus its deviation from its equilibrium value. 
We write
\begin{align}
\left.
\begin{gathered}
\hat{\rho} := \hat{\rho}_0 (1 + \varepsilon \rho), 
\quad
\hat{v}_i := \varepsilon \sqrt{\hat{\theta}_0}\, v_i,
\quad
\hat{\theta} := \hat{\theta}_0 (1 + \varepsilon T), 
\quad
\hat{\sigma}_{ij} := \varepsilon \hat{\rho}_0 \hat{\theta}_0 \sigma_{ij}
\\
\hat{q}_i := \varepsilon \hat{\rho}_0 \hat{\theta}_0^{3/2} q_i,
\quad
\hat{m}_{ijk} := \varepsilon \hat{\rho}_0 \hat{\theta}_0^{3/2} m_{ijk},
\quad
\hat{\Delta} := \varepsilon \hat{\rho}_0 \hat{\theta}_0^{2} \Delta,
\quad
\hat{R}_{ij} := \varepsilon \hat{\rho}_0 \hat{\theta}_0^{2} R_{ij},
\end{gathered}
\right\}
\end{align}
where $\rho$, $v_{i}$, $T$, $\sigma_{ij}$, $q_i$, $m_{ijk}$, $\Delta$ and $R_{ij}$ are the dimensionless deviations of the respective quantities from their equilibrium values.
The microscopic velocity $\hat{c}_i$ is also scaled to give the dimensionless microscopic velocity $\hat{c}_i = \sqrt{\hat{\theta}_0}\, c_i$.

A Taylor series expansion of $\hat{f}_{\mathrm{M}}$, to second order in $\varepsilon$, yields
\begin{align}
\hat{f}_{\mathrm{M}} 
\approx \hat{f}_{M}^{(0)} \left[1+\varepsilon
f_{M}^{(1)}+\varepsilon^{2}f_{M}^{(2)}\right], 
\end{align}%
where%
\begin{align}
\hat{f}_{M}^{(0)}=\frac{\hat{\rho}_{0}}{(2\pi \hat{\theta}_{0})^{3/2}}%
\exp \left( -\frac{c^{2}}{2}\right) 
\end{align}%
is the Maxwellian distribution function in the equilibrium, and%
\begin{align}
f_{M}^{(1)} &=\rho -\frac{3T }{2}
+\frac{T c^{2}}{2}+c_{j}v_{j},
\\
f_{M}^{(2)} 
&=-\frac{3\rho T }{2} 
+ \frac{15T ^{2}}{8}-\frac{v^{2}}{2}
+\left(\rho T - \frac{5T ^{2}}{2}+ v^2 +\frac{T 
^{2}c^{2}}{4}\right) 
\frac{c^{2}}{2}
+\left(\rho - \frac{5T }{2} + \frac{T c^{2}}{2}\right) c_{j}v_{j}.
\end{align}%
Similarly, a Taylor series expansion of $\Phi$, to second order in $\varepsilon$, yields
\begin{align}
\Phi \approx \varepsilon \Phi^{(1)}+\varepsilon^{2}\Phi^{(2)},
\end{align}
where
\begin{align}
\Phi^{(1)} &= \frac{\sigma_{kl}}{2} c_k c_l 
- q_k c_k \left(1-\frac{c^2}{5}\right)
+ \frac{m_{klr}}{6} c_k c_l c_r
\nonumber\\
&\quad\;+ \frac{\Delta}{8} \left(1 - \frac{2 c^2}{3} + \frac{c^4}{15}\right)
-\frac{R_{kl}}{4} \left(1-\frac{c^2}{7}\right) c_k c_l,
\end{align}
\begin{align}
\Phi^{(2)} &= - \frac{\sigma_{kl}}{2} \Big[(\rho + 2 T ) c_k c_l + c_k v_l + c_l v_k 
\Big]
\nonumber\\
&\quad\;+ q_k \left[(\rho + 2 T ) c_k - \frac{1}{5} (\rho + 3 T ) c^2 c_k + \left(1-\frac{c^2}{5}\right) v_k - \frac{2}{5} c_k c_l v_l\right]
\nonumber\\
&\quad\; - \frac{m_{kl r}}{6} \Big[(\rho + 3 T ) c_k c_l c_r
+ c_k c_l v_r + c_k c_r v_l+ c_l c_r v_k
\Big]
\nonumber\\
&\quad\;-\frac{\Delta}{8} \left[\rho + 2 T - \frac{2}{3} (\rho + 3 T ) c^2 + \frac{1}{15} (\rho + 4 T ) c^4 - \frac{4}{3} \left(1-\frac{c^2}{5}\right) c_k v_k\right]
\nonumber\\
&\quad\;\, + \frac{R_{kl}}{4} \left[\left(1-\frac{c^2}{7} \right) \Big\{(\rho + 3 T ) c_k c_l + c_k v_l + c_l v_k
 \Big\} 
-\frac{T}{7} c^2 c_k c_l -\frac{2}{7} c_k c_l c_r v_r\right]. 
\end{align}
Therefore, to second order in $\varepsilon$, $\hat{f}_{\mathrm{G26}}$ can be approximated
as%
\begin{align}
\hat{f}_{\mathrm{G26}} &\approx 
\hat{f}_{M}^{(0)}\left[ 1+\varepsilon \phi^{(1)} +\varepsilon^{2} \phi^{(2)}\right], 
\end{align}%
where
\begin{align}
\phi^{(1)} = \Phi^{(1)}+f_{M}^{(1)}
\quad\text{and}\quad
\phi^{(2)} = \Phi^{(2)} + \Phi^{(1)} f_{M}^{(1)} + f_{M}^{(2)}.
\end{align}
Let $\hat{y} = \hat{f}_{M}^{(0)}$. 
Hence, to second order in $\varepsilon$, 
\begin{align}
\ln{\frac{\hat{f}_{\mathrm{G26}}}{\hat{f}_{M}^{(0)}}}\approx
\ln{\left( 1+\varepsilon \phi^{(1)} +\varepsilon^{2} \phi^{(2)}\right)}
\approx 
\varepsilon \phi^{(1)} 
+\varepsilon^{2} \left(\phi^{(2)}-\frac{ \big\{ \phi^{(1)}\big\}^{2}}{2}\right).
\end{align}
From \eqref{entropy_withdimension}, the dimensionless entropy $\eta = \hat{\eta} / (\hat{k}_B \hat{\rho}_0)$, to second order in $\varepsilon$, is given by
\begin{align}
\eta &= - \frac{1}{\hat{\rho}_0} \int \hat{f}_{\mathrm{G26}}\left( \ln{\frac{\hat{f}_{\mathrm{G26}}}{\hat{f}_{M}^{(0)}}}-1\right) \mathrm{d}\hat{\bm{c}}
\nonumber\\
&\approx - \frac{1}{\hat{\rho}_0}\int \hat{f}_{M}^{(0)}\left( 1 +\varepsilon \phi^{(1)} +\varepsilon^{2} \phi^{(2)}\right) \left[ \varepsilon \phi
^{(1)}+\varepsilon^{2}\left(\phi^{(2)}-\frac{ \big\{\phi^{(1)}\big\}^{2}}{2}\right) -1\right] (\hat{\theta}_0^{3/2} \mathrm{d}\bm{c}) 
\nonumber\\
&= \frac{1}{(2\pi)^{3/2}}%
\int \exp{\left(-\frac{c^{2}}{2}\right)}
\left(1 - \varepsilon^{2}\frac{\big\{\phi^{(1)}\big\}^{2}}{2}\right) \mathrm{d}\bm{c}.
\end{align}
The integration over the velocity space is performed in a straightforward way, which yields
\begin{align}
\eta = 1 + \varepsilon^{2} \left(-\frac{1}{2}\rho ^{2}-\frac{1}{2} v^{2} 
-\frac{3}{4} T^{2}-\frac{1}{4}\sigma ^{2}-\frac{1}{5}q^{2} 
- \frac{1}{12}m^{2} - \frac{1}{56}R^{2} - \frac{1}{240}\Delta^{2}\right).
\end{align}
Substituting $\varepsilon = 1$, one obtains \eqref{entropyansatz} with $a_{0}=1$. 

From \eqref{entropyflux_withdimension}, the dimensionless entropy flux $\Gamma_i = \hat{\Gamma}_i / (\hat{k}_B \hat{\rho}_0 \sqrt{\hat{\theta}_0})$, to second order in $\varepsilon$, is given by
\begin{align}
\Gamma_i &= - \frac{1}{\hat{\rho}_0 \sqrt{\hat{\theta}_0}} \int \hat{f}_{\mathrm{G26}}\left( \ln{\frac{\hat{f}_{\mathrm{G26}}}{\hat{f}_{M}^{(0)}}}-1\right) \hat{c}_i \, \mathrm{d}\hat{\bm{c}}
\nonumber\\
&\approx - \frac{\hat{\theta}_0^{3/2}}{\hat{\rho}_0}\int \hat{f}_{M}^{(0)}\left( 1 +\varepsilon \phi^{(1)} +\varepsilon^{2} \phi^{(2)}\right) \left[ \varepsilon \phi
^{(1)}+\varepsilon^{2}\left(\phi^{(2)}-\frac{ \big\{\phi^{(1)}\big\}^{2}}{2}\right) -1\right] 
 c_i \mathrm{d}\bm{c} 
\nonumber\\
&= \frac{1}{(2\pi)^{3/2}}%
\int \exp{\left(-\frac{c^{2}}{2}\right)}
\left(1 - \varepsilon^{2}\frac{\big\{\phi^{(1)}\big\}^{2}}{2}\right) c_i \, \mathrm{d}\bm{c}.
\end{align}
Again, the integration over the velocity space is performed in a straightforward way, which yields
\begin{align}
\Gamma_i &= - \varepsilon^{2} \left[(\rho + T ) v_{i} + T q_{i} + v_{j}\sigma_{ij} 
+\frac{2}{5}\sigma_{ij} q_j 
+\frac{1}{2} m_{ijk} \sigma_{jk}
\right.
\nonumber\\
&\left. \quad\;
+\frac{1}{5}R_{ij}q_{j} 
+\frac{1}{15}q_{i} \Delta
+\frac{1}{14} m_{ijk}R_{jk}
\right].
\end{align}
Substituting $\varepsilon = 1$ and $\rho+T = p$, one obtains
\begin{align}
\label{G26_entropy_flux}
\Gamma_i &= - \left(p v_{i} + T q_{i} + v_{j}\sigma_{ij} 
+\frac{2}{5}\sigma_{ij} q_j 
+\frac{1}{2} m_{ijk} \sigma_{jk}
+\frac{1}{5}R_{ij}q_{j} 
+\frac{1}{15}q_{i} \Delta
+\frac{1}{14} m_{ijk}R_{jk}\right).
\end{align}
Comparing \eqref{G26_entropy_flux} and \eqref{R26_entropy_flux}, it is evident that except for the terms containing the higher-order moments, namely $\Phi_{ijkl}$, $\Psi_{ijk}$, $\Omega_{j}$, $\Omega_{i}$, all others terms in \eqref{G26_entropy_flux} and \eqref{R26_entropy_flux} are exactly the same.
\section{The entropy density for the linearised R26 equations}
\label{app:entopyCoeff}
Let us assume that the entropy density for the LR26 equations is given by 
\eqref{entropybasicansatz}.
Taking the partial derivative of \eqref{entropybasicansatz} with respect to $t$, one obtains
\begin{align}
\frac{\partial \eta}{\partial t} 
=& a_1 \rho \frac{\partial \rho}{\partial t} 
+a_2 v_i \frac{\partial v_i}{\partial t} 
+ a_3 T \frac{\partial T}{\partial t} 
+ a_4 \sigma_{ij} \frac{\partial \sigma_{ij}}{\partial t} 
+ a_5 q_i \frac{\partial q_i}{\partial t} 
\nonumber\\
&+ a_6 m_{ijk} \frac{\partial m_{ijk}}{\partial t} 
+ a_7 R_{ij} \frac{\partial R_{ij}}{\partial t} 
+ a_8 \Delta \frac{\partial \Delta}{\partial t}. 
\end{align}
Replacing the time derivatives of field variables in the above equation using the G26 equations \eqref{massBal}--\eqref{HFBal} (with $F_i = 0$ and $p \approx \rho + T$) and \eqref{mBal}--\eqref{DeltaBal}, the time derivative of $\eta$ can be simplified as follows:
\begin{align}
\frac{\partial \eta}{\partial t} 
=& -a_1 \rho \frac{\partial v_i}{\partial x_i} 
-a_2 v_i \left(\frac{\partial \sigma_{ij}}{\partial x_j} + \frac{\partial \rho}{\partial x_i} + \frac{\partial T}{\partial x_i}\right) 
-\frac{2}{3} a_3 T \left(\frac{\partial q_i}{\partial x_i} + \frac{\partial v_i}{\partial x_i}\right)
\nonumber\\
&- a_4 \sigma_{ij} \left(\frac{\partial m_{ijk}}{\partial x_k}  
+ \frac{4}{5} \frac{\partial q_{\langle i}}{\partial x_{j\rangle}} 
+\frac{1}{\mathrm{Kn}}
\sigma_{ij} + 2 \frac{\partial v_{\langle i}}{\partial x_{j\rangle}}\right) 
\nonumber\\
&- a_5 q_i \left(\frac{1}{2} \frac{\partial R_{ij}}{\partial x_j}  
+\frac{1}{6} \frac{\partial \Delta}{\partial x_i}
+\frac{\partial \sigma_{ij}}{\partial x_j}
+\frac{\mathrm{Pr}}{\mathrm{Kn}}
q_{i} + \frac{5}{2} \frac{\partial T}{\partial x_i} \right)
\nonumber\\
&- a_6 m_{ijk} \left(\frac{\partial \Phi_{ijkl}}{\partial x_l}  
+\frac{3}{7} \frac{\partial R_{\langle ij}}{\partial x_{k\rangle}} + \frac{\mathrm{Pr}_{m}}{\mathrm{Kn}} m_{ijk} + 3 \frac{\partial \sigma_{\langle ij}}{\partial x_{k\rangle}} \right)
\nonumber\\
&- a_7 R_{ij} \left(\frac{\partial \Psi_{ijk}}{\partial x_k}  
+2 \frac{\partial m_{ijk}}{\partial x_k}  
+\frac{2}{5} \frac{\partial \Omega_{\langle i}}{\partial x_{j\rangle}} + \frac{\mathrm{Pr}_{R}}{\mathrm{Kn}}%
 R_{ij} + \frac{28}{5} \frac{\partial q_{\langle i}}{\partial x_{j\rangle}}\right)
\nonumber\\
&- a_8 \Delta \left(\frac{\partial \Omega_i}{\partial x_i}  
+ \frac{\mathrm{Pr}_{\Delta }}{\mathrm{Kn}}  \Delta + 8 \frac{\partial q_i}{\partial x_i}\right) 
\end{align}%
or
\begin{align}
\label{detadtwithunknowns}
\frac{\partial \eta}{\partial t} 
=& - \left(a_1 \rho \frac{\partial v_i}{\partial x_i} + a_2 v_i \frac{\partial \rho}{\partial x_i} \right) 
- \left( a_2 v_i\frac{\partial T}{\partial x_i} + \frac{2}{3} a_3 T \frac{\partial v_i}{\partial x_i}\right) 
- \left(a_2 v_i \frac{\partial \sigma_{ij}}{\partial x_j} + 2 a_4 \sigma_{ij} \frac{\partial v_i}{\partial x_j}\right)
\nonumber\\
&- \left(\frac{2}{3} a_3 T \frac{\partial q_i}{\partial x_i}  + \frac{5}{2} a_5 q_i \frac{\partial T}{\partial x_i} \right)
- \left(a_4 \sigma_{ij} \frac{\partial m_{ijk}}{\partial x_k}  
+ 3 a_6 m_{ijk} \frac{\partial \sigma_{ij}}{\partial x_k}\right)
\nonumber\\
&- \left(\frac{4}{5} a_4 \sigma_{ij} \frac{\partial q_i}{\partial x_j} 
+ a_5 q_i \frac{\partial \sigma_{ij}}{\partial x_j}\right) 
- \left(\frac{1}{2} a_5 q_i \frac{\partial R_{ij}}{\partial x_j} + \frac{28}{5} a_7 R_{ij} \frac{\partial q_i}{\partial x_j}\right)
\nonumber\\
&+\left(\frac{1}{6} a_5 q_i \frac{\partial \Delta}{\partial x_i}  + 8 a_8 \Delta \frac{\partial q_i}{\partial x_i}\right)
- \left(\frac{3}{7} a_6 m_{ijk} \frac{\partial R_{ij}}{\partial x_k} + 2 a_7 R_{ij} \frac{\partial m_{ijk}}{\partial x_k}  
\right)
\nonumber\\
&- a_6 m_{ijk} \frac{\partial \Phi_{ijkl}}{\partial x_l}
- a_7 R_{ij} \frac{\partial \Psi_{ijk}}{\partial x_k}  
- \frac{2}{5} a_7 R_{ij} \frac{\partial \Omega_{\langle i}}{\partial x_{j\rangle}}
- a_8 \Delta \frac{\partial \Omega_i}{\partial x_i}
\nonumber\\
&- \frac{a_4}{\mathrm{Kn}} \sigma^2
- a_5 \frac{\mathrm{Pr}}{\mathrm{Kn}} q^2
- a_6 \frac{\mathrm{Pr}_{m}}{\mathrm{Kn}} m^2
- a_7 \frac{\mathrm{Pr}_{R}}{\mathrm{Kn}} R^2
- a_8 \frac{\mathrm{Pr}_{\Delta}}{\mathrm{Kn}} \Delta^2.
\end{align}%
The goal is to write the above equation in the form of \eqref{entropylawform}, which can be accomplished by taking
\begin{align}
\left.
\begin{gathered}
a_1 = a_2, 
\quad
a_2 = \frac{2}{3} a_3,
\quad
a_2 = 2 a_4,
\quad
\frac{2}{3} a_3 = \frac{5}{2} a_5,
\quad
a_4 = 3 a_6,
\\
\frac{4}{5} a_4 = a_5,
\quad
\frac{1}{2} a_5 = \frac{28}{5} a_7,
\quad
\frac{1}{6} a_5 = 8 a_8,
\quad
\frac{3}{7} a_6 = 2 a_7.
\end{gathered}
\right\}
\end{align}
These are nine equations in eight variables. Nevertheless, they are consistent, and only seven of them are linearly independent. Thus, on taking $a_1=-1$, the other coefficients are
\begin{align}
a_2 = -1, 
\quad
a_3 = -\frac{3}{2}, 
\quad
a_4 = -\frac{1}{2}, 
\quad
a_5 = -\frac{2}{5}, 
\quad
a_6 = -\frac{1}{6}, 
\quad
a_7 = -\frac{1}{28}, 
\quad
a_8 = -\frac{1}{120}. 
\end{align}
Note that the constant $a_1$ has been taken as negative in order to get the non-negative entropy production rate.
Consequently, the entropy \eqref{entropybasicansatz} becomes \eqref{entropyansatz}.
\section{Entropy flux at the interface}
\label{app:entopyfluxI}
The entropy flux at the interface is given by \eqref{entropyproductionsurface}. 
Substituting $\Gamma_{i}^{\mathrm{(gas)}} = \Gamma_{i}$ from \eqref{R26_entropy_flux} and $\Gamma_{i}^{\mathrm{(liquid)}}$ from \eqref{entropyflux_out} into \eqref{entropyproductionsurface},  the entropy flux at the interface reads
\begin{align}
\label{Sigma_app_eq1}
\Sigma^{I}&= - \underline{\left(p v_{i} - p^{\ell} v_{i}^{\ell}
+ T q_{i} - T^{\ell} q_{i}^{\ell}
+ v_{j}\sigma_{ij} - v_{j}^{\ell}\sigma_{ij}^{\ell} 
\right) n_i}
\nonumber\\
&\quad\,
- \left(\frac{2}{5}\sigma_{ij} q_j 
+\frac{1}{2} m_{ijk} \sigma_{jk}
+\frac{1}{5}R_{ij}q_{j} 
+\frac{1}{15}q_{i} \Delta
+\frac{1}{14} m_{ijk}R_{jk}
\right.
\nonumber\\
&\left.\quad\,
+\frac{1}{6}\Phi_{ijkl}m_{jkl}
+\frac{1}{28}\Psi_{ijk}R_{jk}
+\frac{1}{70}R_{ij}\Omega_{j}
+\frac{1}{120} \Omega_{i} \Delta\right)n_{i}.
\end{align}%
Let us refer the underlined term in \eqref{Sigma_app_eq1} by $\underline{\Sigma}^{I}$, i.e.%
\begin{align}
\underline{\Sigma}^{I} &= \left(p v_{i} - p^{\ell} v_{i}^{\ell}
+ T q_{i} - T^{\ell} q_{i}^{\ell}
+ v_{j}\sigma_{ij} - v_{j}^{\ell}\sigma_{ij}^{\ell} 
\right) n_i
\nonumber\\
&=p v_{i} n_i - p^{\ell} v_{i}^{\ell} n_i
+ T q_{i} n_i - T^{\ell} q_{i}^{\ell} n_i
+ v_i \sigma_{ij} n_j - v_i^{\ell}\sigma_{ij}^{\ell} n_j
\label{Sigma_underline0}.
\end{align}
Equation~\eqref{Sigma_underline0} is simplified using the conservation laws at the interface.
For a simple interface (assumed here), the conservation laws for the mass, momentum and energy apprise the continuity of the normal fluxes \citep{SBRF2017}. 
With the gas (or vapour) properties on the left-hand sides and the liquid properties on the right-hand sides, the conservation laws at the interface in linear-dimensionless form read \citep{Muller1985, SBRF2017}
\begin{align}
\label{massBal_I}
\left(v_i -v_i^{I}\right) n_i &=\rho^{\ell} \left(v_i^{\ell}-v_i^{I}\right) n_i, 
\\
\label{momBal_I}
\left(p \delta_{ij}+\sigma_{ij}\right) n_j 
&= \left( p^{\ell} \delta_{ij} +\sigma_{ij}^{\ell}\right) n_j ,
\\
\label{energyBal_I}
\left[h_0 \left(v_i - v_i^{I}\right) + q_i\right] n_i 
&= \left[\rho^{\ell} h_0^{\ell} \left( v_i^{\ell} - v_i^{I}\right) + q_i^{\ell}\right] n_i,
\end{align}
where $v_i^{I}$ is the velocity of the interface, and $h_0$ and $h_0^{\ell}$ are the dimensionless enthalpies of the vapour and liquid, respectively, at temperature $\hat{T}_0$. 
The variables in \eqref{massBal_I}--\eqref{energyBal_I} have been non-dimensionalised using $\hat{\rho}_0$, $\hat{\theta}_0$ and their appropriate combinations.  
Substituting $\sigma_{ij}^{\ell} n_j$ and $q_i^{\ell} n_i$ from \eqref{momBal_I} and \eqref{energyBal_I} in \eqref{Sigma_underline0}, the underlined term in \eqref{Sigma_app_eq1} becomes
\begin{align}
\underline{\Sigma}^{I}
&=p \left(v_i - v_i^{\ell}\right) n_i
+ \left(T- T^{\ell}\right) q_{i} n_i - T^{\ell} \left(h_0 - h_0^{\ell}\right) \left(v_i -v_i^{I}\right) n_i 
+ \left(v_j - v_j^{\ell}\right) \sigma_{ij} n_i
\nonumber\\
&=p \mathscr{V}_i n_i
+ \mathcal{T} q_{i} n_i - T^{\ell} \left(h_0 - h_0^{\ell}\right) \mathcal{V}_i n_i 
+ \mathscr{V}_j \sigma_{ij} n_i.
\label{Sigma_underline1}
\end{align}
where $\mathscr{V}_i = v_i - v_i^{\ell}$, $\mathcal{T} = T- T^{\ell}$, $\mathcal{V}_i = v_i -v_i^{I}$. 
On using \eqref{massBal_I}, it turns out that
\begin{align}
\label{normalslipvelocity}
\mathscr{V}_i \, n_i
=\left(v_i - v_i^{\ell}\right) n_i
= \left[\mathcal{V}_i - \left(v_i^{\ell} - v_i^{I}\right)\right] n_i
= \left(1 - \frac{1}{\rho^{\ell}}\right) \mathcal{V}_i n_i
\approx
\mathcal{V}_i \, n_i
\end{align}
because the density of a gas/vapour is usually much smaller than the density of a liquid/solid, and hence $\rho^{\ell} \gg 1$ or $1/\rho^{\ell} \ll 1$.
The enthalpy term in \eqref{Sigma_underline1} is eliminated using the definition of the (dimensionless) saturation pressure $p_{\mathrm{sat}}$ at temperature $T^{\ell}$ \citep{RanaPRL2019},
\begin{align}
\label{psat}
p_{\mathrm{sat}} \equiv p_{\mathrm{sat}} (T^{\ell}) = T^{\ell} \left(h_0 - h_0^{\ell}\right).
\end{align}
Using \eqref{normalslipvelocity} and \eqref{psat}, equation \eqref{Sigma_underline1} reduces to 
\begin{align}
\underline{\Sigma}^{I} 
&=p \mathcal{V}_i \, n_i
+ \mathcal{T} q_{i} n_i - p_{\mathrm{sat}} \mathcal{V}_i \, n_i 
+ \mathscr{V}_j \sigma_{ij} n_i
= \left(\mathcal{P} \mathcal{V}_i 
+ \mathcal{T} q_{i}
+ \mathscr{V}_j \sigma_{ij}\right) n_i,
\label{Sigma_underline2}
\end{align}
where $\mathcal{P} = p - p_{\mathrm{sat}}$. 
Replacing the underlined terms in \eqref{Sigma_app_eq1} with the value of $\underline{\Sigma}^{I}$ obtained in \eqref{Sigma_underline2}, the entropy flux at the interface is given by \eqref{entropyProdwall}.
%
\section{Decomposition of symmetric-tracefree tensors}\label{app:Decomp}
In the following, we decompose vectors and symmetric-tracefree tensors of rank two and three into their components in the direction of a given normal $\bm{n}$ and in the directions perpendicular to it (mutually perpendicular tangential directions, let us say $\bm{t}_1$ and $\bm{t}_2$). Obviously, $n_i {t_1}_i = n_i {t_2}_i = {t_1}_i {t_2}_i = 0$. When it is not required to distinguish between $\bm{t}_1$ and $\bm{t}_2$, both are referred to with a common notation $\bm{t}$ for simplicity, and $\bm{t}$ could either be $\bm{t}_1$ and $\bm{t}_2$.
\begin{itemize}
\item 
A vector $a_i$ is decomposed as 
\begin{align}
\label{decomvector}
a_{i} = a_{n} n_{i} + \bar{a}_{i},
\end{align}
where $a_{n} n_{i}$ with $a_n := a_k n_k$ is the normal component of $a_i$
while $\bar{a}_{i} := a_{i} - a_{n} n_{i}$ is the tangential component of $%
a_i $. By definition, $\bar{a}_{i}$ is such that $\bar{a}_{i} n_i = 0$. 
Furthermore, multiplication of \eqref{decomvector} with $t_i \in \{{t_1}_i,{t_2}_i\}$ gives
\begin{align}
a_t=\bar{a}_t,
\end{align}
where $a_t:=a_i t_i$, $\bar{a}_t:=\bar{a}_{i} t_i$.
\item
A symmetric-tracefree rank-2 tensor $A_{ij}$ is decomposed as \citep{RanaStruchtrup2016,RGS2018} 
\begin{align}  \label{decompsigma}
A_{ij} = \frac{3}{2} A_{nn} \, n_{\langle i}n_{j \rangle} + \bar{A}_{ni}
n_{j} + \bar{A}_{nj} n_{i} + \tilde{A}_{ij},
\end{align}
where Einstein summation never applies to the indices `$n$', and 
\begin{align}
\label{decomtensor2}
\left.
\begin{aligned}
A_{nn} &:=
A_{ij} n_i n_j,
\\
\bar{A}_{ni} &:= A_{ij} n_{j} - A_{nn} n_{i},
\\
\tilde{A}_{ij} &:= A_{ij} - \frac{3}{2} A_{nn} \, n_{\langle i}n_{j \rangle} - 
\bar{A}_{ni} n_{j} - \bar{A}_{nj} n_{i}.
\end{aligned}
\right\}
\end{align}
It is easy to verify that $\bar{A}_{ni} n_{i} = 0$ and $\tilde{A}_{kk} = \tilde{A}_{ij} n_{i} = 
\tilde{A}_{ij} n_{j} = 0$. 
Furthermore, multiplication of \eqref{decomtensor2}$_2$ with $t_i \in \{{t_1}_i,{t_2}_i\}$ gives
\begin{align}
\bar{A}_{nt} = A_{nt},
\end{align}
where $\bar{A}_{nt} := \bar{A}_{ni} t_i$, $A_{nt}=A_{tn}:= A_{ij} t_i n_{j}$; multiplication of \eqref{decomtensor2}$_3$ with $t_i t_j$ gives
\begin{align}
\tilde{A}_{tt} &= A_{tt} + \frac{1}{2} A_{nn}, 
\end{align}
where $\tilde{A}_{tt} := \tilde{A}_{ij} t_i t_j$ and $A_{tt} := A_{ij} t_i t_j$;
multiplication of \eqref{decomtensor2}$_3$ with ${t_1}_i {t_2}_j$ gives
\begin{align}
\tilde{A}_{{t_1}{t_2}} &= A_{{t_1}{t_2}}, 
\end{align}
where $\tilde{A}_{{t_1}{t_2}}: = \tilde{A}_{ij} {t_1}_i {t_2}_j$ and $A_{{t_1}{t_2}} := A_{ij} {t_1}_i {t_2}_j$.
\item
A symmetric-tracefree rank-3 tensor $A_{ijk}$ is decomposed as 
\begin{align}
A_{ijk} &= \frac{5}{2} A_{nnn} n_{\langle i}n_{j}n_{k\rangle}
+\frac{15}{4}\bar{A}_{nn\langle i}n_{j}n_{k\rangle}
+ 3\tilde{A}_{n\langle ij}n_{k\rangle} + \check{A}_{ijk}.
\end{align}
where 
\begingroup 
\allowdisplaybreaks
\begin{align}
\label{decomtensor3}
\left.
\begin{aligned}
A_{nnn} &:= A_{ijk} n_{i} n_{j}n_{k},
\\
\bar{A}_{nni} &:= A_{ijk}n_{j}n_{k}-A_{nnn}n_{i},
\\
\tilde{A}_{nij} &:= A_{ijk} n_{k} - \frac{3}{2} A_{nnn} n_{\langle i}n_{j\rangle} - \bar{A}_{nni} n_{j} - \bar{A}_{nnj} n_{i},
\\
\check{A}_{ijk} &:= A_{ijk} - \frac{5}{2} A_{nnn} n_{\langle i}n_{j}n_{k\rangle}
- \frac{15}{4}\bar{A}_{nn\langle i}n_{j}n_{k\rangle}
- 3\tilde{A}_{n\langle ij}n_{k\rangle}.
\end{aligned}
\right\}
\end{align}
\endgroup
It is again easy to verify that 
$\bar{A}_{nni}n_{i}=\tilde{A}_{nij}n_{j}=\tilde{A}_{nii}=\check{A}_{ijk}n_{k}=\check{A}_{ikk}=0$. 
Furthermore, multiplication of \eqref{decomtensor3}$_2$ with $t_i \in \{{t_1}_i,{t_2}_i\}$ gives
\begin{align}
\bar{A}_{nnt} = A_{nnt},
\end{align}
where $\bar{A}_{nnt} := \bar{A}_{nni} t_i$, $A_{nnt}=A_{tnn}:= A_{ijk}t_i n_{j}n_{k}$; 
multiplication of \eqref{decomtensor3}$_3$ with $t_i t_j$ gives
\begin{align}
\tilde{A}_{ntt} &= A_{ntt} + \frac{1}{2} A_{nnn}, 
\end{align}
where $\tilde{A}_{ntt}: = \tilde{A}_{nij} t_i t_j$ and $A_{ntt} = A_{ttn} := A_{ijk} t_i t_j n_{k}$;
multiplication of \eqref{decomtensor3}$_3$ with ${t_1}_i {t_2}_j$ gives
\begin{align}
\tilde{A}_{n{t_1}{t_2}} &= A_{n{t_1}{t_2}}, 
\end{align}
where $\tilde{A}_{n {t_1}{t_2}}: = \tilde{A}_{nij} {t_1}_i {t_2}_j$ and $A_{n{t_1}{t_2}} = A_{{t_1}{t_2}n} := A_{ijk} {t_1}_i {t_2}_j n_{k}$;
multiplication of \eqref{decomtensor3}$_4$ with $t_i t_j t_k$ gives
\begin{align}
\check{A}_{ttt} = A_{ttt} 
+ \frac{3}{4} A_{n nt}, 
\end{align}
where $\check{A}_{ttt} := \check{A}_{ijk} t_i t_j t_k$ and $A_{ttt} := A_{ijk} t_i t_j t_k$;
multiplication of \eqref{decomtensor3}$_4$ with ${t_1}_i {t_1}_j {t_2}_k$ gives
\begin{align}
\check{A}_{{t_1} {t_1} {t_2}} &= A_{{t_1} {t_1} {t_2}} + \frac{1}{4} A_{n n{t_2}}, 
\end{align}
where $\check{A}_{{t_1} {t_1} {t_2}} := \check{A}_{ijk} {t_1}_i {t_1}_j {t_2}_k$ and $A_{{t_1} {t_1} {t_2}} := A_{ijk} {t_1}_i {t_1}_j {t_2}_k$.
\end{itemize}
\section{Analytic solution for the unknowns in ansatz \eqref{eqn:vectorfields}--\eqref{eqn:tensorrfields}}
\label{app:sol}
The analytic solution for the unknowns appearing in \eqref{eqn:vectorfields}--\eqref{eqn:tensorrfields} are as follows:
\begingroup
\allowdisplaybreaks
\begin{align}
\label{eqn:vectors}
a_{1}(r) &= \frac{C_1}{2r}+\frac{C_2}{3r^3}-K_1 \Theta_1(r) \left(\frac{0.975701 \mathrm{Kn}^3}{r^3}+\frac{0.497885 \mathrm{Kn}^2}{r^2}\right)
\nonumber\\
&\quad - K_2 \Theta_2(r) \left(\frac{1.4893 \mathrm{Kn}^3}{r^3}+\frac{1.88528 \mathrm{Kn}^2}{r^2}\right),
\\
a_{2}(r) &= \frac{C_1}{4r}-\frac{C_2}{6r^3}+K_1 \Theta_1(r) \left(\frac{0.487851 \mathrm{Kn}^3}{r^3}+\frac{0.248943 \mathrm{Kn}^2}{r^2}+\frac{0.127032 \mathrm{Kn}}{r}\right)
\nonumber\\
&\quad +K_2 \Theta_2(r) \left(\frac{0.74465 \mathrm{Kn}^3}{r^3}+\frac{0.942641 \mathrm{Kn}^2}{r^2}+\frac{1.19327 \mathrm{Kn}}{r}\right),
\\
\alpha_{1}(r) &= \frac{\mathrm{Kn}^2 }{r^3}\left(\frac{C_3}{6\mathrm{Kn}}-\frac{3 C_1}{2}\right)
+K_1 \Theta_1(r) \left(\frac{1.50825 \mathrm{Kn}^3}{r^3}+\frac{0.769634 \mathrm{Kn}^2}{r^2}\right)
\nonumber\\
&\quad -K_2 \Theta_2(r) \left(\frac{0.80299 \mathrm{Kn}^3}{r^3}+\frac{1.01649 \mathrm{Kn}^2}{r^2}\right),
\\
\alpha_{2}(r) &= \frac{\mathrm{Kn}^2 }{r^3}\left(\frac{3C_1}{4}-\frac{C_3}{12 \mathrm{Kn}}\right)
\nonumber\\
&\quad-K_1 \Theta_1(r) \left(\frac{0.754123 \mathrm{Kn}^3}{r^3}+\frac{0.384817 \mathrm{Kn}^2}{r^2}+\frac{0.196366 \mathrm{Kn}}{r}\right)
\nonumber\\
&\quad+ K_2 \Theta_2(r)  \left(\frac{0.401495 \mathrm{Kn}^3}{r^3}+\frac{0.508246 \mathrm{Kn}^2}{r^2}+\frac{0.643381 \mathrm{Kn}}{r}\right),
\end{align}
\begin{align}
\label{eqn:scalars}
c(r) &= \frac{C_3}{45 r^2}-K_3 \Theta_3(r) \left(\frac{0.0172036 \mathrm{Kn}^2}{r^2}+\frac{0.0200113 \mathrm{Kn}}{r}\right)
\nonumber\\
&\quad -K_4 \Theta_4(r) \left(\frac{0.0629047 \mathrm{Kn}^2}{r^2}+\frac{0.0426083 \mathrm{Kn}}{r}\right)
\nonumber\\
&\quad -K_5 \Theta_5(r) \left(\frac{0.443094 \mathrm{Kn}^2}{r^2}+\frac{0.200538 \mathrm{Kn}}{r}\right),
\\
d(r) &=\frac{C_1\mathrm{Kn}}{2r^2}-K_3\Theta_3(r) \left(\frac{0.931108 \mathrm{Kn}^2}{r^2}+\frac{1.08307 \mathrm{Kn}}{r}\right)
\nonumber\\
&\quad-K_4 \Theta_4(r) \left(\frac{0.254173 \mathrm{Kn}^2}{r^2}+\frac{0.172163 \mathrm{Kn}}{r}\right)
\nonumber\\
&\quad-K_5 \Theta_5(r) \left(\frac{0.0501203 \mathrm{Kn}^2}{r^2}+\frac{0.0226838 \mathrm{Kn}}{r}\right),
\\
\delta(r) &=K_3 \Theta_3(r)  \left(\frac{14.0055 \mathrm{Kn}^2}{r^2}+\frac{16.2913 \mathrm{Kn}}{r}\right)
-K_4 \Theta_4(r) \left(\frac{1.71119 \mathrm{Kn}^2}{r^2}+\frac{1.15907 \mathrm{Kn}}{r}\right)
\nonumber\\
&\quad+K_5 \Theta_5(r)\left(\frac{3.75265 \mathrm{Kn}^2}{r^2}+\frac{1.6984 \mathrm{Kn}}{r}\right).
\end{align}
\begin{align}
\label{eqn:tensors}
b_1(r) &=\frac{C_1 \mathrm{Kn}}{r^2}+\frac{\mathrm{Kn}^4}{r^4}\left(-\frac{10 C_1}{\mathrm{Kn}}+\frac{2 C_2}{\mathrm{Kn}^3}+\frac{2 C_3}{5\mathrm{Kn}^2}\right)
\nonumber\\
&\quad+K_3 \Theta_3(r) \left(\frac{6.19341 \mathrm{Kn}^4}{r^4}+\frac{7.20421 \mathrm{Kn}^3}{r^3}+\frac{3.72443 \mathrm{Kn}^2}{r^2}+\frac{1.08307 \mathrm{Kn}}{r}\right)
\nonumber\\
&\quad+K_4 \Theta_4(r) \left(\frac{4.98597 \mathrm{Kn}^4}{r^4}+\frac{3.37723 \mathrm{Kn}^3}{r^3}+\frac{1.01669 \mathrm{Kn}^2}{r^2}+\frac{0.172163 \mathrm{Kn}}{r}\right)
\nonumber\\
&\quad+K_5 \Theta_5(r) \left(\frac{2.20218 \mathrm{Kn}^4}{r^4}+\frac{0.996676 \mathrm{Kn}^3}{r^3}+\frac{0.200481 \mathrm{Kn}^2}{r^2}+\frac{0.0226838 \mathrm{Kn}}{r}\right),
\\
b_2(r) &=\frac{\mathrm{Kn}^4}{r^4}\left(-\frac{5 C_1}{\mathrm{Kn}}+\frac{C_2}{\mathrm{Kn}^3}+\frac{\text{C3}}{5\mathrm{Kn}^2}\right)
\nonumber\\
&\quad+K_3 \Theta_3(r)\left(\frac{3.09671 \mathrm{Kn}^4}{r^4}+\frac{3.6021 \mathrm{Kn}^3}{r^3}+\frac{1.39666 \mathrm{Kn}^2}{r^2}\right)
\nonumber\\
&\quad+K_4 \Theta_4(r) \left(\frac{2.49299 \mathrm{Kn}^4}{r^4}+\frac{1.68862 \mathrm{Kn}^3}{r^3}+\frac{0.381259 \mathrm{Kn}^2}{r^2}\right)
\nonumber\\
&\quad+K_5 \Theta_5(r) \left(\frac{1.10109 \mathrm{Kn}^4}{r^4}+\frac{0.498338 \mathrm{Kn}^3}{r^3}+\frac{0.0751805 \mathrm{Kn}^2}{r^2}\right),
\\
\beta_1(r) &=\frac{\mathrm{Kn}^4}{r^4}\left(\frac{12C_3}{5\mathrm{Kn}^2}-\frac{228 C_1}{7\mathrm{Kn}}\right)
\nonumber\\
&\quad+K_1 \Theta_1(r) \left(\frac{46.338 \mathrm{Kn}^4}{r^4}+\frac{23.6455 \mathrm{Kn}^3}{r^3}+\frac{4.02199 \mathrm{Kn}^2}{r^2}\right)
\nonumber\\
&\quad-K_2 \Theta_2(r) \left(\frac{4.00877 \mathrm{Kn}^4}{r^4}+\frac{5.07465 \mathrm{Kn}^3}{r^3}+\frac{2.14131 \mathrm{Kn}^2}{r^2}\right)
\nonumber\\
&\quad-K_3 \Theta_3(r)\left(\frac{42.868 \mathrm{Kn}^4}{r^4}+\frac{49.8643 \mathrm{Kn}^3}{r^3}+\frac{25.7788 \mathrm{Kn}^2}{r^2}+\frac{7.49652 \mathrm{Kn}}{r}\right)
\nonumber\\
&\quad+K_4\Theta_4(r) \left(\frac{7.38704 \mathrm{Kn}^4}{r^4}+\frac{5.00359 \mathrm{Kn}^3}{r^3}+\frac{1.5063 \mathrm{Kn}^2}{r^2}+\frac{0.255071 \mathrm{Kn}}{r}\right)
\nonumber\\
&\quad+K_5\Theta_5(r) \left(\frac{37.9774 \mathrm{Kn}^4}{r^4}+\frac{17.1881 \mathrm{Kn}^3}{r^3}+\frac{3.45738 \mathrm{Kn}^2}{r^2}+\frac{0.391191 \mathrm{Kn}}{r}\right),
\\
\beta_2(r) &=\frac{\mathrm{Kn}^4}{r^4}\left(\frac{6C_3}{5\mathrm{Kn}^2}-\frac{114 C_1}{7\mathrm{Kn}}\right)
\nonumber\\
&\quad+K_1\Theta_1(r) \left(\frac{23.169 \mathrm{Kn}^4}{r^4}+\frac{11.8228 \mathrm{Kn}^3}{r^3}+\frac{3.01649 \mathrm{Kn}^2}{r^2}+\frac{0.51309 \mathrm{Kn}}{r}\right)
\nonumber\\
&\quad-K_2\Theta_2(r)\left(\frac{2.00439 \mathrm{Kn}^4}{r^4}+\frac{2.53732 \mathrm{Kn}^3}{r^3}+\frac{1.60598 \mathrm{Kn}^2}{r^2}+\frac{0.677662 \mathrm{Kn}}{r}\right)
\nonumber\\
&\quad-K_3 \Theta_3(r)\left(\frac{21.434 \mathrm{Kn}^4}{r^4}+\frac{24.9321 \mathrm{Kn}^3}{r^3}+\frac{9.66706 \mathrm{Kn}^2}{r^2}\right)
\nonumber\\
&\quad+K_4\Theta_4(r)\left(\frac{3.69352 \mathrm{Kn}^4}{r^4}+\frac{2.50179 \mathrm{Kn}^3}{r^3}+\frac{0.564861 \mathrm{Kn}^2}{r^2}\right)
\nonumber\\
&\quad+K_5\Theta_5(r) \left(\frac{18.9887 \mathrm{Kn}^4}{r^4}+\frac{8.59403 \mathrm{Kn}^3}{r^3}+\frac{1.29652 \mathrm{Kn}^2}{r^2}\right),
\end{align}
\begin{align}
\label{eqn:kappa1}
\kappa_1(r) &=\frac{24 C_1 \mathrm{Kn}^2}{5 r^3}+\frac{\mathrm{Kn}^5}{r^5}\left(-\frac{7344 C_1}{49 \mathrm{Kn}}+\frac{16 C_2}{\mathrm{Kn}^3}+\frac{208 C_3}{35 \mathrm{Kn}^2}\right)
\nonumber\\
&\quad+K_1\Theta_1(r) \left(\frac{64.3578 \mathrm{Kn}^5}{r^5}+\frac{32.8408 \mathrm{Kn}^4}{r^4}+\frac{6.70326 \mathrm{Kn}^3}{r^3}+\frac{0.570095 \mathrm{Kn}^2}{r^2}\right)
\nonumber\\
&\quad+K_2 \Theta_2(r)\left(\frac{50.8419 \mathrm{Kn}^5}{r^5}+\frac{64.3599 \mathrm{Kn}^4}{r^4}+\frac{32.5889 \mathrm{Kn}^3}{r^3}+\frac{6.87563 \mathrm{Kn}^2}{r^2}\right)
\nonumber\\
&\quad+K_3\Theta_3(r)\left(\frac{30.5159 \mathrm{Kn}^5}{r^5}+\frac{35.4962 \mathrm{Kn}^4}{r^4}+\frac{18.5802 \mathrm{Kn}^3}{r^3}+\frac{5.60327 \mathrm{Kn}^2}{r^2}+\frac{0.931108 \mathrm{Kn}}{r}\right)
\nonumber\\
&\quad+K_4 \Theta_4(r)\left(\frac{72.4497 \mathrm{Kn}^5}{r^5}+\frac{49.0736 \mathrm{Kn}^4}{r^4}+\frac{14.9579 \mathrm{Kn}^3}{r^3}+\frac{2.62674 \mathrm{Kn}^2}{r^2}+\frac{0.254173 \mathrm{Kn}}{r}\right)
\nonumber\\
&\quad+K_5 \Theta_5(r)\left(\frac{71.6731 \mathrm{Kn}^5}{r^5}+\frac{32.4383 \mathrm{Kn}^4}{r^4}+\frac{6.60653 \mathrm{Kn}^3}{r^3}+\frac{0.775193 \mathrm{Kn}^2}{r^2}+\frac{0.0501203 \mathrm{Kn}}{r}\right),
\\
\label{eqn:kappa2}
\kappa_2(r) &=\frac{4 C_1 \mathrm{Kn}^2}{5 r^3}+\frac{\mathrm{Kn}^5}{r^5}\left(-\frac{3672 C_1}{49 \mathrm{Kn}}+\frac{8 C_2}{\mathrm{Kn}^3}+\frac{104 C_3}{35 \mathrm{Kn}^2}\right)
\nonumber\\
&\quad+K_1 \Theta_1(r)\left(\frac{32.1789 \mathrm{Kn}^5}{r^5}+\frac{16.4204 \mathrm{Kn}^4}{r^4}+\frac{3.91023 \mathrm{Kn}^3}{r^3}+\frac{0.570095 \mathrm{Kn}^2}{r^2}+\frac{0.0484851 \mathrm{Kn}}{r}\right)
\nonumber\\
&\quad+K_2\Theta_2(r)\left(\frac{25.4209 \mathrm{Kn}^5}{r^5}+\frac{32.18 \mathrm{Kn}^4}{r^4}+\frac{19.0102 \mathrm{Kn}^3}{r^3}+\frac{6.87563 \mathrm{Kn}^2}{r^2}+\frac{1.45063 \mathrm{Kn}}{r}\right)
\nonumber\\
&\quad+K_3\Theta_3(r)\left(\frac{15.2579 \mathrm{Kn}^5}{r^5}+\frac{17.7481 \mathrm{Kn}^4}{r^4}+\frac{8.25788 \mathrm{Kn}^3}{r^3}+\frac{1.60094 \mathrm{Kn}^2}{r^2}\right)
\nonumber\\
&\quad+K_4\Theta_4(r)\left(\frac{36.2249 \mathrm{Kn}^5}{r^5}+\frac{24.5368 \mathrm{Kn}^4}{r^4}+\frac{6.64796 \mathrm{Kn}^3}{r^3}+\frac{0.750496 \mathrm{Kn}^2}{r^2}\right)
\nonumber\\
&\quad+K_5\Theta_5(r)\left(\frac{35.8366 \mathrm{Kn}^5}{r^5}+\frac{16.2192 \mathrm{Kn}^4}{r^4}+\frac{2.93623 \mathrm{Kn}^3}{r^3}+\frac{0.221484 \mathrm{Kn}^2}{r^2}\right).
\end{align} 
\endgroup
In \eqref{eqn:vectors}--\eqref{eqn:kappa2}, the coefficients 
\begin{align}
\left.
\begin{gathered}
\Theta_1(r)  =  \exp\left[-\dfrac{0.510285}{\mathrm{Kn}} (r-1)\right],
\quad
\Theta_2(r)=\exp\left[-\dfrac{1.26588}{\mathrm{Kn}}(r-1)\right],
\\
\Theta_3(r)  =  \exp\left[-\frac{1.16321}{\mathrm{Kn}} (r-1)\right],
\quad
\Theta_4(r)  =  \exp\left[-\frac{0.677347}{\mathrm{Kn}} (r-1)\right]
\\
\textrm{and}\quad
\Theta_5(r)  =  \exp\left[-\frac{0.452587}{\mathrm{Kn}} (r-1)\right]
\end{gathered}
\right\}
\end{align}
describe the Knudsen layer functions that vanish as $r \to \infty$.
%

\bibliography{refer}
\bibliographystyle{jfm_doi}

\end{document}


\maketitle
\vspace*{-2mm}

The purpose of this supplementary material is to illustrate that the solutions of the LR26 equations for the problems, namely (i) a steady gas flow past a rigid sphere and (ii) a steady vapour flow past an evaporating spherical droplet, obtained with the PBC and  the MBC differ only negligibly.
\begin{figure}[!b]
\centering
\includegraphics[height=45mm]{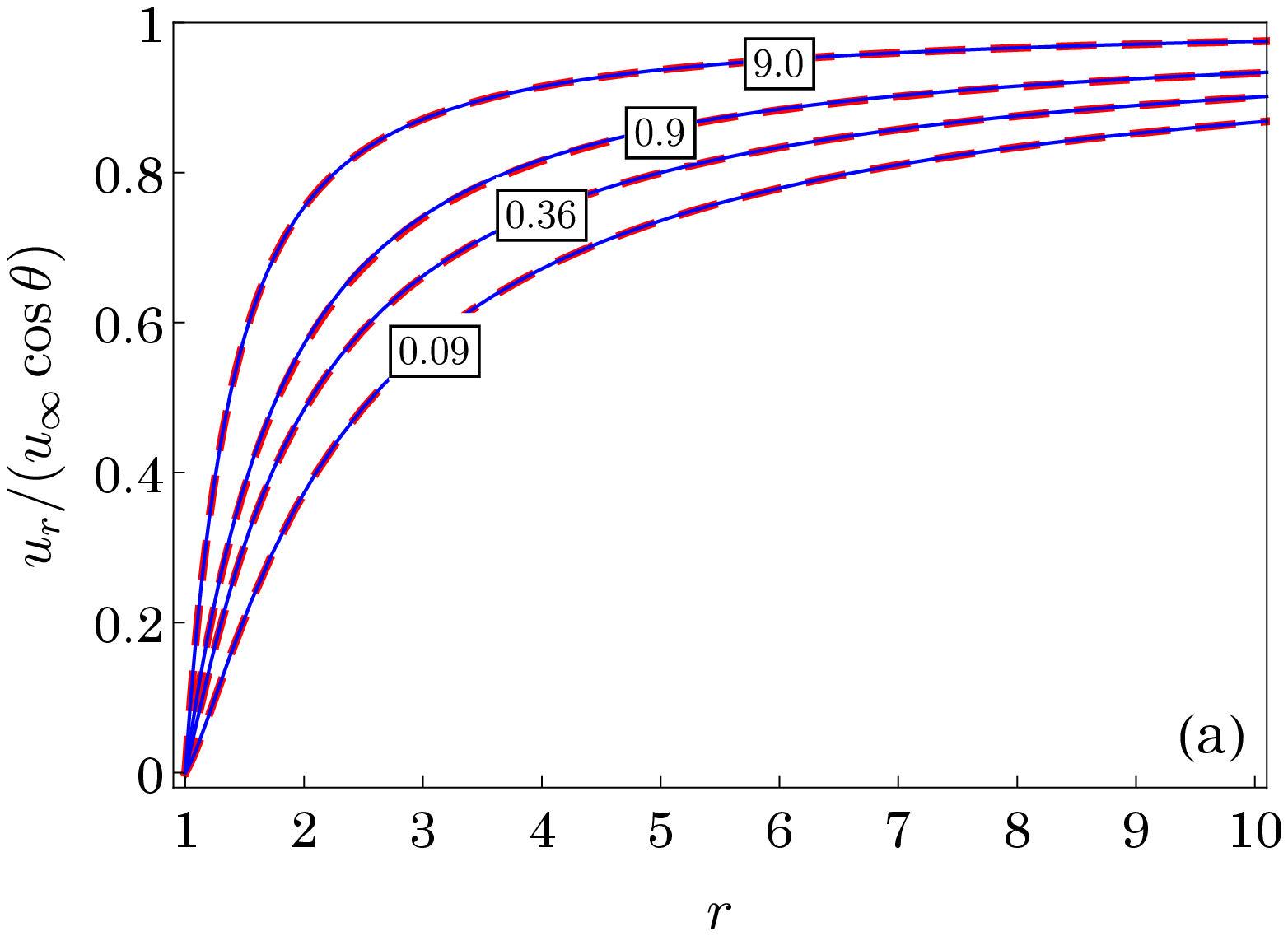}
\hfill
\includegraphics[height=45mm]{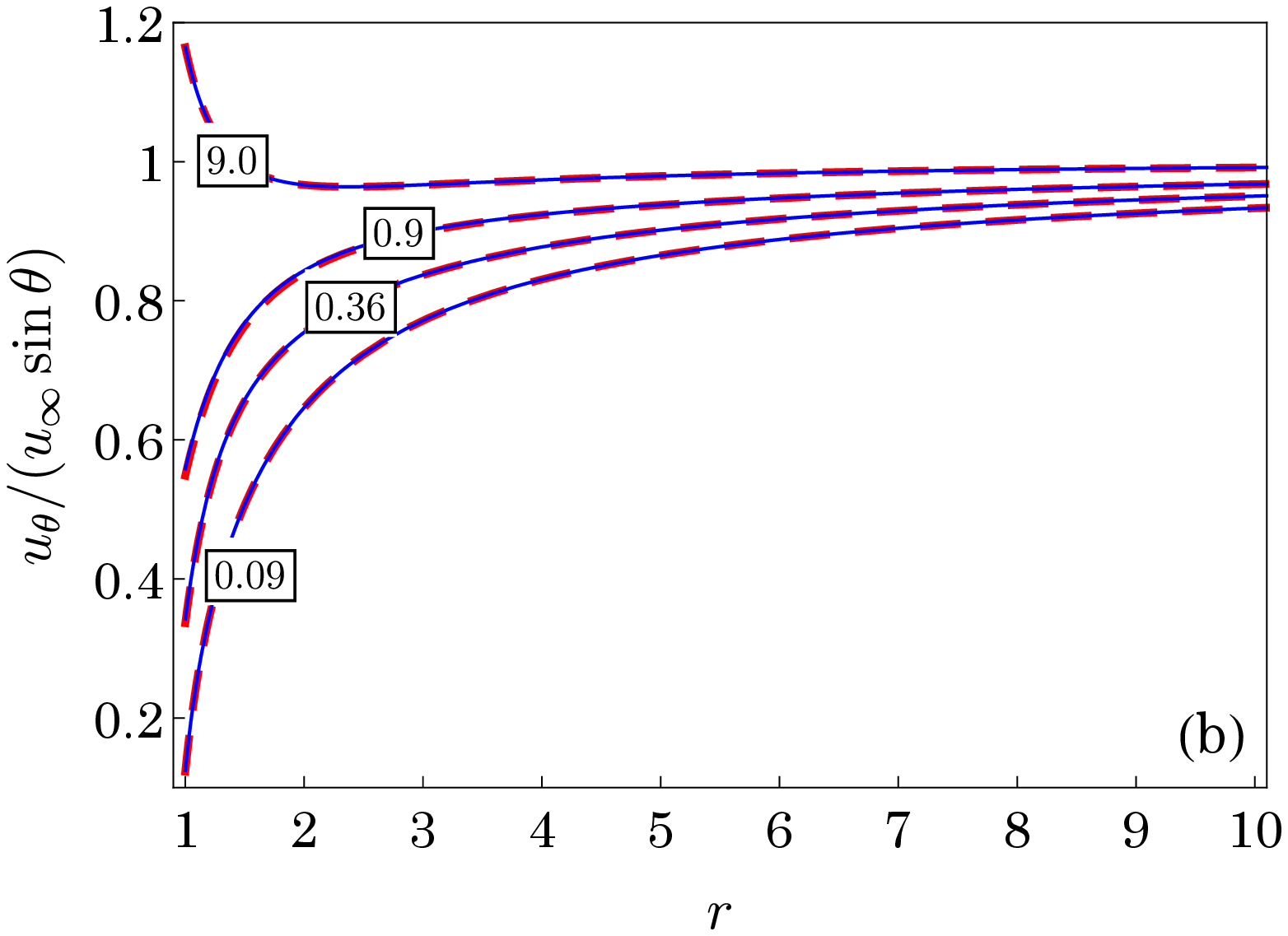}
\\
\includegraphics[height=45mm]{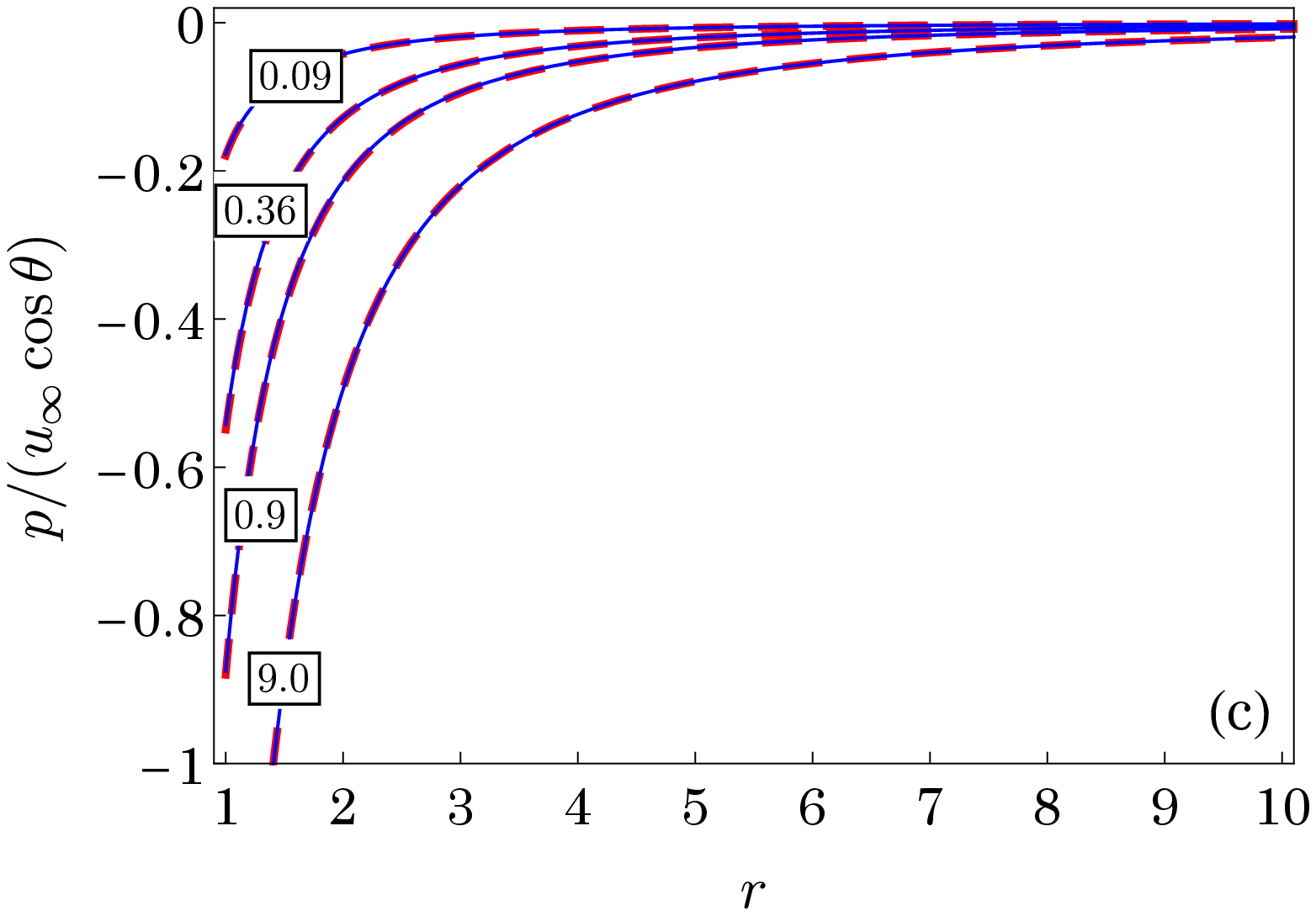}
\hfill
\includegraphics[height=45mm]{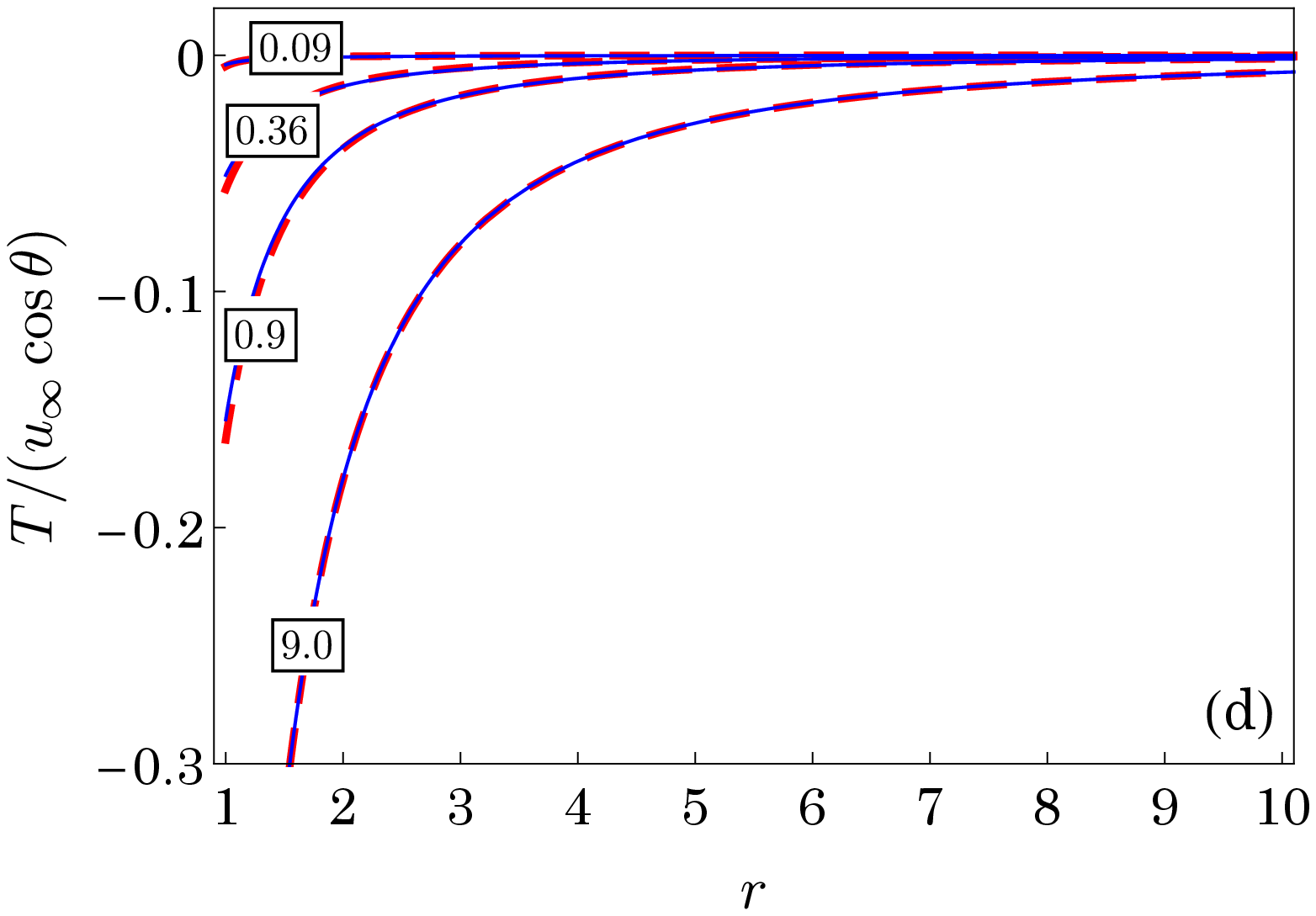}
\caption{\label{fig:profiles_non_evp_R26MBCvsPBC}Profiles of the (a) radial velocity, (b) polar velocity, (c) pressure and (d) temperature of the gas in a steady gas flow past a rigid sphere as functions of the radial distance from the surface of the sphere for various Knudsen numbers (depicted in the boxes above the curves). 
Results are obtained from the LR26 theory with the PBC (solid blue lines) and with the MBC (dashed red lines).}
\end{figure}

\begin{figure}[!t]
\centering
\includegraphics[height=45mm]{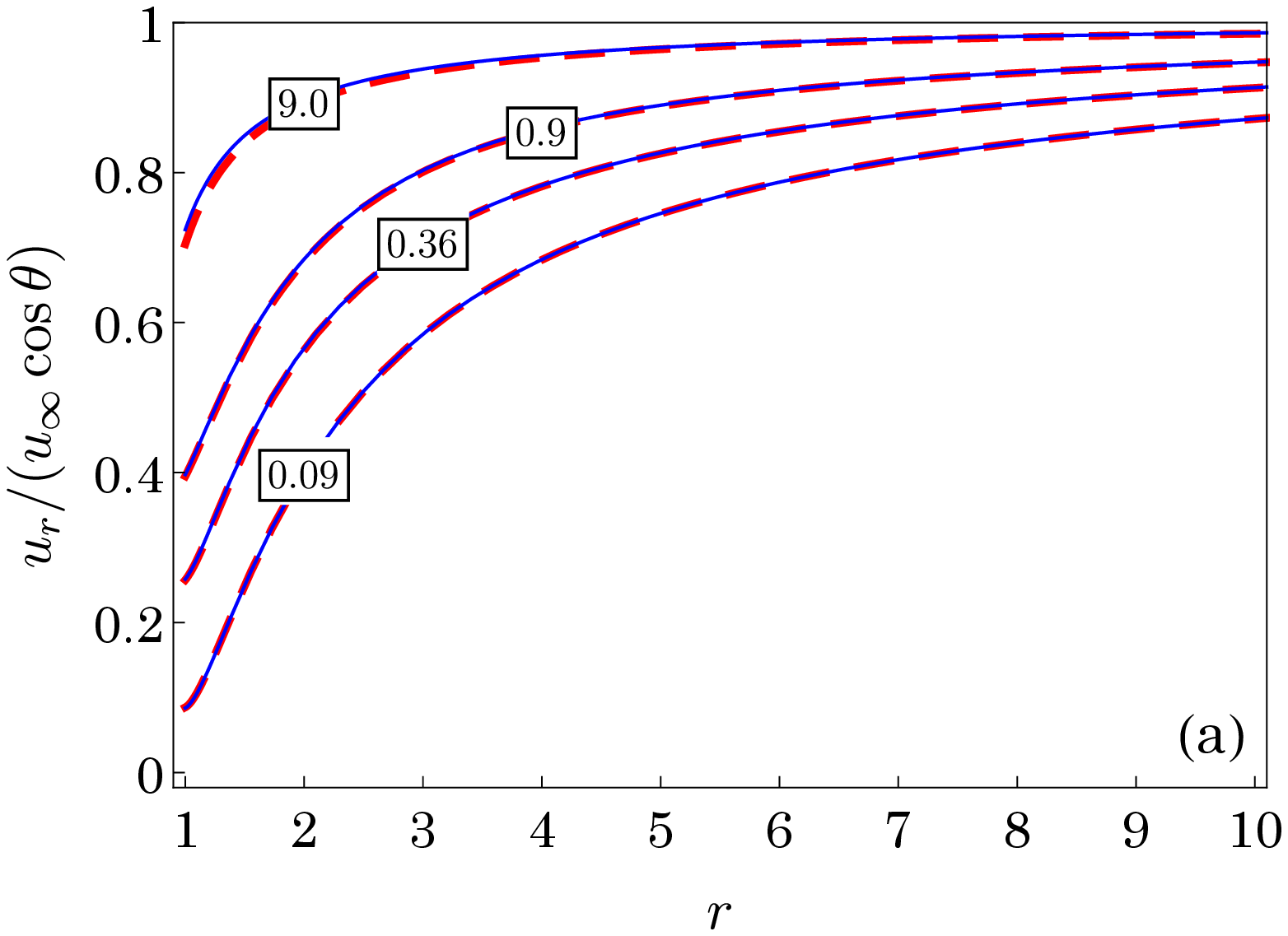}
\hfill
\includegraphics[height=45mm]{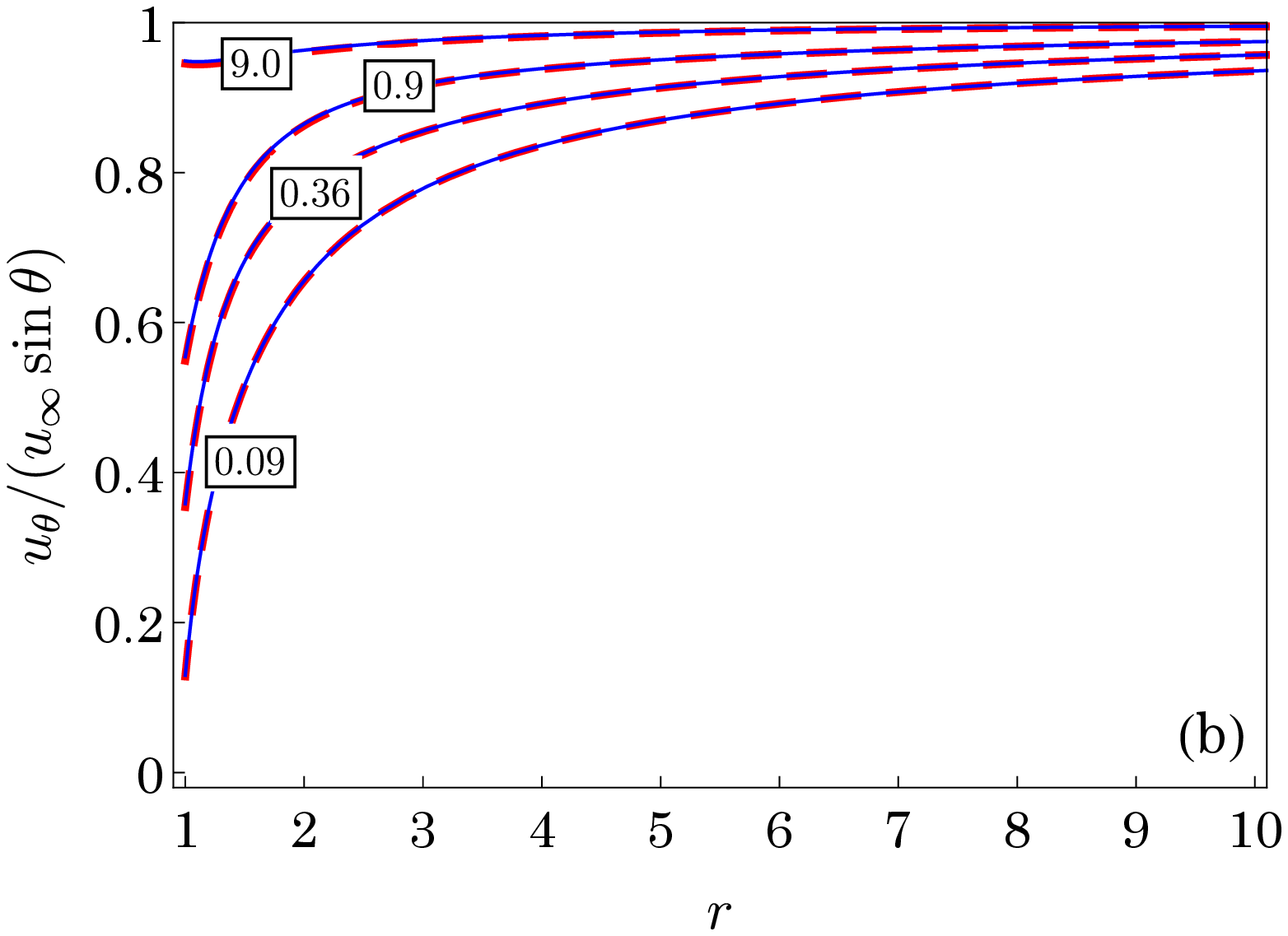}
\\
\includegraphics[height=45mm]{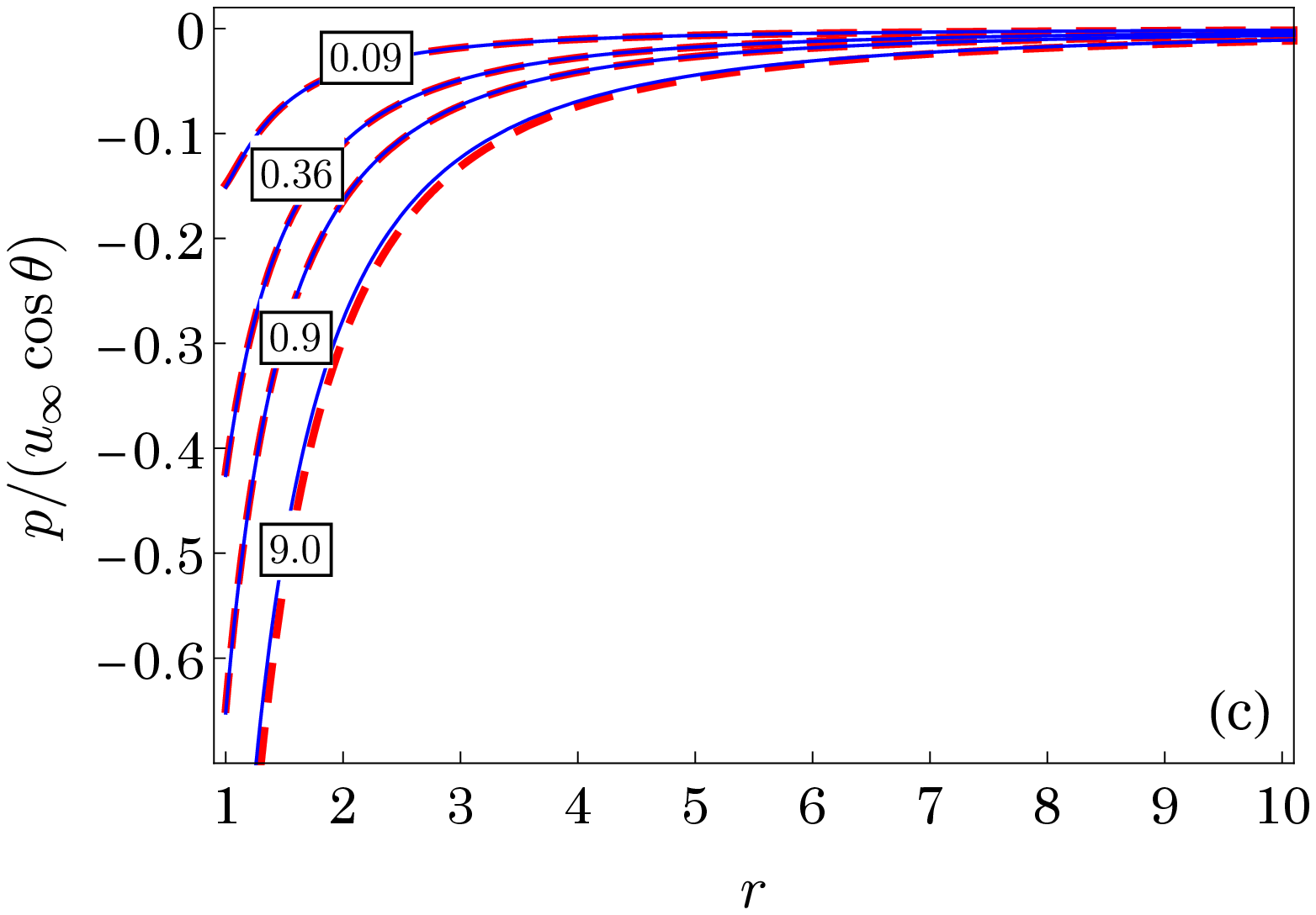}
\hfill
\includegraphics[height=45mm]{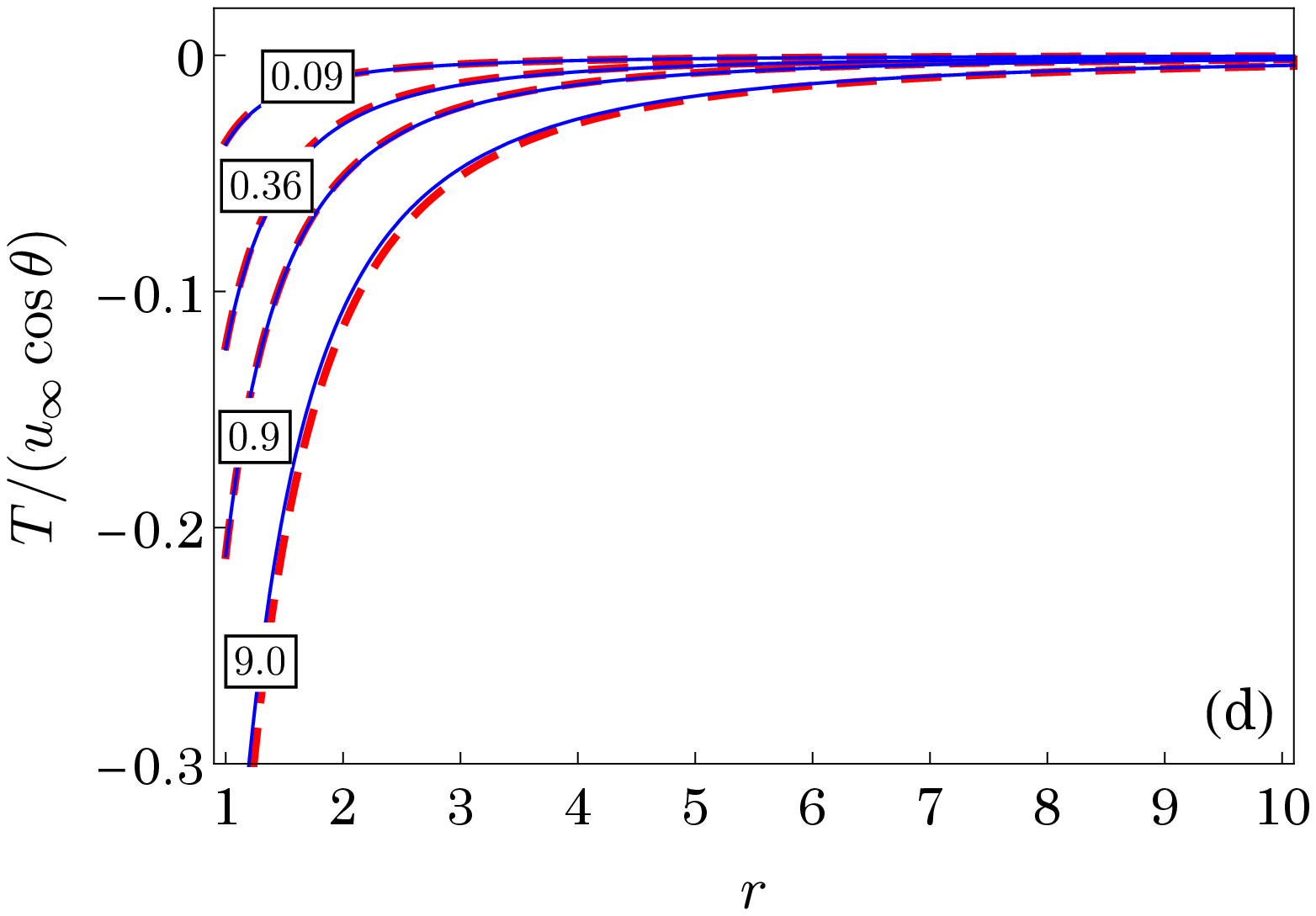}
\caption{\label{fig:profiles_evp_R26MBCvsPBC}
Same as figure~\ref{fig:profiles_non_evp_R26MBCvsPBC}
but in a steady vapour flow past an evaporating spherical droplet.}
\end{figure}

Figures \ref{fig:profiles_non_evp_R26MBCvsPBC} and \ref{fig:profiles_evp_R26MBCvsPBC} exhibit the (a) radial velocity, (b) polar velocity, (c) pressure and (d) temperature profiles obtained with the PBC (solid blue lines) and with the MBC (dashed red line) for different values of the Knudsen number (depicted in the boxes above the curves) for the aforementioned problems (respectively). 
Clearly, for both problems, the difference between the results with the PBC and MBC is negligibly small.

\begin{figure}[!h]
\centering
\includegraphics[height=45mm]{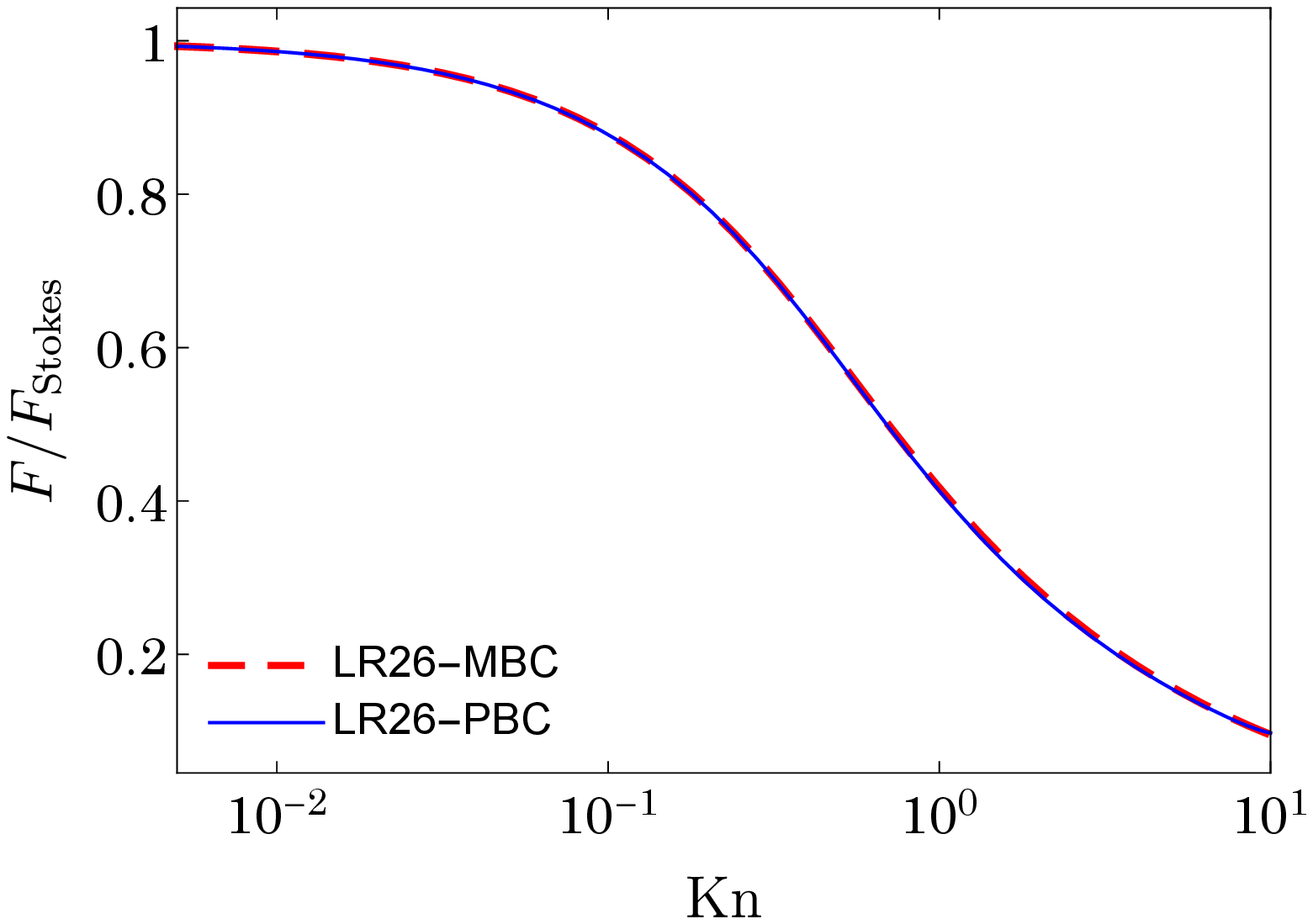}
\hfill
\includegraphics[height=45mm]{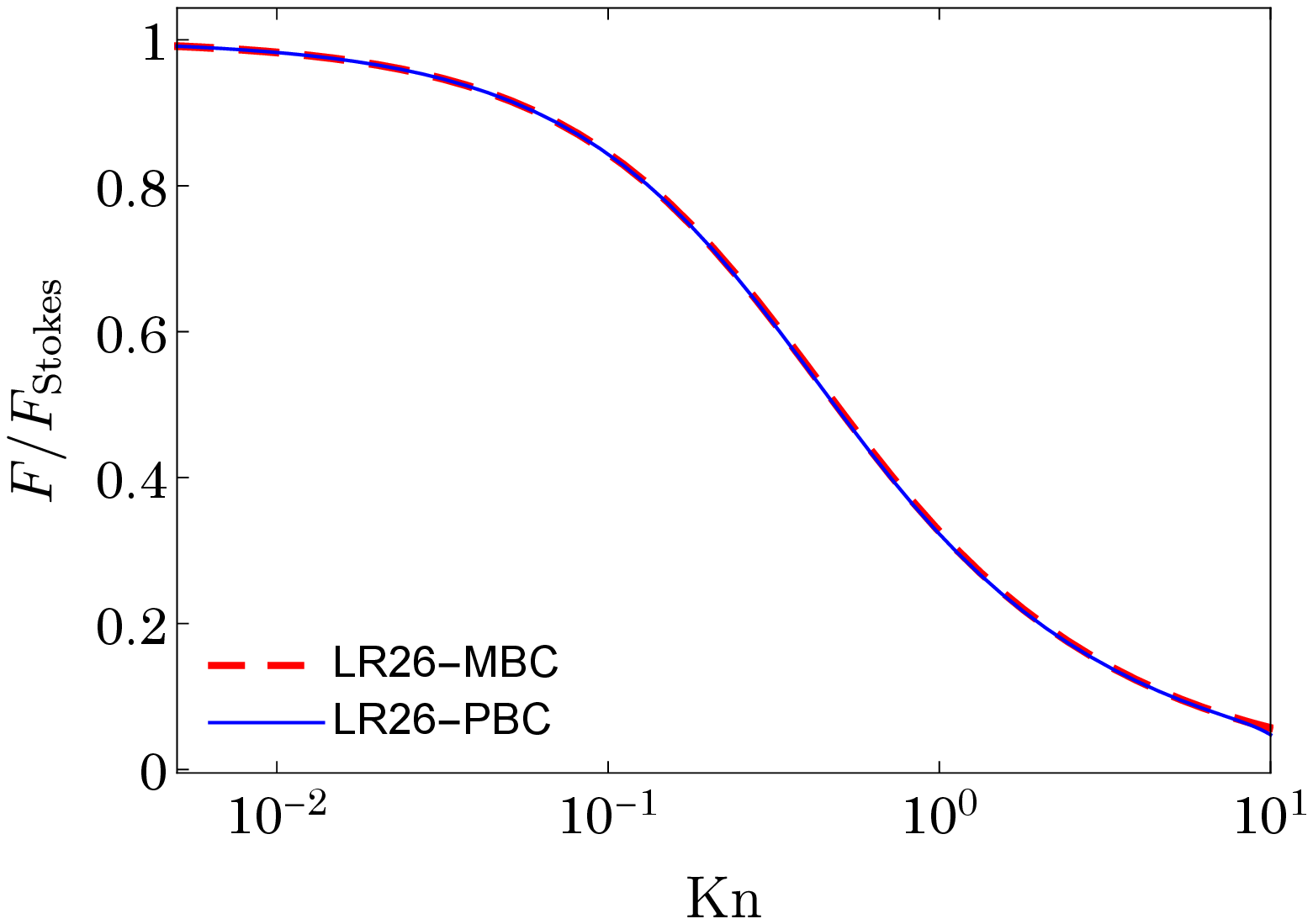}
\caption{\label{drag_R26MBCvsPBC} Normalized drag force from the LR26 theory with the PBC (solid blue line) and MBC (dashed red lines) as a function of the Knudsen number in the case of (left) a gas flow past a rigid sphere and (right) a vapour flow past an evaporating droplet. 
The results are normalized with the Stokes drag $F_{\mathrm{Stokes}}=6\pi\mathrm{Kn} u_\infty$.}
\end{figure}

Figure~\ref{drag_R26MBCvsPBC} displays the normalized (with the Stokes drag $F_{\mathrm{Stokes}}=6\pi\mathrm{Kn} u_\infty$) drag force obtained from the LR26 theory with the PBC (solid blue line) and MBC (dashed red lines) as a function of the Knudsen number for the aforementioned problems. 
Also, the drag force obtained from the LR26 equations with the PBC and MBC are virtually indistinguishable for both problems.
